%% file: CTA_paper_for_arxiv.tex
\documentclass{article}
\usepackage{geometry}
\usepackage[margin=1.5cm]{caption}

\usepackage{makeidx}         % allows index generation
\usepackage{graphicx}        % standard LaTeX graphics tool
\usepackage[bottom]{footmisc}% places footnotes at page bottom

\usepackage{rotating}
\usepackage{epsfig}
\usepackage{enumitem}

\usepackage{amsmath}
\usepackage{amssymb}

\usepackage[square,numbers]{natbib}
\makeindex             % used for the subject index
\begin{document}
%\tableofcontents{}
\title{The Cherenkov Telescope Array}
% Use \titlerunning{Short Title} for an abbreviated version of
% your contribution title if the original one is too long
\author{Werner Hofmann\thanks{corresponding author} ~and Roberta Zanin}
% Use \authorrunning{Short Title} for an abbreviated version of 
% your contribution title if the original one is too long
%\institute{Werrner Hofmann \at Max-Planck-Institut f\"ur Kernphysik, Postfach 103980, 69029 Heidelberg, Germany, \email{werner.hofmann@mpi-hd.mpg.de}
%\and Roberta Zanin \at CTA Observatory, Via P. Gobetti 93, 40129 Bologna, Italy, \email{roberta.zanin@cta-observatory.org}}
%
% Use the package "url.sty" to avoid
% problems with special characters
% used in your e-mail or web address
%
\maketitle
\abstract{The Cherenkov Telescope Array Observatory (CTAO) is a next-generation facility for ground-based very high energy gamma ray astronomy.
CTAO will be operated as an open observatory.
With two sites, in the northern and southern hemispheres, the Cherenkov Telescope Array CTA will provide full-sky coverage, improving sensitivity by an order of magnitude
over current instruments, with a wide gamma ray energy coverage from 20 GeV to 300 TeV. CTA will use telescope arrays composed 
of three types of telescopes, optimized to cover different energy ranges. The large telescopes covering the lowest energies provide rapid 
slewing capability, for follow-up of transients. Key Science Projects (KSPs) are developed to form a significant part of the CTAO observing program
during the first decade of operation, providing legacy data sets such as surveys or deep observations of key targets. }

%\section{Keywords} 
%Gamma rays, Imaging Atmospheric Cherenkov Telescopes, Cherenkov Telescope Array Observatory, very %high energies, astronomical observatory,  transient follow up, Key Science Projects.

\tableofcontents

%\chapter{The Cherenkov Telescope Array}

\section{Science motivation for  the Cherenkov Telescope Array (CTA)}
\label{SecMotivation}
\input{SecMotivation.tex}

\section{CTA concept and history}
\label{SecConceptHistory}
This section serves to introduce the concepts and strategies underlying the design of CTA, and to summarize the evolution of the project.
\input{SubSecConcept.tex}
\input{SubSecHistory.tex}

\section{Telescope arrays}
\label{SecArrays}
\input{SecArrays.tex}

\section{CTA Observatory}
\label{SecObservatory}
\input{SecObservatory.tex}

\section{CTA science performance and key science}
\label{SecPerfScience}

\subsection{Instrument performance}
\label{SubSecPerformance}
\input{SubSecPerformance.tex}

\subsection{Key Science Projects}
\label{SubSecKSP}
\input{SubSecKSP.tex}

\subsection{Science Performance -- selected topics}
\label{SubSecScPerf}

CTA science capabilities are illustrated here for few selected examples, based on  \cite{CTAConsortium:2017dvg} and on in-depth follow-up studies such as  \cite{2021JCAP...02..048A,2021JCAP...01..057A}.

\input{SubSubSecGPS.tex}
\input{SubSubSecAGN.tex}
\input{SubSubSecEBL.tex}

\input{SubSubSecDM.tex}

\section{Conclusions}
\input{SecConclusions.tex}

\section*{Acknowledgements}
\input{Acknowledgements}

\newcommand{\apjs}{{\it Astrophys. J. Supp. Series}}
\newcommand{\aap}{A\&A}
\newcommand{\jcap}{Journal of Cosmology and Astroparticle Physics}
\bibliographystyle{unsrt}
\bibliography{Citations_WH.bib}

\end{document}

%% file: SecMotivation.tex
% need to add references

The Cherenkov Telescope Array Observatory (CTAO) is a next-generation facility for observing very high energy cosmic gamma rays. `Very high energy' here refers to the Teraelectronvolt (TeV) energy domain, give or take 2 orders of magnitude in energy. With energies well above any thermal energy scales in the current Universe, cosmic gamma rays in this energy domain are produced by non-thermal mechanisms, when high-energy cosmic particles interact with interstellar gas or with radiation fields, or when heavy relics left over from the Big Bang annihilate or decay. The spectrum of gamma rays reflects that of the primary particles; for nuclei interacting with target gas, the spectrum of gamma rays resembles that of the primary particles, shifted down in energy by one order of magnitude. Beyond interacting nuclei, a second main mechanism for gamma ray production is the interaction of high-energy electrons or positrons with radiation fields via the Inverse-Compton mechanism; here, a power-law spectrum of electrons with index $\Gamma$ results in a gamma ray spectrum with index $(\Gamma +1)/2$, up to the point when the Klein-Nishina domain is reached (the center of mass energy of the interaction significantly exceeds $m_e c^2$), when the gamma-ray spectrum steepens. Very high energy gamma rays hence trace populations of high-energy particles in the Universe, and serve to locate cosmic particle accelerators, where the density of particles is high, or locations where relic particles accumulate. The charged particles impinging on Earth cannot be used to locate cosmic accelerators, since they propagate diffusively in interstellar and intergalactic magnetic fields, and are significantly deflected except at ultra-high energies well beyond $10^{20}$\,eV.

Starting with the discovery of very high energy gamma rays from the Crab Nebula in 1989 by the Whipple telescope \cite{CrabWhipple}, Cherenkov telescopes such as the H.E.S.S. \cite{HESS}, MAGIC \cite{MAGIC} and VERITAS \cite{VERITAS} telescope systems -- in the following referred to as `current IACTs' -- and lately particle arrays such as HAWC \cite{HAWC} and LHAASO \cite{LHAASO} have driven the rapid evolution of very high energy gamma ray astronomy. Over 200 sources of very high energy gamma rays were detected, associated with objects such as supernova remnants, pulsar wind nebulae, binary systems, novae, stellar clusters and star-forming regions, star-forming galaxies, active galactic nuclei, or gamma-ray bursts \cite{tevcat}. Very high energy particle acceleration has proven to be ubiquitous throughout the Universe. Today's instruments provide gamma ray sky maps with 5' angular resolution, energy spectra that cover three decades in gamma-ray energy, and light curves showing variability on all scales from sub-minute scales to years.

Despite this remarkable evolution of very high energy gamma ray astronomy, current IACTs are not sensitive enough and versatile enough to decisively answer some of the key questions. For example, a quantitative understand of cosmic ray origin -- which sources contribute at which level -- is lacking, and the origin of PeV-scale cosmic nuclei is completely open. One reason could be that current IACTs probe typical sources only up to distances of few kpc, so only a small fraction of the Galaxy is explored and rare source types are likely to be missed. For particle acceleration around the black holes of Active Galactic Nuclei (AGN), and in their jets, the acceleration mechanism and in particular the origins of the observed rapid variability remain to be understood. Gamma rays also serve to probe annihilation signatures of WIMP Dark Matter -- with the Galactic Center as the most promising region -- but limits on annihilation rates of TeV-scale particles are still above the predicted rates (as of 2021), and cannot confirm or exclude models.

The design of the Cherenkov Telescope Array Observatory as a next generation facility is driven by the following main science themes \cite{CTAConsortium:2017dvg}:
\begin{itemize}
\item {\bf Theme 1: Understanding the Origin and Role of Relativistic Cosmic Particles,} addressing questions such as:
What are the sites of high-energy particle acceleration in the universe?
What are the mechanisms for cosmic particle acceleration?
What role do accelerated particles play in feedback on star formation and galaxy evolution?
\item {\bf Theme 2: Probing Extreme Environments,} with topics including: 
What physical processes are at work close to neutron stars and black holes?
What are the characteristics of relativistic jets, winds and explosions?
How intense are radiation fields and magnetic fields in cosmic voids, and how do these evolve over cosmic time?
\item {\bf Theme 3: Exploring Frontiers in Physics,} aiming at answering:
What is the nature of dark matter? How is it distributed?
Are there quantum gravitational effects on photon propagation?
Do axion-like particles exist?
\end{itemize}
The `Science with CTA' book \cite{CTAConsortium:2017dvg} provides details on these themes, and on the Key Science Projects (KSPs) that serve to provide the data to address them.
%Examples of the science performance of CTA are given in Section~\ref{},  KSPs are discussed in Section~\ref{}.

Towards addressing these science themes, the following targets for the design of CTA were identified \cite{CTAConsortium:2017dvg}:
\begin{itemize}
\item improve the sensitivity level of current IACTs by an order of magnitude at TeV energies,
\item significantly boost detection area, and hence photon rate, providing access to the shortest timescale phenomena,
\item substantially improve angular resolution and field of view and hence ability to image extended sources,
\item provide energy coverage for photons from 20 GeV to at least 300 TeV, to give CTA reach to high-redshifts and extreme accelerators,
\item dramatically enhance surveying capability, monitoring capability, and flexibility of operation, enabling simultaneous observations of objects in multiple fields,
\item serve a wide user community, with provision of data products and tools suitable for non-expert users, and
\item provide access to the entire sky, with sites in two hemispheres. 
\end{itemize}

In the following, the concept of CTA and the historical evolution of CTA are summarised (Section~\ref{SecConceptHistory}). Section \ref{SecArrays} describes the layout of the telescope arrays, the characteristics of the telescopes, and the sites where arrays are located. CTA will be operated as an open observatory; relevant aspects and CTAO organisation are presented in Section \ref{SecObservatory}. Section~\ref{SecPerfScience} finally addresses CTA instrument performance, science performance and KSPs.

%% file: SubSecConcept.tex
\subsection{CTA concept}
\label{SubSecConcept}

The CTA concept is based on an evolution of the proven and highly successful concept of arrays of Cherenkov telescopes, imaging several views of an air shower to allow the three-dimensional reconstruction of the shower trajectory, the measurement of the energy of the incident particle  initiating the shower, and -- within limits -- the identification of its type, in particular distinguishing electromagnetic showers from hadron-induced showers.

Systems of Cherenkov telescopes such as H.E.S.S., MAGIC or VERITAS \cite{HESS,MAGIC,VERITAS} consist of few telescopes, with a spacing comparable to the radius of the Cherenkov light pool. This limits the effective detection area to roughly the size of the light pool, once a coincidence between telescopes is required. It also limits the stereo angle under which showers are viewed, and only for a modest fraction of events all telescopes see the shower. The CTA concept is based on a grid of Cherenkov telescopes that covers an area large compared to the Cherenkov light pool, while maintaining a spacing of the scale of the light pool radius. This has several consequences:
\begin{itemize}
\item the effective detection area is determined by the area of the telescope grid, rather than the size of the light pool, and grows linearly with the number of telescopes;
\item a shower whose light pool is contained in the telescope grid is viewed by a larger number of telescopes, providing improved reconstruction compared to telescope systems like H.E.S.S. MAGIC, or VERITAS, improved angular resolution and improved background rejection;
\item for a shower whose light pool is contained in the grid, there is always one or more telescopes in the region of maximal light intensity, lowering the effective energy threshold. 
\end{itemize}
In the regime where sensitivity is limited by the signal-to-background ratio -- typically up to a few TeV -- the sensitivity asymptotically grows with the square root of the array area, and hence -- for fixed telescope spacing -- with the square root of the number of telescopes. In the transition regime from few to many telescopes, however, the various effects combine to provide a faster growth of sensitivity with telescope number.

The wide energy range targeted for CTA -- from 20 GeV to beyond 300 TeV -- places a further challenge. To detect 20 GeV showers, telescopes of the 20 m class are required. On the other hand, to detect a sufficient number of 300 TeV gamma rays, array areas in the 10\,km$^2$ range are needed. Towards a cost-effective solution, the CTA energy range is covered by three types of telescopes, with 23-m Large-Sized Telescopes (LST) detecting the lowest energies over a relatively modest area of about 10$^5$\,m$^2$, 12-m Medium-Sized Telescopes covering the core energy range from about 100\,GeV to several TeV for a km$^2$-sized detection area, and 4-m Small-Sized Telescopes detecting multi-TeV showers of a multi-km$^2$ area. The LSTs -- in the baseline design four on each site -- work like the H.E.S.S. or VERITAS telescope systems, but with lower energy threshold due to their larger size. The MSTs and SSTs aim to realize the above-discussed concept of large grids, with corresponding gains in sensitivity. The combined array provides a detection area that steadily increases with energy, compensating for the gamma flux that is rather steeply falling with energy, for a very wide usable energy range over which sources can be detected. 

Another key ingredient towards CTA performance are improvements in the design and performance of individual telescopes, in particular
\begin{itemize}
\item photo sensors with improved quantum efficiency, such as photomultiplier tubes with about 40\% peak quantum efficiency (LST, MST), or silicon photo sensors (SST),
\item an increased field of view compared to current telescopes (for MST and SST), using (for the SST) dual-mirror optics for improved imaging over a large field of view,
\item very high slewing speeds for the LST, for rapid follow-up of transients,
\item improved reliability and maintainability of the telescopes, for efficient observatory operation.
\end{itemize}
The layout of the arrays and the design of the different telescopes is detailed in Section~\ref{SecArrays}.

Essential element of the concept is that -- a first in ground-based gamma ray astronomy -- CTA will be operated as an observatory, open to external users, and providing users with fully reconstructed air showers and science tools to generate spectra, sky maps and light curves. CTA will be able to respond to alerts from other instruments, and will be able to issue alerts on its own, both on sub-minute time scales.

%% file: SubSecHistory.tex
\subsection{CTA history}
\label{SecHistory}

Already in 1992, at the time when just the second TeV gamma-ray source -- Mrk 421 -- was discovered by the 10m-diameter Whipple imaging Cherenkov telescope, the workshop `Towards a Major Atmospheric Cherenkov Detector' at Palaiseau/France \cite{1992tmac.conf.....F} discussed future powerful Cherenkov instruments; contributions included the discussion of large arrays of imaging Cherenkov telescopes. Instead of a single large instrument, however, three major new instruments emerged in the coming years: MAGIC \cite{MAGIC}, a 17\,m telescope on La Palma inaugurated in 2003, later upgraded with a 2nd 17\,m telescope; H.E.S.S. \cite{HESS}, a system of four Wipple-sized telescopes in Namibia inaugurated in 2004, originally planned as the first stage of a 16-telescope system but later upgraded by adding a 28\,m telescope; and VERITAS \cite{VERITAS}, also a system of four Whipple-sized telescopes in Arizona inaugurated in 2007, originally planned as a 7-telescope system. After very successful operation and science with these instruments, activities towards a next-generation system in Europe were triggered by the APPEC
\footnote{APPEC -- the Astroparticle Physics European Consortium -- is a consortium of 19 funding agencies, national government institutions, and institutes from 17 European countries (Status 2021)} roadmapping process for astroparticle physics in Europe, and by the 2004 announcement of plans for the ESFRI\footnote{ESFRI -- the European Strategy Forum on Research Infrastructures -- is a forum of government representatives to develop the scientific integration of Europe} European Roadmap of large-scale research infrastructures. In late 2005,  H.E.S.S. and MAGIC scientists submitted to ESFRI a project concept for the Cherenkov Telescope Array (CTA). CTA was indeed included as an `emerging proposal' in the 2006 ESFRI Roadmap, was highly ranked in the first APPEC Roadmap in 2008/9, and was listed as a regular project in the 2008 update of the ESFRI Roadmap. After initial community meetings in 2007, the CTA Consortium (CTAC) was formally established in 2008, and over the years grew to more than 1500 members from more than 200 institutes in over 30 countries, in 2022. 

A parallel process started in the US around 2005, discussing an array of telescopes using the novel Schwarzschild-Couder dual-mirror optics for improved Cherenkov imaging \cite{2007APh....28...10V}, and resulting in 2007 in the proposal for the Advanced Gamma-ray Imaging System (AGIS) \cite{2010cosp...38.2317O}.
 Based on a recommendation by the 2010 Decadal Survey of Astronomy and Astrophysics in the US, the AGIS team merged with CTA, creating a word-wide unified effort towards a next-generation instrument.

The CTA Consortium both developed the science case for CTA and, jointly with the CTA project office (established in 2010), evolved the design and prototyping of the instrument. Major milestones towards documenting the science case were the CTA Astroparticle Physics special volume in 2013 \cite{2013CTA} and the `Science with CTA' book in 2017 \cite{CTAConsortium:2017dvg}.
Following on an initial presentation of design concepts in 2011 \cite{CTAConsortium:2010umy}, a CTA Preliminary Technical Design Report (PTDR) was released in 2013,  and a Technical Design Report (TDR) in 2015. As already mentioned, the design combined three different telescope sizes, 23-m Large-Sized Telescopes (LST), 12-m Medium-Sized Telescopes (MST), and 4-m Small-Sized Telescopes (SST), with 4 LSTs on both the Northern and Southern sites, 25 MSTs in the South and 15 MSTs in the North, and 70 SSTs in the South (see Section \ref{SecArrays}), for a total of 99 telescopes in the South and 19 in the North. The SSTs are mostly relevant for Galactic science, and are concentrated in the South. In reaction to funding limitations, in 2016 a `threshold' design for the arrays was defined, with 4 LSTs and 5 MSTs (rather than 4 + 15)  in the North, and 15 MSTs and 50 SSTs (rather than 4 LSTs + 25 + 70) telescopes in the South. The lack of LSTs in the South implied a temporary specialisation on extragalactic science in the North, and on higher-energy Galactic  sources in the South.

A parallel effort served to secure funding for CTA, and to establish a legal entity to build and operate the future observatory. Starting from late 2010, the CTA Preparatory Phase was supported by a grant from the European Commission, allowing the creation of a central project office located in Heidelberg. The CTA Resource Board (RB) of agency representatives was founded in 2011; in July 2012 representatives from 13 countries signed a Declaration of Intent (DOI) where signatories expressed their common interest in participating in the construction and operation of CTA. In July 2014, agencies founded the CTA Observatory (CTAO) gGmbH with seat in Heidelberg, to provide a legal entity preparing the construction of CTA. CTAO gGmbH is governed by its Council. RB and CTAO Council co-existed for over a year, until late 2015.  The formal CTA array site selection process was initiated in 2012 by the RB; in July 2015 the RB decided to enter into detailed contract negotiations for hosting CTA on the European Southern Observatory (ESO) Paranal grounds in Chile and at the Roque de los Muchachos site in La Palma, Canary Islands. Site negotiations were concluded in December 2018 with the signature of the hosting agreement for the Chilean CTA site; the hosting agreement for the La Palma site was already signed in 2016. Details are given below in Section\,\ref{SubSecSites}. The search for the site of the future CTAO Headquarters and of the Science Data Management Centre was managed by the Council, with an call posted in late 2015. The decision in June 2016 to locate the CTAO Headquarters in Bologna/Italy was accompanied by the decision to create the CTAO European Research Infrastructure Consortium (ERIC) as final legal entity to build and operate the Observatory.  The decision to move to an ERIC was based on the necessity to have a legal entity recognized in all EU countries, still allowing the participation to non-EU countries, but with an establishing process faster than that of an international organisation. 
A Board of Government Representatives (BGR) was funded in 2018 towards preparing the ERIC. After a number of iterations, contributions and funding for the first stage of CTA -- the `alpha configuration', with 4 LSTs and 9 MSTs in the north, and 14 MSTs and 37 SSTs in the south, see Section~\ref{SubSecAlpha} -- were agreed in June 2021. The ERIC documents were submitted to the European Commission in mid-2022; establishment of the ERIC -- as of writing this chapter -- is expected for mid-2023. In parallel, pre-construction activities regarding site infrastructure started on both the Northern and Southern CTAO array sites, and the first LST telescope was deployed and commissioned on the Northern array site. In late 2022, as part of a special Italian program, funding was approved to provide two LSTs and 5 SSTs envisioned as additional telescopes for the Southern array, allowing it to cover the full energy range from 20 GeV to 300 TeV.

%% file: SecArrays.tex
This section describes the choice and optimisation of the layout of the telescope arrays, the telescope design and technology, the scheme for monitoring and calibration of the arrays, the array sites and their selection, and the `alpha' array configuration chosen for the initial implementation, with space for future enhancement of the arrays.

\subsection{Simulation and layout optimisation}

\begin{figure}[htbp]
\begin{center}
\includegraphics[width=11cm]{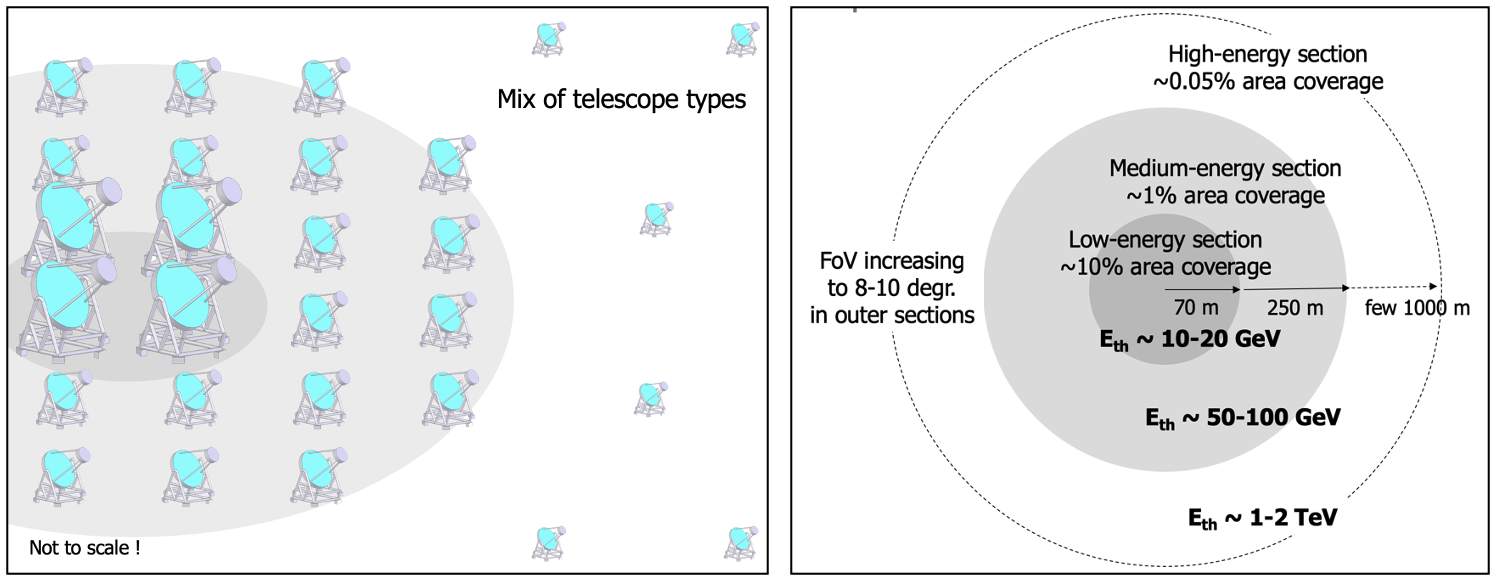}
\caption{Early concept of CTA, presented at an ESFRI meeting 2006 in Brussels.}
\label{fig_CTA_2006}
\end{center}
\end{figure}

Already the very first sketches of an array layout for CTAO bear significant similarity with the final layouts; a 2006 concept drawing (Fig.~\ref{fig_CTA_2006}) shows three zones with telescope of different sizes, the inner compact zone of very large telescopes providing very low energy threshold (10-20 GeV), the middle zone with 10-m class telescopes providing 50-100 GeV threshold over an area of about 200000 m$^2$, and the outer, multi-km$^2$ zone with small telescopes providing 1-2 TeV threshold. This concept was based on the experience with then existing telescope systems, coupled with the knowledge that covering more than one decade in energy with a given telescope type would likely not be cost-effective.

However, while designs look similar, it was a long effort to evolve from the concept to fully cost- and performance optimised array layouts and telescope designs. Details are given in \cite{2013APh....43..171B, 2017APh....93...76H, 2019APh...111...35A}, from which the following discussion is taken. 

In the initial design steps, a generic telescope cost model was used with mirror area, field of view and pixel size as primary parameters, allowing to compose different array layouts at constant total cost. 
As a first step, semi-analytical performance estimations were carried out using parameterisations for the responses of each telescope type. These studies allowed quick estimates of gamma-ray and cosmic-ray detection rates for a wide variety of telescope configurations and arrangements. Simulations of regular square grids of telescopes were performed to quantify the impact of parameters such as mirror area, field of view, pixel size or telescope spacing.

To validate and fine tune the telescope configurations calculated with these simplified approaches, a series of five large-scale MC simulations were performed, in each step refining the modelling of telescopes based on the concurrently evolving telescope designs and telescope prototypes. The full-scope design number of telescopes (4 LSTs, 25 MSTs and 70 SSTs for the Southern array and 4 LSTs and 15 MSTs for the Northern array) was fixed after a combined effort involving the production of large-scale MC simulations, evaluation of the performance of very different array layouts, and study of the effect of this diverse set of layouts over a large variety of key scientific cases  (see \cite{2019APh...111...35A} and references given there). Once telescope numbers were fixed, and given the arrangement of telescope types in roughly concentric circles, the main free parameter is the telescope spacing, with a large spacing optimising high-energy detection rates at the expense of low-energy response and number of telescopes viewing a given shower, relevant for angular resolution and cosmic-ray background rejection. To support the cross-calibration between different types of telescopes, one/two units are added almost at the center of the closer telescope-type sub-array. 

\begin{figure}[htbp]
\begin{center}
\includegraphics[width=3.7cm]{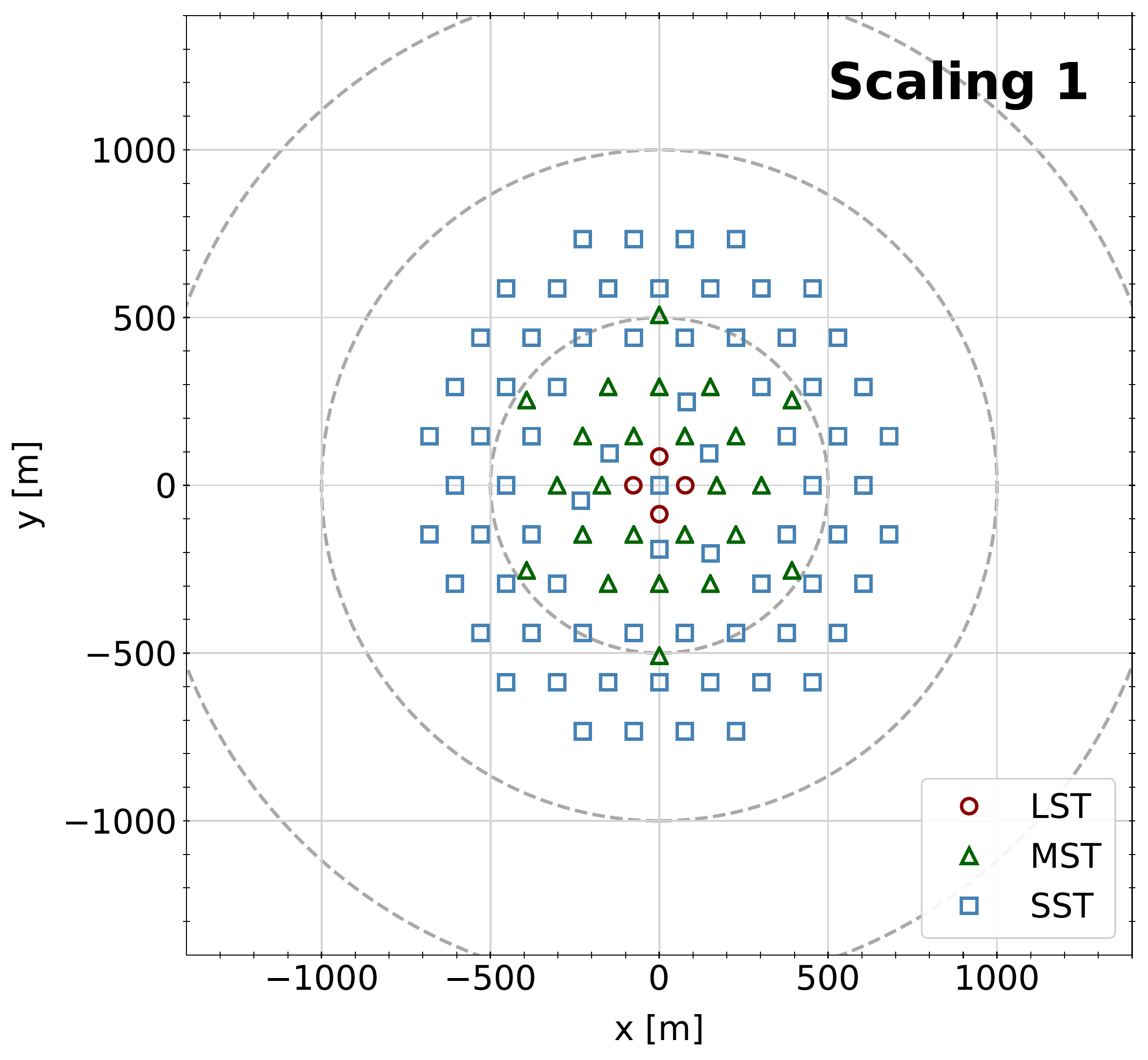}
\includegraphics[width=3.7cm]{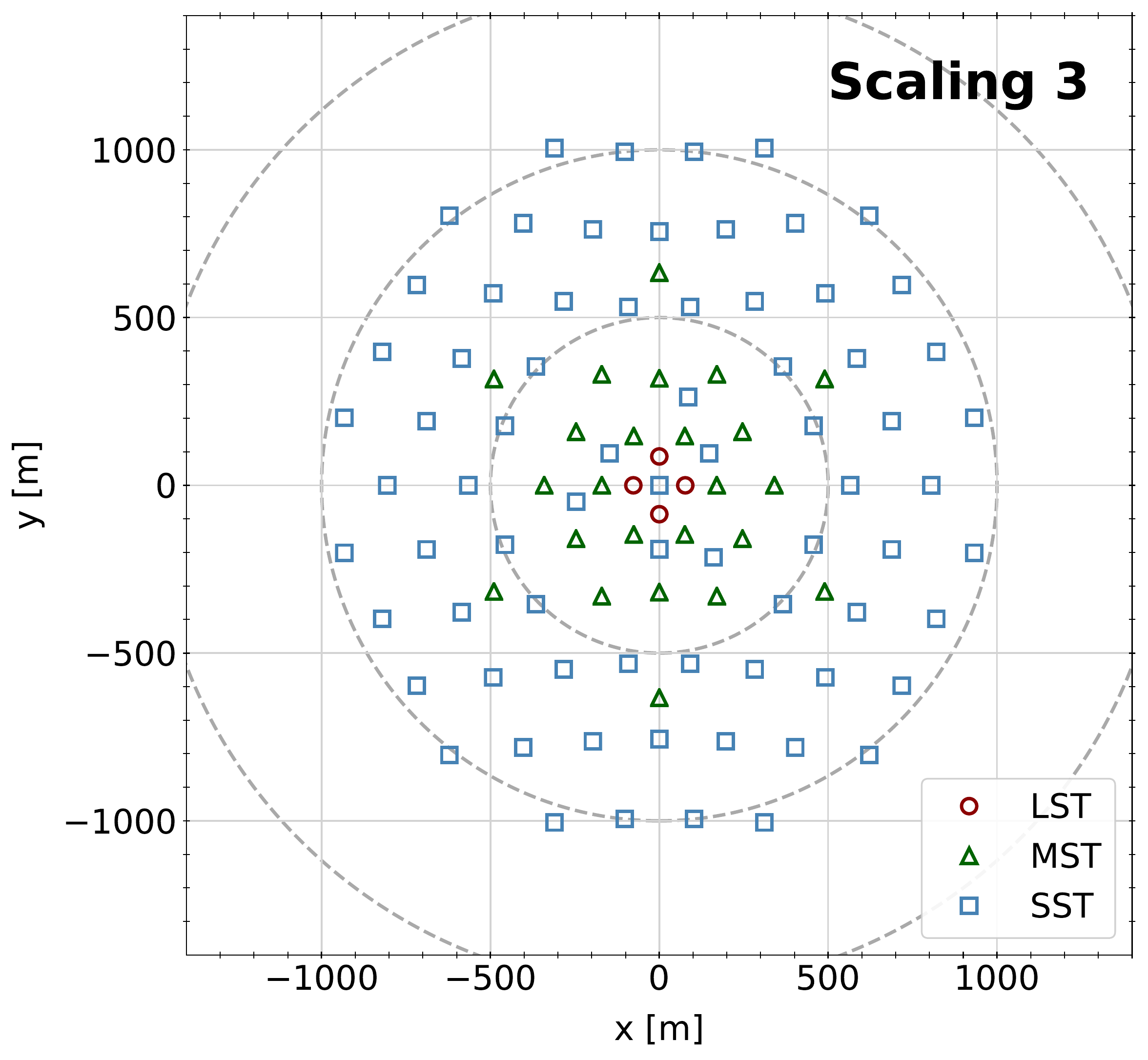}
\includegraphics[width=3.7cm]{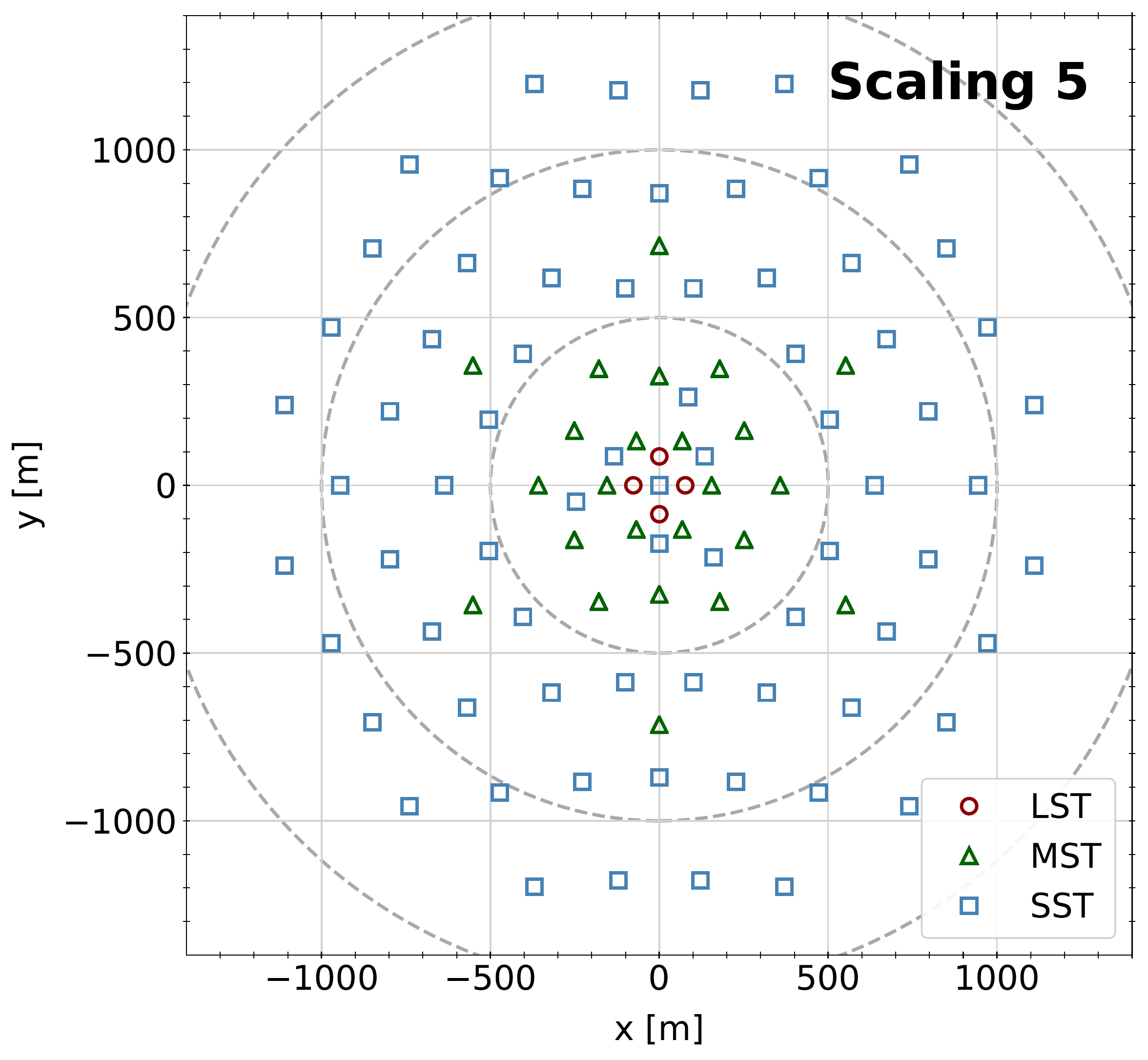}
\caption{ Examples of the three scaled layouts for one of the Paranal site candidates (“S1”). LST positions are indicated by red circles, MSTs by green triangles, and SSTs by blue squares. From \cite{2017ICRC...35..811C}.}
\label{fig_scaling}
\end{center}
\end{figure}

Each of the large-scale Monte Carlo simulations used `master layouts' with 100s of telescopes, distributed over areas up to 6 km$^2$, designed to contain numerous possible CTA layouts of equivalent cost. To identify the optimal arrangement, telescope subsets corresponding to plausible layouts were extracted, analysed and their performance was compared with respect to each other. Air showers initiated by gamma rays, cosmic-ray nuclei
and electrons were simulated using the CORSIKA package \cite{1998cmcc.book.....H}. The telescope response was simulated using the {\it sim\_telarray} package \cite{2008APh....30..149B}. The performance of each telescope layout was estimated by analysing these data using reconstruction methods  developed for the H.E.S.S., MAGIC and VERITAS IACT systems, and adapted for analysis of the CTAO arrays. The first large-scale production (Prod1, 2009) covered a wide range of different layouts \cite{2013APh....43..171B}, from very compact ones, focused on low energies, to very extended ones, focused on multi-TeV energies. The evaluation of these layouts, studying their impact on a range of science cases, resulted in a clear preference for intermediate layouts with a balanced performance over a wide energy range. The second large-scale production (Prod2, 2012-2014) refined the layout optimisation studies while placing additional emphasis on assessing the effect of site-related parameters on performance at the proposed sites to host the CTA Observatory; details are discussed below in the context of CTA site selection. The third large-scale production (Prod3, 2016-2020) was carried out for the primary CTA site candidates, Paranal (Chile) and La Palma (Spain). Telescope design configurations were updated and a significantly larger and more realistic set of available telescope positions were included. The aim of this production was to refine the optimisation, defining the final telescope layout for both CTA arrays by reducing the optimisation uncertainty to the few percent level. For the Paranal site alone this simulation required $\approx$ 120 million HEP-SPEC06 CPU hours and $\approx$ 1.4 PB of disk storage. Fig.~\ref{fig_scaling} illustrates three different scalings of radial telescope positions.

\begin{figure}[htbp]
\begin{center}
\includegraphics[width=7cm]{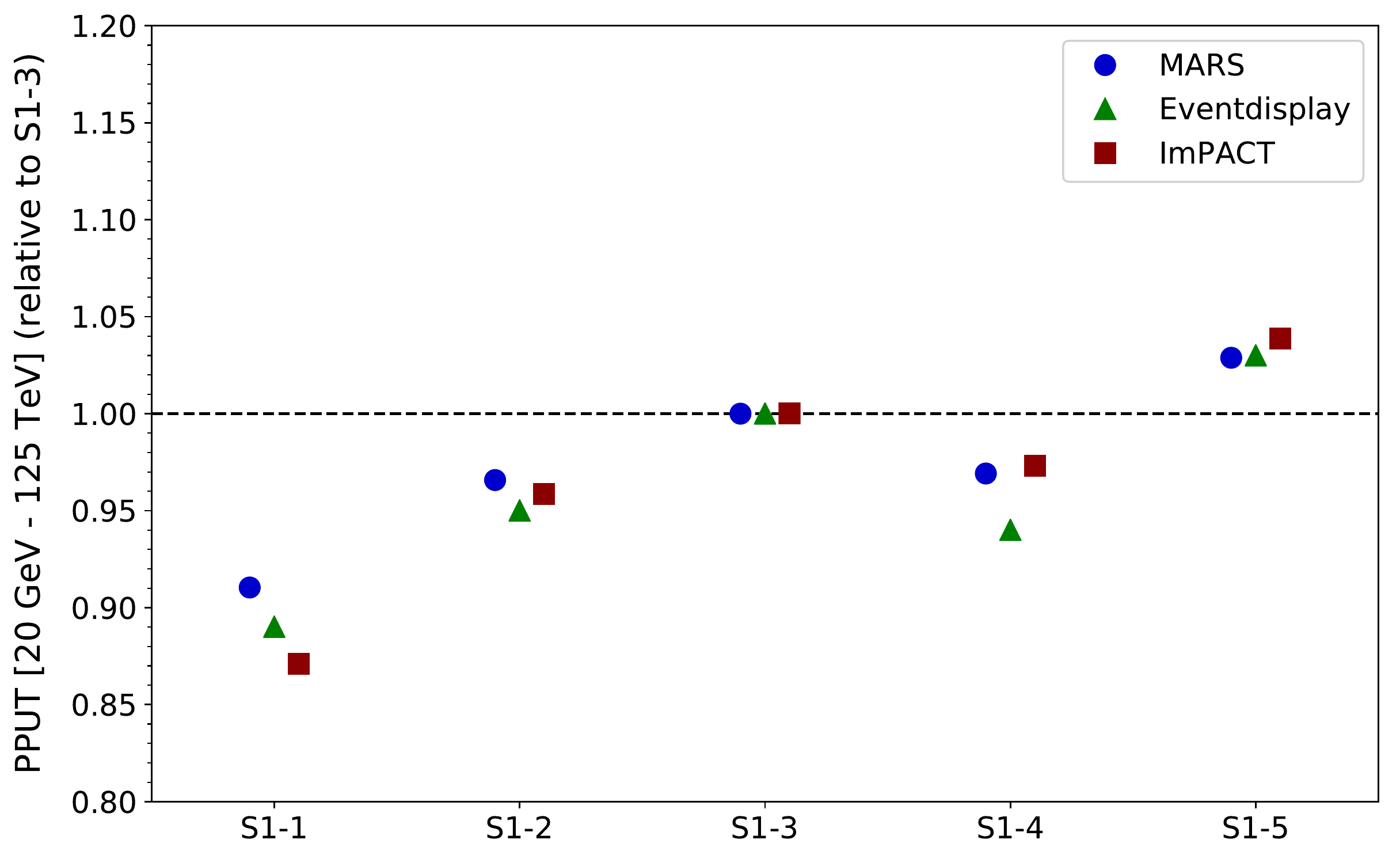}
\includegraphics[width=7cm]{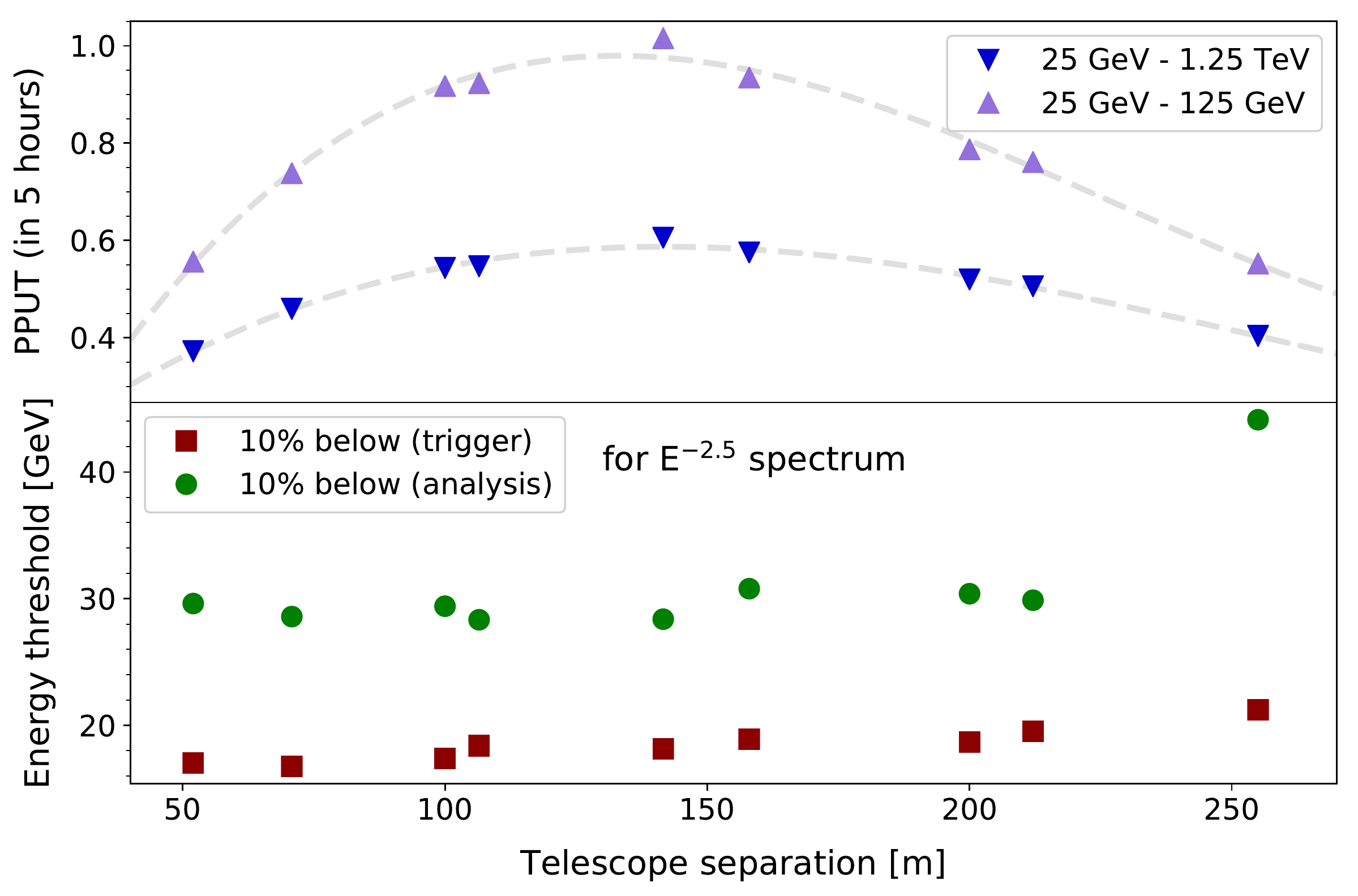}
\caption{Top: Spectrum-averaged sensitivity (PPUT) for five different array scalings (three of which are shown in Fig.~\ref{fig_scaling}, and for different analysis chains (``MARS'',``Eventdisplay",``ImPACT"). Bottom: PPUT and energy threshold of pairs of LSTs as a function of their separation. From \cite{2019APh...111...35A}.}
\label{fig_LST_separation}
\end{center}
\end{figure}

The primary parameter to evaluate the performance of array layouts was the differential sensitivity, defined as the minimal detectable flux in 0.2-decade energy bands, calculated from the simulated detection rates for gamma rays in a source region of size defined by the (energy-dependent) angular resolution, and the corresponding background rates from cosmic ray nucleon and electron-initiated air showers. As a single figure of merit, the spectrum-averaged sensitivity was used (Performance per unit time, PPUT). PPUT is defined as the (suitably normalized) geometric mean of the sensitivities in the individual 0.2-decade bands. PPUT may be calculated for the whole CTA-required energy range to estimate the overall performance, i.e. from 20 GeV up to 300 (50) TeV for the southern (northern) array, or for energy sub-ranges, to evaluate specific telescope sub-system capabilities. PPUT is defined such that a larger number corresponds to better performance \cite{2019APh...111...35A}. Independent analysis chains were used to process the full Monte Carlo production for a large number of telescope configurations for both the Paranal and La Palma sites. Fig. \ref{fig_LST_separation} illustrates how PPUT varies with scaling of telescope distances in the array, and with the separation between pairs of LSTs; also shown is the LST energy threshold. The figure illustrates that the array performance shows a very flat optimum; even vastly different layouts such as `Scaling 1' and `Scaling 5' (Fig.~\ref{fig_scaling}) differ in performance by only 15\%. Different analysis packages provide results consistent within few \%. The spacing of the LST telescopes is rather uncritical; there is a flat optimum between about 100\,m and 170\,m.

\begin{figure}[htbp]
\begin{center}
\includegraphics[width=5.5cm]{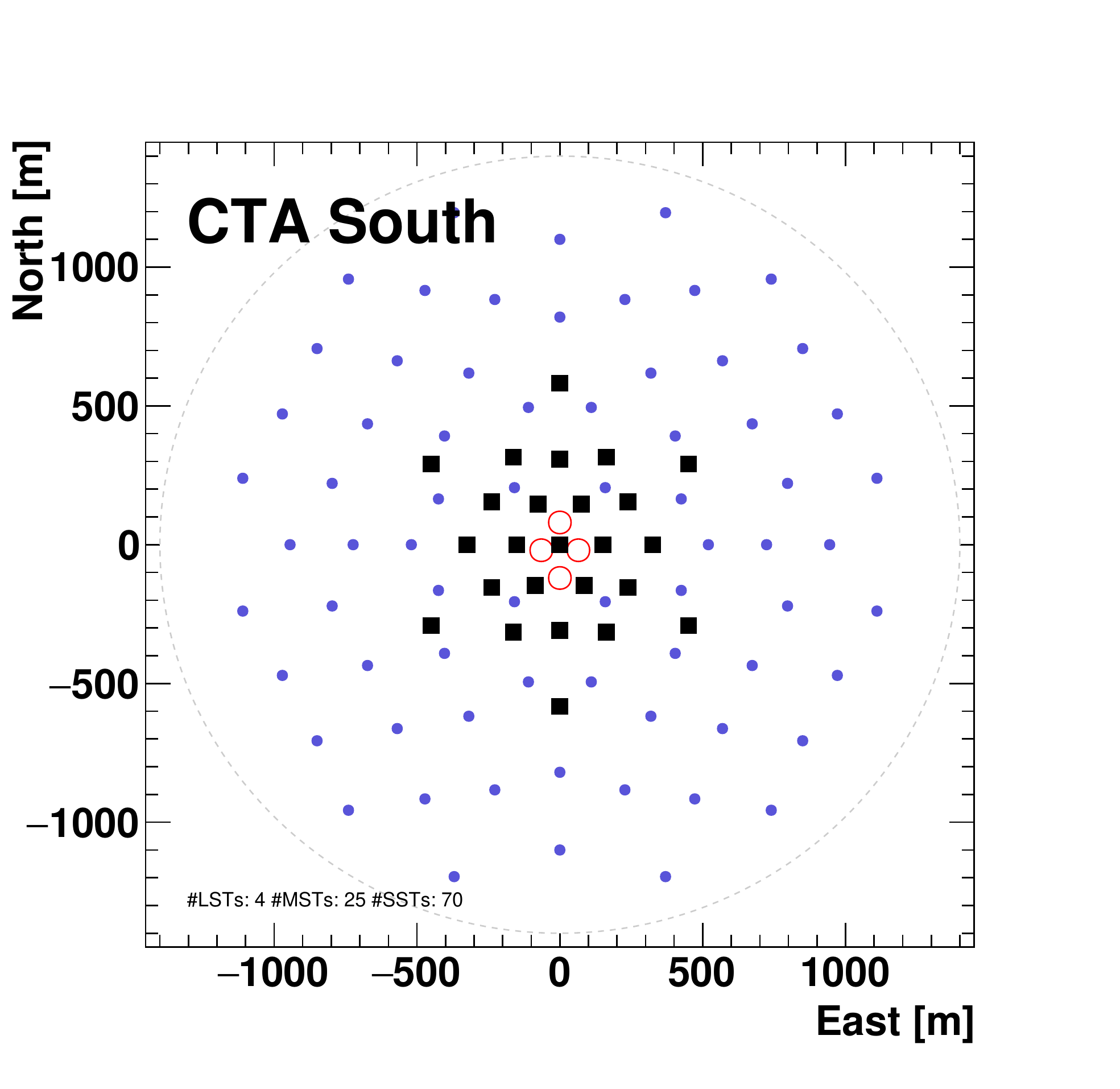}
\includegraphics[width=5.5cm]{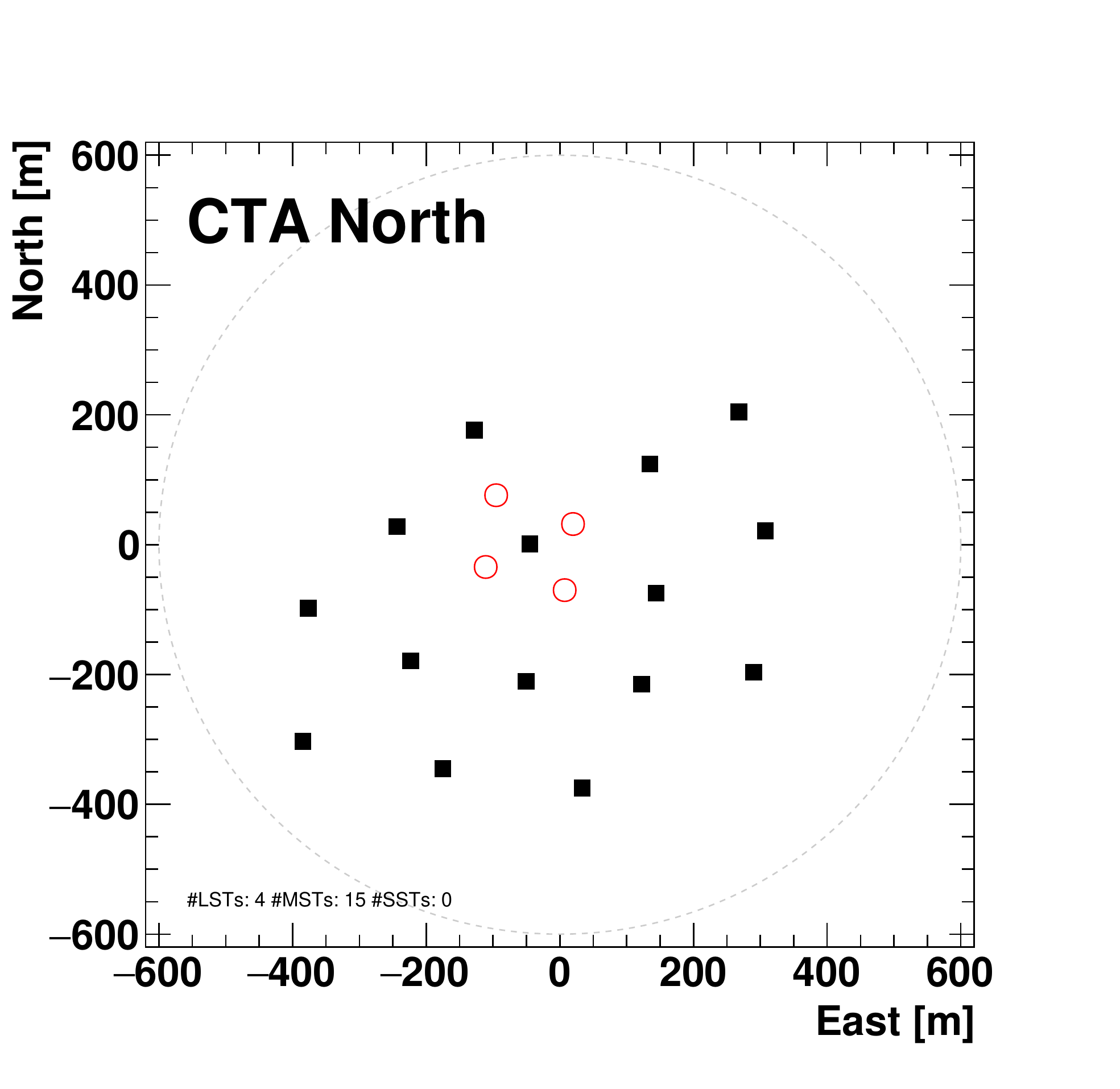}
\caption{Left: `omega configuration' array layout for the southern CTA array at Paranal with 4 LSTs (open circles), 25 MSTs (filled squares) and 70 SSTs (filled circles). Right: Layout for the northern CTA array on La Palma. From \cite{2019ICRC...36..733M}.}
\label{fig_prod3layouts}
\end{center}
\end{figure}

\begin{figure}[htbp]
\begin{center}
\includegraphics[width=4.5cm]{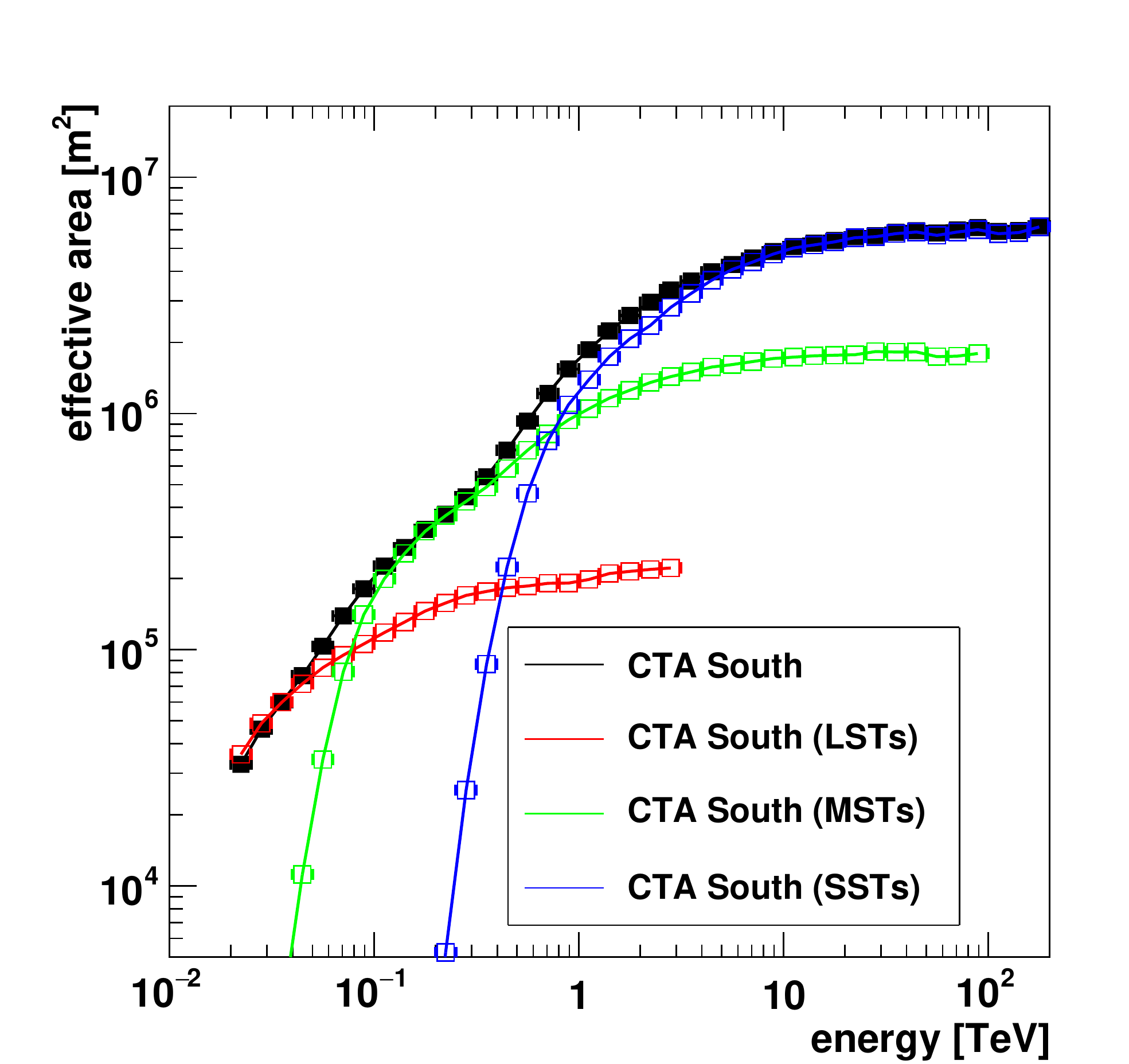}
\includegraphics[width=6.5cm]{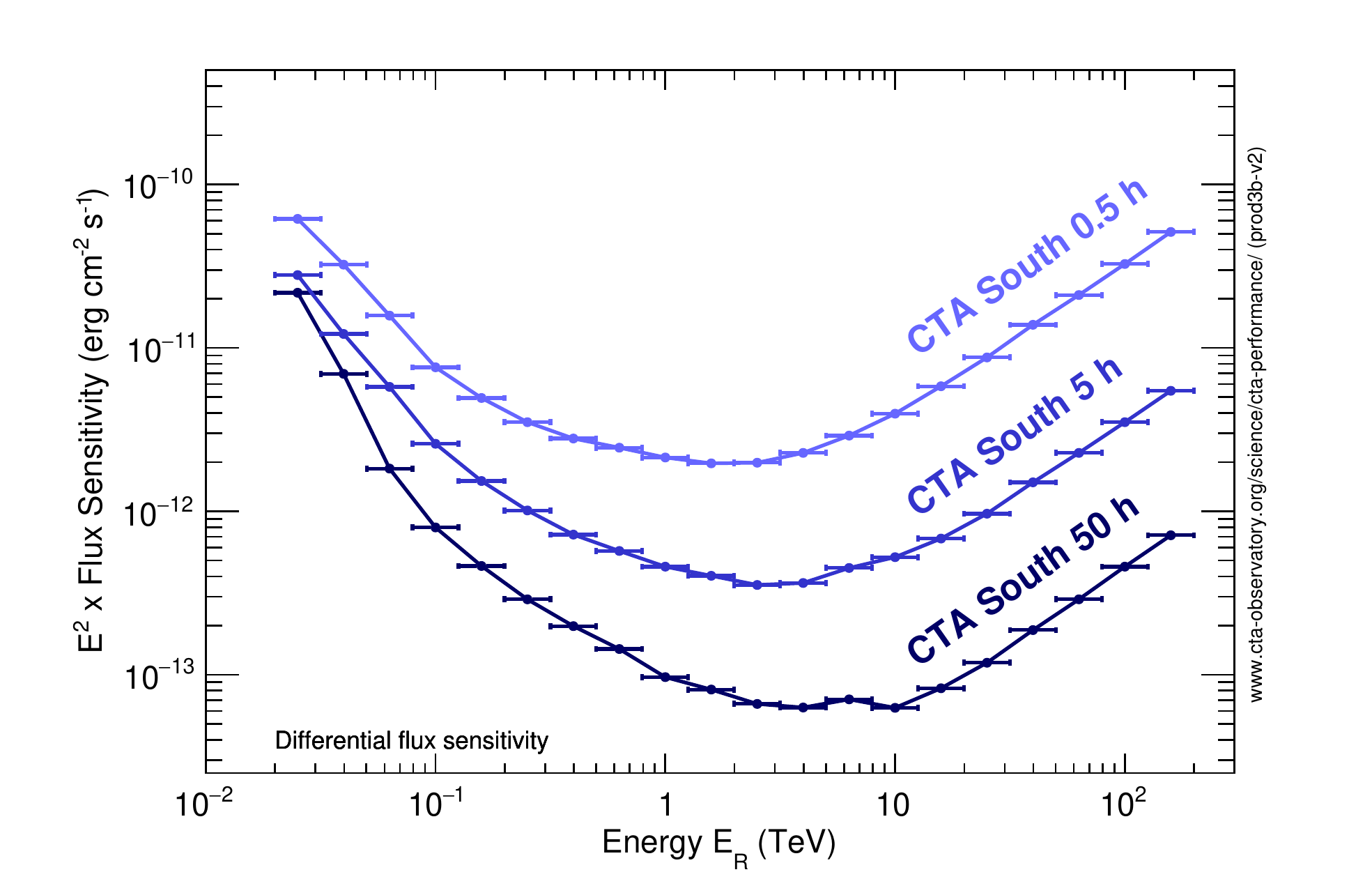}
\caption{Left: Effective gamma-ray detection area of the southern CTA array in its `omega' configuration, showing the contributions of the different telescope types. From \cite{2017ICRC...35..846M}.
Right: Corresponding differential flux sensitivity, for observing times of 0.5 h, 5 h and 50 h, at $20^\circ$ zenith angle. From \cite{2019ICRC...36..733M}.}
\label{Fig_effarea}
\end{center}
\end{figure}

Based on Prod3 simulations, optimal array layouts for the full scope configuration (later termed `omega configuration') were defined, and the resulting instrument response functions were made publicly available, allowing widespread simulation of CTA science cases. Fig.~\ref{fig_prod3layouts} shows the resulting array layouts. Fig.~\ref{Fig_effarea} (left) illustrates how the effective detection area of the southern array grows with energy, compensating the decline in the flux of typical gamma-ray sources and providing a large dynamic range in energy over which gamma-ray spectra can be followed. The LSTs provide low threshold but saturate at $2 \cdot 10^5$\,m$^2$ detection area; the MSTs kick in at about 100 GeV and saturate slightly above $10^6$\,m$^2$; SSTs take over above 1 TeV and at high energy provide detection areas in excess of $5 \cdot 10^6$\,m$^2$. Fig.~\ref{Fig_effarea} (right) shows the corresponding differential flux sensitivity as a function gamma-ray energy and observing time.

Monte Carlo production Prod4 (2018-2019) used refined models in particular for the SST telescopes, and served to derive the performance of  three different SST designs considered at that time (see Section \ref{SubsubsecSST}), towards a decision which design or combination of designs to implement.

The Prod5 Monte Carlo production (2020-2021) finally served to optimize the layout of the `alpha configuration' arrays with initially reduced numbers of telescopes (see below), also accounting for updated terrain-related constraints in the placement of the telescopes on the northern site. Telescope simulations were improved and tuned, taking into account e.g. the measured wavelength-dependent mirror reflectivities and their point spread functions; measured lightguide efficiencies; the measured sensor quantum efficiencies, pulse shapes, and afterpulse rates; improved simulations of electronics chains and digitisation and trigger systems; calibration procedures and uncertainties in resulting calibration coefficients. Between 50 and 100 parameters per telescope were determined from telescope prototypes, documented, and implemented in the simulations \cite{zenodoProd5}.

\subsection{Telescopes}
\begin{figure}[htbp]
\begin{center}
\includegraphics[width=11cm]{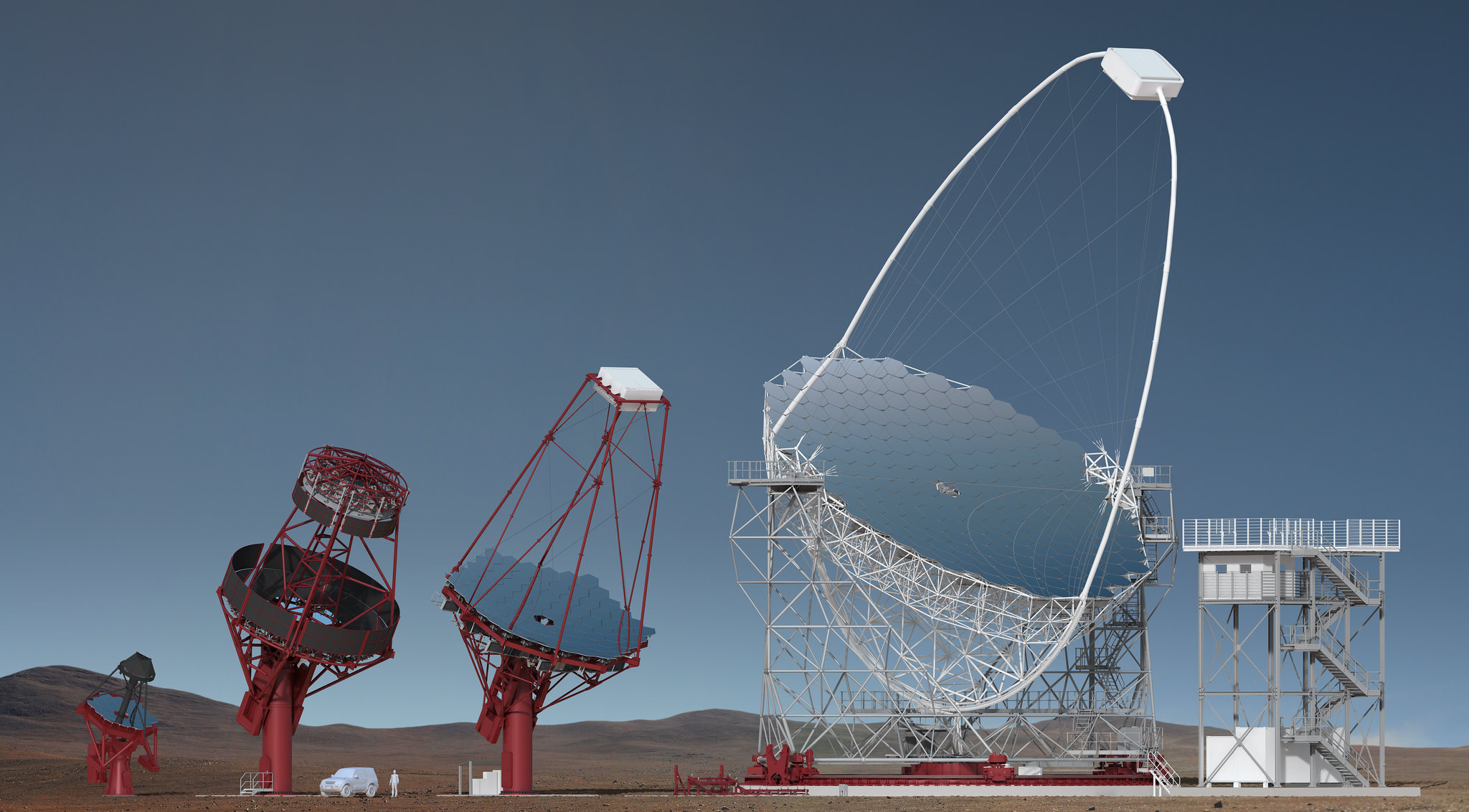}
\caption{CAD models of the different telescopes: SST, SCT, MST, LST (left to right). SSTs, MSTs and LSTs are included in the initial `alpha configuration' arrays, SCTs are planned for the enhancement stage. Credit: CTAO.}
\label{fig_telescopes}
\end{center}
\end{figure}

\begin{figure}[htbp]
\begin{center}
\includegraphics[width=11cm]{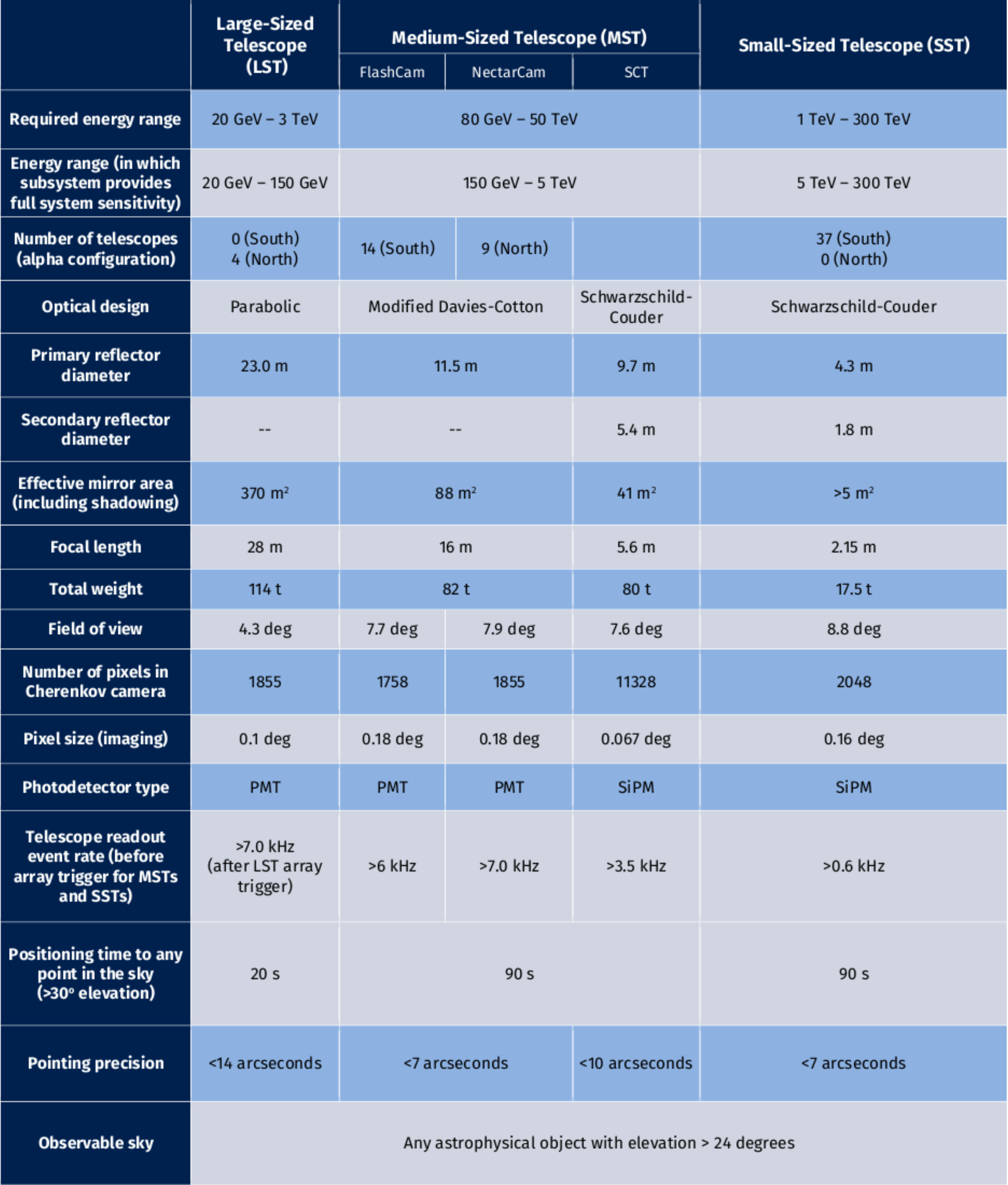}
\caption{Characteristics of CTA telescopes.}
\label{fig_telescopespecs}
\end{center}
\end{figure}

\begin{figure}[htbp]
\begin{center}
\includegraphics[width=5cm]{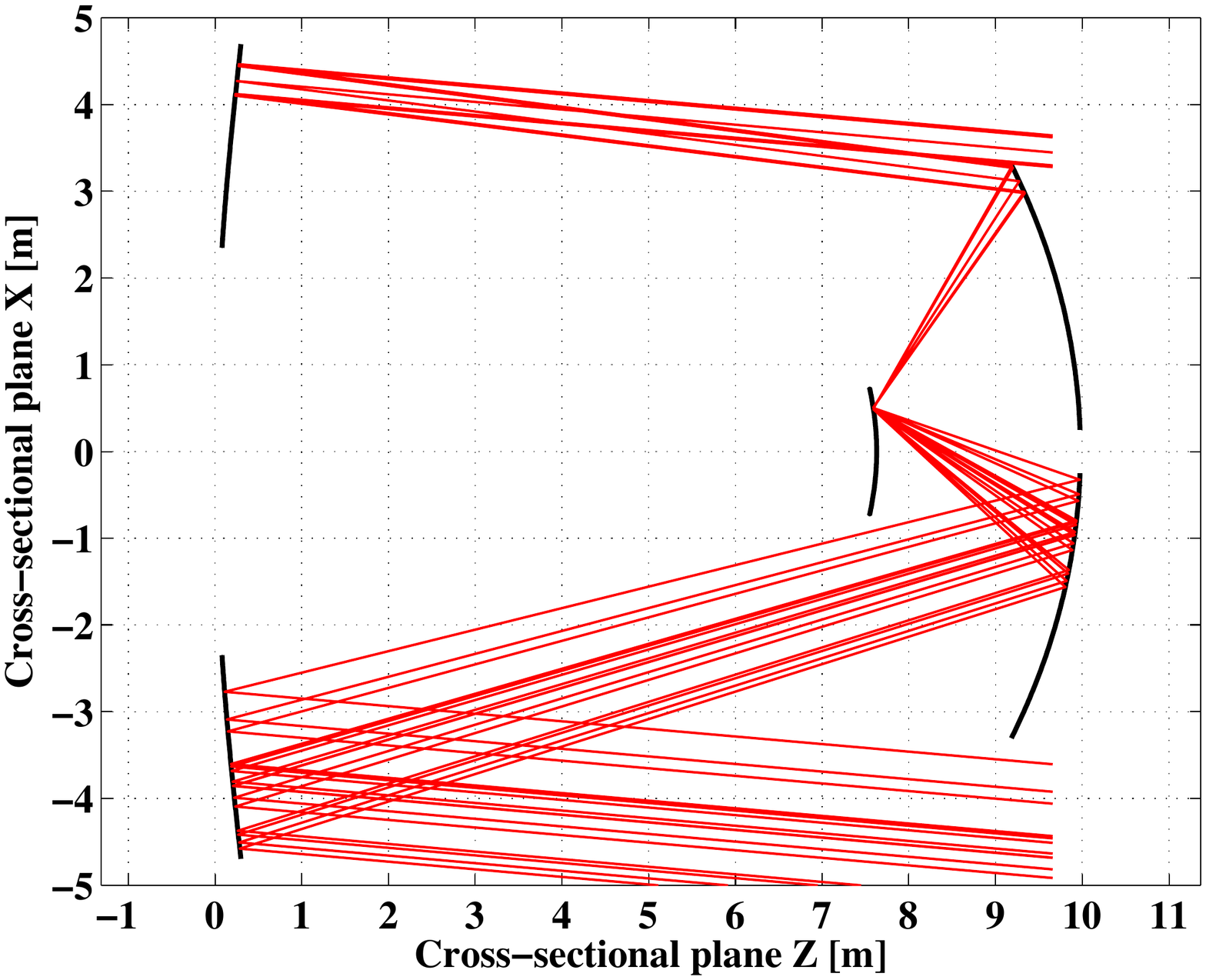}
\caption{Schwarzschild-Couder telescope optics with a primary (left) and secondary (right) mirror and compact focal plane, as used in the SST and SCT. From \cite{2007APh....28...10V}.}
\label{fig_SchwarzschildCouder}
\end{center}
\end{figure}

The design of the CTA project is based on the use of three different
types of telescopes, each of which optimised for its specific energy
range: the Large-Sized Telescopes (LST), the Medium-Sized Telescopes (MST) and the Small-Sized Telescopes (SST).
The different CTA telescopes and their sizes are illustrated in Fig.~\ref{fig_telescopes}.
%Several prototypes were designed for the MSTs and the SSTs and they
%will be illustrated in Fig.~\ref{fig_telescopes} and will be described in the following sections.
The key telescope specifications are summarized in Fig.~\ref{fig_telescopespecs}. 
% The different CTA telescopes and their sizes are illustrated in Fig.~\ref{fig_telescopes}. In their initial (`alpha') configurations, the arrays will include Large-Sized Telescopes (LST), Medium-Sized Telescopes (MST) and Small-Sized Telescopes (SST). Schwarzschild-Couder Telescopes (SCT) are planned to be added during a later enhancement phase. Key telescope specifications are summarized in Fig.~\ref{fig_telescopespecs}.
While MSTs and LSTs included in the `alpha configuration' represent an evolution of the single-mirror telescopes as employed in the H.E.S.S., MAGIC and VERITAS telescope arrays, the SSTs use an innovative Schwarzschild-Couder dual-mirror optics (Fig.~\ref{fig_SchwarzschildCouder}), that provides improved imaging and smaller point spread function over a large field of view by correcting comatic aberrations, and that results in a very compact focal plane instrumentation (`camera') tugged underneath the secondary mirror. To reduce astigmatism, the focal plane is curved. The small plate scale of the dual mirror telescope enables the use of silicon photosensors. The medium-sized Schwarzschild-Couder Telescopes (SCT) also use dual-mirror optics and are developed towards a later enhancement phase.

Compared to  IACTs such as H.E.S.S., MAGIC or VERITAS, the design of the CTA telescopes emphasises high reliability and good maintainability; the specification for telescope uptime requires about 98\% availability. 

\subsubsection{Large-Sized Telescope (LST)}

\begin{figure}[htbp]
\begin{center}
\includegraphics[width=10cm]{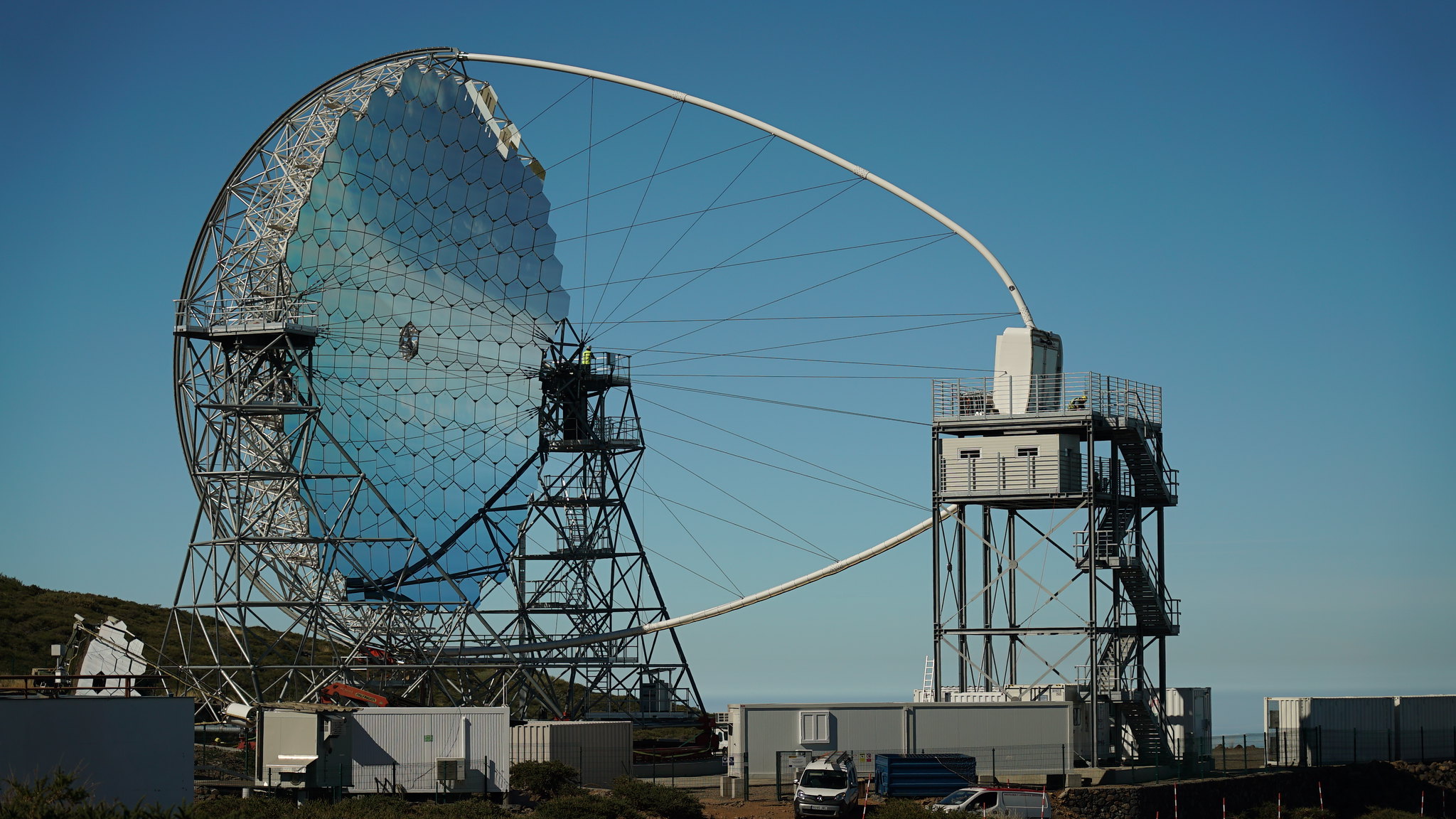}
\caption{Left: The LST prototype on La Palma (2018). Credit: Iván Jiménez (IAC).
%Credit: Tomohiro Inada. 
%Right: (looking for nice image of the camera)
}
\label{fig_LST}
\end{center}
\end{figure}

The LSTs determine CTA sensitivity in the energy range from 20 GeV to 150 GeV. The LST \cite{2019ICRC...36..653C} is an alt-azimuth telescope, the mount rotating on a 23.9-m diameter rail (Fig.~\ref{fig_LST}). The telescope has a 23 m diameter parabolic reflective surface composed of 198 hexagonal facets, that are supported by a tubular structure made of reinforced carbon fibre and steel tubes. The resulting low weight of the telescope of only 114 t allows rapid repositioning, within less than 20 s to any point in the sky; this feature is key for follow-up in particular of GRB alerts. An active mirror control system keeps the mirror facets aligned, compensating deformations of the mirror support structure.

The LST camera weighs about two tons, and contains 1855 photomultiplier tubes (PMTs) arranged in 265 modules, providing a $4.3^\circ$ optical field of view, with $0.1^\circ$ pixels, matched to the angular scale of low-energy gamma-ray showers. High-quantum efficiency PMTs are used, with a peak quantum efficiency of 42\%, for efficient detection of the Cherenkov light. To maximise the light throughput, each photosensor is equipped with an optical light concentrator. The PMTs are read out by the Domino Ring Sampler (DRS4) analog pipeline integrated circuit. The camera trigger strategy is based on the shower topology and the temporal evolution of the Cherenkov signal produced in the camera. The four LST cameras are interconnected in order to form an on-line coincidence trigger among the telescopes, which helps to suppress accidental triggers.

A prototype LST telescope -- actually the first LST -- was inaugurated in October 2018 on La Palma and -- as of early 2023 -- is in the process of commissioning and science verification, prior to hand-over to the CTA Observatory (see \cite{PoS(ICRC2021)872} and references given there). The LST prototype has demonstrated its performance by detecting steady TeV gamma-ray emission from the Crab Nebula, pulsed emission from the Crab pulsar, and also TeV emission from a range of AGNs, including Mrk 501, Mrk 421, 1ES 1959+650, 1ES 0647+250 and PG 1553+113 \cite{PoS(ICRC2021)806}. Last but not least the LST prototype has detected also the recurrent nova Rho Ophiuchi. 

\subsubsection{Medium-Sized Telescope (MST)}

\begin{figure}[htbp]
\begin{center}
\includegraphics[width=6cm,angle=0,origin=c]{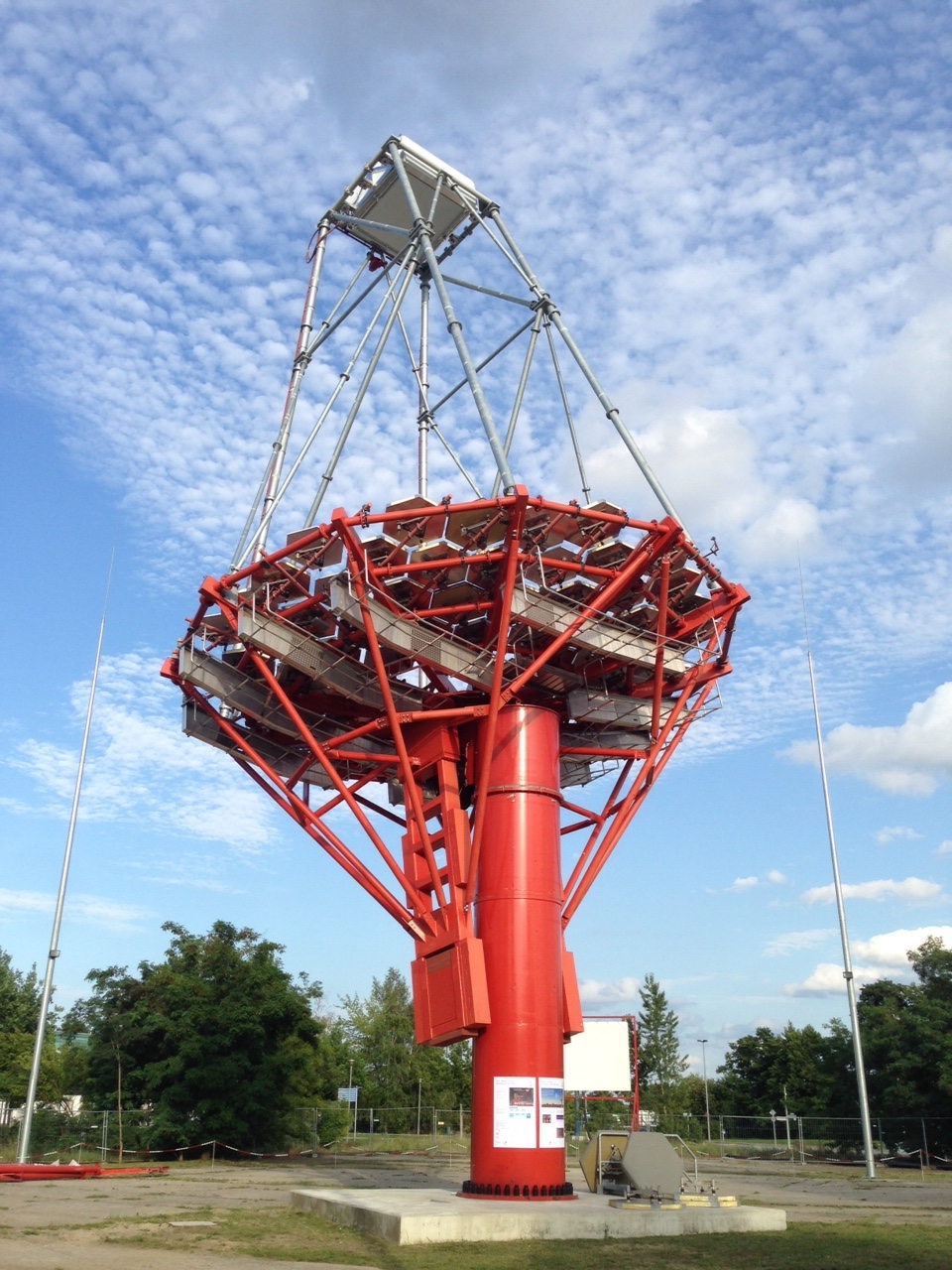}
\caption{The MST prototype in Berlin-Adlershof (2015). Credit: CTA Collaboration.}
\label{fig_MST}
\end{center}
\end{figure}

{\bf Single-mirror MST}\\
The MST telescopes (Fig.~\ref{fig_MST}) govern CTA sensitivity in the core energy range from 150 GeV to 5 TeV. The telescopes consist of an alt-az positioner, a reflector of approximately 12 m diameter and 16 m focal length, and a photomultiplier camera. The telescope structure is mostly fabricated in steel, with a space-frame dish carrying 86 identical hexagonal mirror facets of 1.2 m size and 32.14 m radius of curvature. Facets are arranged in a modified Davies-Cotton design, on a sphere of 19.2 m radius, representing a tradeoff between good PSF
over a large field of view, and isochronicity of the reflector.  Mirror facets  are fabricated using a cold slumping technology, with thin glass sheets on both sides of an aluminium honeycomb structure. The facets are surface-aluminised, with a multilayer protective coating. Two actuators for each facet serve for facet alignment.  Azimuth and elevation drives of the MST telescope are located in the head of the positioner tower, with large-diameter bearings and two servo motors for each axis, acting over worm gears. To ensure high availability and low long-term maintenance efforts most of the electrical components are installed in cabinets inside the positioner tower, with its 2 m diameter and 9 m height.

The Cherenkov cameras provide $0.17^\circ$ pixels over a $\approx 8^\circ$ optical field of view. Two versions of the camera were developed: NectarCAM will be used on the telescopes of the northern array, and FlashCam in the southern array. The two cameras, with 1855 and 1764 pixels, respectively, provide near-identical performance. 
NectarCAM uses the ‘Nectar’ analog pipeline ASIC for signal capture with GSample/s sampling rate and shares design features and components with the LST camera. NectarCAM is composed of 265 individual and easily removable modules. Each module is composed of seven photo-detectors (photomultipliers) with their associated readout and local trigger electronics. Each photo-detector is associated with an individual high voltage and preamplifier board.
The FlashCAM design follows a horizontal architecture with the photon detector plane (PDP), the readout electronics and the data acquisition system as key building blocks. The PDP contains photomultiplier tubes (PMTs) arranged in a hexagonal structure. 12 PMTs are combined in a PDP module, which provides high voltage and contains pre-amplifiers, as well as a micro controller for slow control, monitoring and safety functions. The signals are then transmitted via cables to the readout electronics, the design of which is based on a fully digital approach with continuous signal digitisation and digital trigger processing. Signals are digitised continuously with 12-bit Flash-ADCs with a sampling frequency of 250 MS/s. The readout system uses custom-developed high performance Ethernet transmission from the front-end into the camera server.

 A prototype MST was operated in Berlin-Adlershof, demonstrating the basic functionality of telescope structure and cameras.

\noindent{\bf Schwarzschild-Couder Telescope}\\
The Schwarzschild-Couder Telescopes (SCTs) are medium-sized telescopes with an innovative design, similar to the SST one.  The SCTs were proposed as an alternative to the single-mirror MST telescopes; while not fully ready for large-scale production at the time of this writing, they are planned for a future enhancement stage of CTAO. \\
Dual-mirror optics provides an improved point spread function over a wide field of view, permitting more precise imaging and reconstruction of air showers and hence better angular resolution for gamma rays. Unlike the SSTs, the medium-sized SCT telescopes were designed to fully exploit these features, with a camera of $0.067^\circ$ pixel size, resulting in a total of about 11000 pixels covering a $7.6^\circ$ optical field of view. 
The optical system of SCTs is significantly more complex than that of MSTs, with a 9.7\,m diameter primary mirror M1 segmented in 48 aspherical mirror panels, and a 5.4\,diameter secondary M2 segmented into 24 panels. The focal length is 5.6\,m. The camera uses modules very similar to the modules used in the SST cameras, with square 6\,mm x 6\,mm pixels. To achieve a point spread function compatible with this pixel size, sub-mm and sub-mrad alignment of the mirror panels is required. The motion of each of the 72 mirror panels is controlled in six degrees of freedom by six actuators assembled into a Stewart Platform. A panel-to-panel alignment system measures the relative misalignment between neighbouring panels, and the global alignment system  measures the relative positions of the main optical elements. The SCT optical system is carried by an alt-az positioner adapted from, and mostly identical to the MST positioner.

\begin{figure}[htbp]
\begin{center}
\includegraphics[width=6cm]{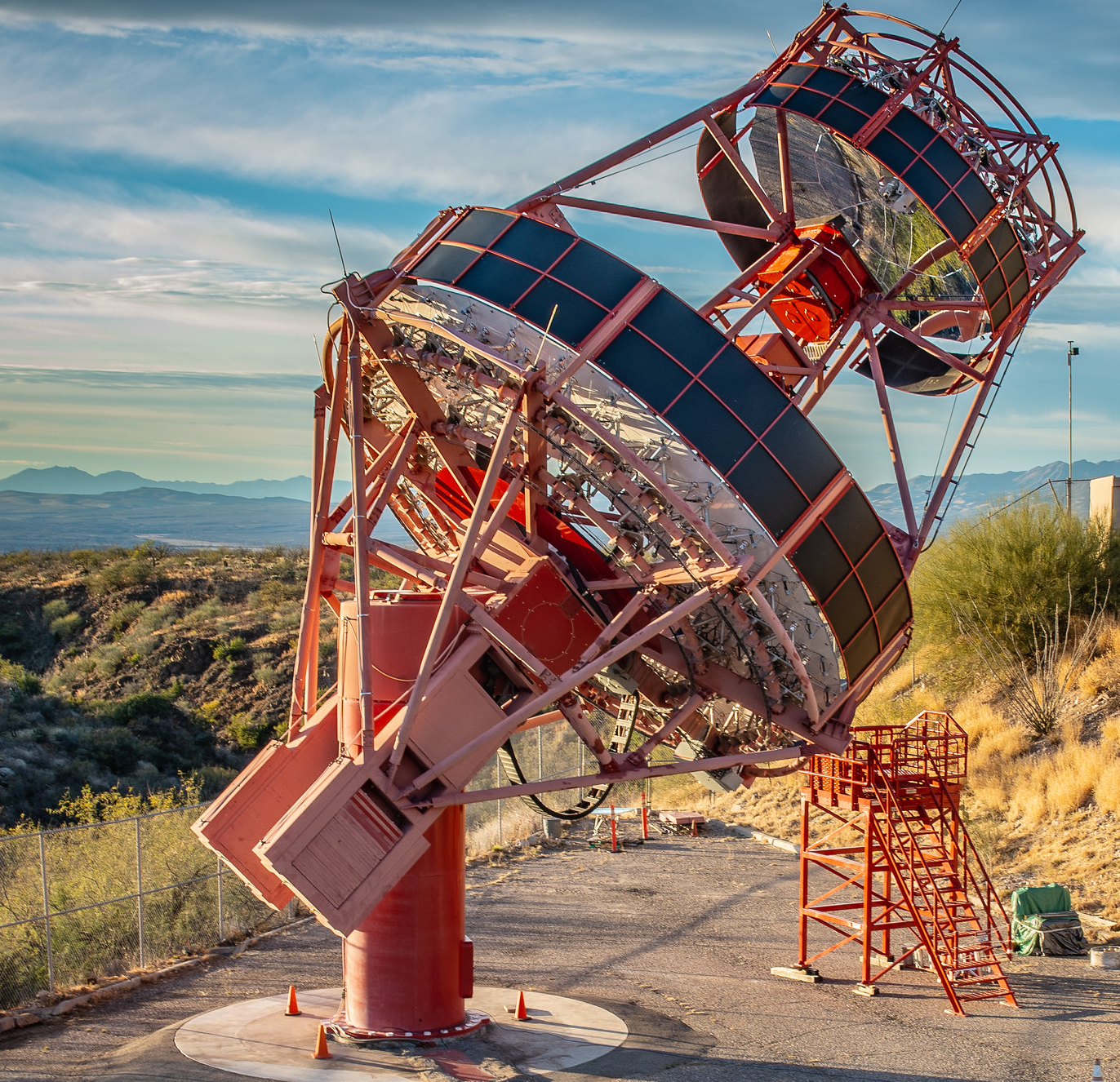}
\caption{Prototype of the medium-sized SCT telescope, at the Whipple Observatory in Arizona (2019). Credit: Deivid Ribeiro, Columbia University.}
\label{fig_SCT}
\end{center}
\end{figure}

The pSCT (Fig.~\ref{fig_SCT})  is an SCT prototype at the Whipple Observatory in Arizona (see \cite{2021APh...12802562A} and references given there).  The telescope was inaugurated, and recorded first light, in January 2019. A first successful alignment of the full system was completed in December 2019. The $2 \sigma$ containment radius of the optical psf was measured to 2.8 arc-min, which meets the design specification of 3.6 arc-min. Commissioning and first observations continued until February 2020, shortly before the observatory was temporarily closed due to the COVID-19 pandemic. Observations of the Crab Nebula with the pSCT began on January 18, 2020, for a total on-source exposure of 22 h, resulting in a $7 \sigma$ detection \cite{CTASCTConsortium:2021mtm}. The ultimate goal of the SCT team is to deliver at least 10 SCTs for the enhancement of the CTA observatory.

\subsubsection{Small-Sized Telescope (SST)}
\label{SubsubsecSST}

\begin{figure}[htbp]
\begin{center}
\includegraphics[height=4.0cm]{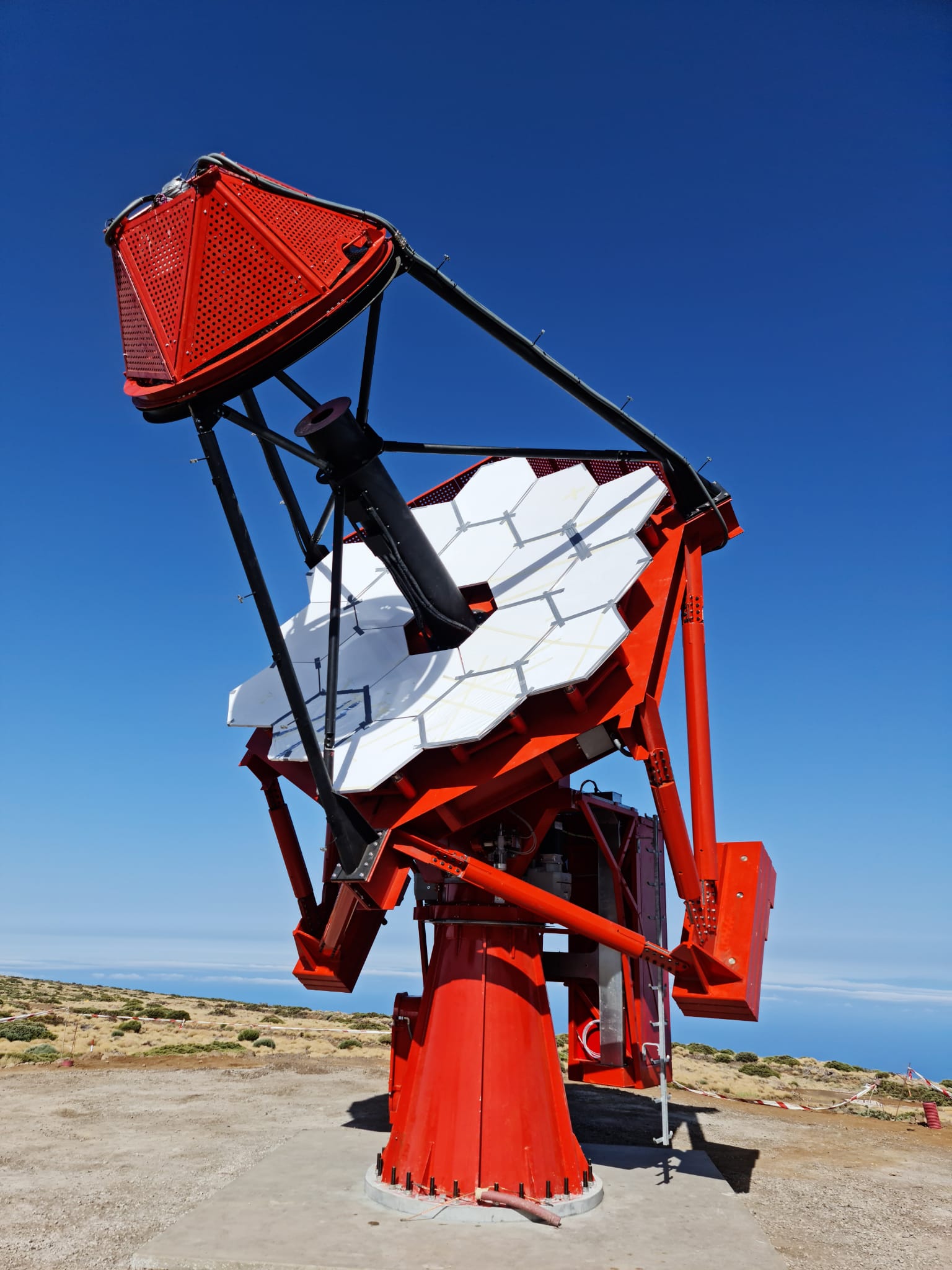}
\includegraphics[height=4.0cm]{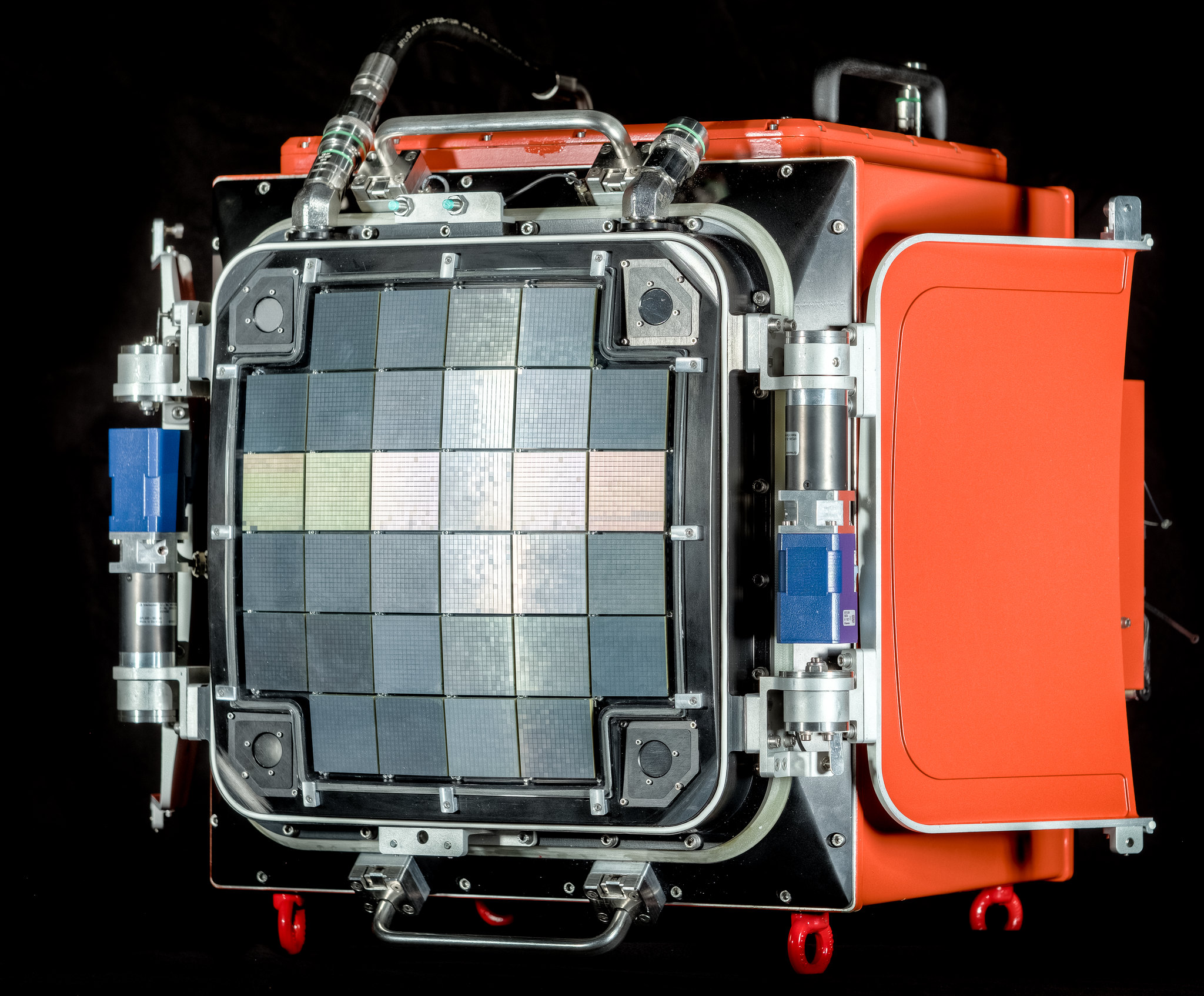}
\caption{Left: The first telescope of the ASTRI-MiniArray located in Tenerife  (2022). The telescope structure is equivalent to that of the SST one, whereas the camera of the ASTRI telescope is different from the one to be installed on the SSTs. Credit: ASTRI collaboration and the EIE GROUP srl. Right: Prototype of the Compact High Energy Camera (CHEC-S) with its SiPM. Credit: Christian Foehr (MPIK)}
\label{fig_SST}
\end{center}
\end{figure}

The SSTs (see \cite{PoS(ICRC2021)728} and references given there) govern the sensitivity of the southern CTA array at energies above 5 TeV; an important aspect in the development of SSTs was modest cost and the ability to mass-produce the telescopes. Various options to implement SSTs were studied: initially, a 6-m single mirror telescope with a photomultiplier camera was investigated, but was not pursued due to the high cost of the PMT-equipped camera. Significantly reduced costs were possible by using compact cameras with silicon sensors; two implementations of a dual-mirror telescope with Schwarzschild-Couder optics were proposed and prototyped -- ASTRI \cite{2020A&A...634A..22L} and GCT \cite{2017NIMPA.845..355D}, and one single-mirror implementation, also using a silicon sensor camera \cite{2019ICRC...36..694H}. Following a harmonisation process in 2019, a formal decision was made to base the SST on the dual-reflector design adopted by the ASTRI telescope, and the CHEC-S camera prototypes (Fig.~\ref{fig_SST}).

The SST telescope has an alt-az positioner carrying a primary mirror of 4.3\,m diameter, a 1.8\,m secondary at a distance of 3 m from the primary, and the camera with an active area of 35\,cm, positioned 52 cm from the secondary. The effective focal length is 2.15\,m. The primary mirror is composed of 18 aspherical hexagonal panels, the secondary is a monolithic 19\,mm thick glass shell bent to the required curvature. In 2017 the ASTRI-Horn prototype installed at the Catania Observatory on Sicily provided a robust optical validation of a Schwarzschild-Couder telescope; the telescope provides an approximately constant point spread function across the full field of view, consistent with optics simulations. Equipped with a camera prototype, the telescope successfully detected the Crab Nebula.

The CHEC-S camera contains 2048 pixels tiling the curved focal plane, resulting in a $8.8^\circ$ field of view and $0.19^\circ$ pixel size. Pixels are arranged in 32 identical camera modules connected to a single backplane; each module contains a silicon PMT (SiPM) tile with 64 pixels each of active size 6\,mm x 6\,mm. The TARGET set of ASICs provides SiPM bias control, waveform sampling at 1 GSamples/s and digitisation, and first-level triggering. Trigger signals from the modules are combined in a gate array (FPGA) on the backplane. As for all CTA cameras, a White Rabbit system provides absolute timing. The camera and in particular the focal plane with the SiPMs are liquid-cooled. The CHEC-S camera prototype was successfully tested on the ASTRI-Horn telescope structure in 2019, and designs of telescope and camera are being finalised towards mass production.

\subsubsection{Triggering}

For air shower detection, the telescope arrays use multi-level trigger schemes: each telescope derives a local trigger by identifying local and simultaneous concentrations of Cherenkov light in the camera. This is achieved either by thresholding pixel signals and searching for coincidences in neighbouring pixels, or by summing the signals from groups of pixels and comparing the sum to a threshold. For events passing the telescope trigger, images are buffered locally within a telescope, and a message is sent to a global array trigger (implemented in software). Only in case of a multi-telescope coincidence, telescopes are requested to transmit the event information. This scheme allows lower telescope trigger thresholds, that would otherwise saturate the telescope readout if all events with a local trigger were stored.

\subsection{Monitoring and calibration}

Compared to IACTs such as H.E.S.S., MAGIC or VERITAS, CTAO aims to provide significantly reduced systematic errors on the measured gamma ray fluxes and source locations. In particular, the systematic error on the overall energy scale is specified to less than 10\%, and effective areas should be known to better than 5\%. Source localisation should be possible with systematic errors of less than 3 arc-sec, above few hundreds of GeV and under favourable environmental conditions. These requirements, as well as the required high availability of telescopes, imply efficient monitoring and calibration of both telescope parameters and atmospheric parameters \cite{2014SPIE.9149E..19G}.

Local muons generating ring images provide a `calibrated' light source that can be used for continuous calibration of telescope response and for flat-fielding of cameras, during science operations. The relative response of pixels can be monitored within 3\% (MST, LST) on 20-minute time scales, the absolute response with 4\% (MST, LST) to 5\% (SST), and the relative optical throughput within 1\% (MST, LST) to 5\% (SST), on minute time scales \cite{2019ApJS..243...11G}. From the width of muon rings, the point spread function of the telescopes can be monitored. Camera triggers are set up to accept and transmit local muon images, without requiring a coincidence of multiple telescopes as required for air shower images. One limitation on this method is that the Cherenkov spectrum of local muons differs from the (partly-absorbed) Cherenkov spectrum of air shower particles; the Cherenkov emission from air showers detected by different telescopes provides an additional tool to cross-calibrate the response of telescopes and to trace possible systematics. All these techniques have the advantage that they come `for free' as part of normal data taking. They are complemented by special calibration light sources, in particular by flat-fielding light sources on each telescope, and absolutely calibrated light sources -- the Illuminators -- that can illuminate individual telescopes with known light intensity.

Atmospheric characteristics influence the Cherenkov light yield from shower particles and need to be monitored; this concerns molecular density profiles, molecular absorption and aerosol extinction. Raman LIDARs at both sites will serve to characterise both ground-level aerosols and the vertical structure of higher altitude aerosols and clouds. Requirements include the ability to characterise aerosol extinction at two wavelengths to distances of 30 km with an accuracy better than 5\%, within time scales of about a minute. Corresponding LIDAR systems operating at laser wavelengths of 335\,nm, 532\,nm and 1064\,nm were developed and successfully tested; incoming light is analysed at two elastic (355\,nm, 532\,nm) and two Raman (387\,nm, 607\,nm) wavelengths. The LIDARs will be accompanied by wide-angle optical telescopes, the FRAMs, designed to produce atmospheric extinction maps of starlight, through various optical filters with high angular resolution, yet covering a large field of view. Maps of integral atmospheric optical depth are continuously produced during observations. Interlaced Raman LIDAR measurements serve to provide the vertical structure of optical depth. Atmospheric density profiles will be calibrated using radiosonde measurements, but will largely rely on atmospheric profile prediction by global atmospheric data assimilation models (GDAS), or by the European Center Mid-term Weather Forecast (ECMWF) . For details and references, see e.g. \cite{2017ICRC...35..833E, 2019arXiv190909342B}.

\subsection{Sites}
\label{SubSecSites}

The activities to search for sites for CTA started in 2008. Steps included the identification and characterisation of promising sites, resulting in a short list of sites for the final selection. A formal call for proposals for CTA sites was launched in November 2011. Results of the site evaluation were summarised in the October 2013 Site Selection Summary report \footnote{CTA internal, unpublished}, that provided the basis for the site recommendations by the CTA Consortium and by the Site Selection Committee (SSC) appointed by the CTA Resource Board (RB). 

Requirements for candidate sites included a latitude between $20^\circ$ and $40^\circ$ for the Northern array, and between $-20^\circ$ and $-40^\circ$ for the Southern array; an elevation between 1500 to 3800 m asl; a sufficiently large and flat area for deployment of the arrays (O(5 km$^2$) for the Southern site); criteria on ambient conditions such as temperature range, temperature gradient, precipitation, average wind speeds; at least 70\% of completely cloud-free nights; limits on night sky brightness reflecting ambient light levels, and restrictions regarding events that could damage the facility, such as earthquakes, hail or extreme wind speeds.

Nine sites were investigated in detail. The five southern hemisphere locations included two sites in Namibia (Aar and H.E.S.S.), a site on the ESO Paranal area in Chile, and two sites in Argentina (Leoncito and San Antonio de los Cobres - SAC). The four northern hemisphere locations were in Arizona/USA (Yavapai and Meteor Crater), on the Canary Island of Teneriffe (Teide), and in Mexico (San Pedro Martir). The Teide site was later replaced by the Roque de los Muchachos site on the nearby island La Palma, providing better protection from light pollution. The final ESO Paranal site was moved south by about 10\,km compared to the initially considered location, primarily to avoid interference with ELT construction but also to increase the distance to ELT's laser guide beams.

\begin{figure}[htbp]
\begin{center}
\includegraphics[width=5.5cm]{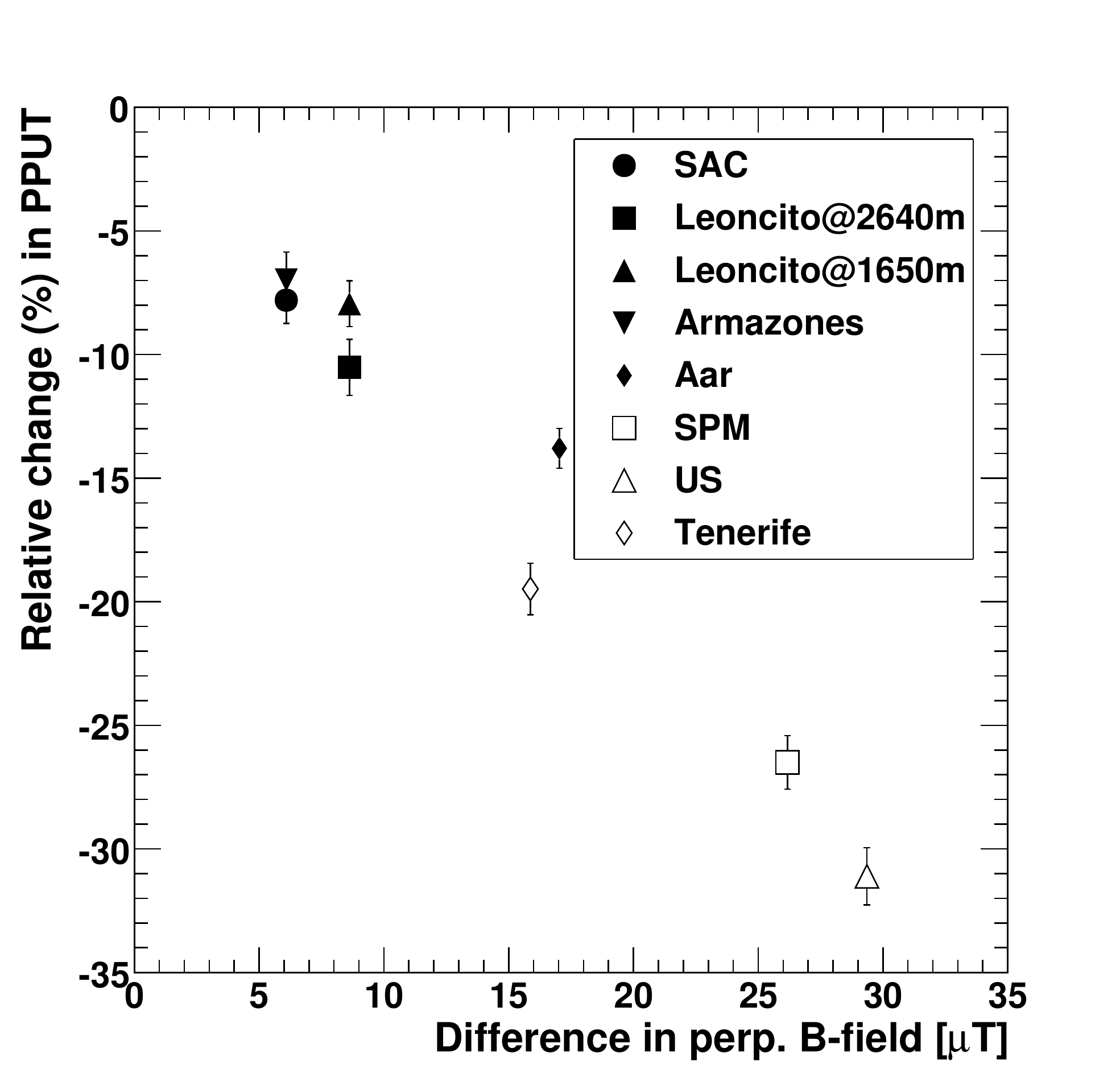}
\includegraphics[width=5.5cm]{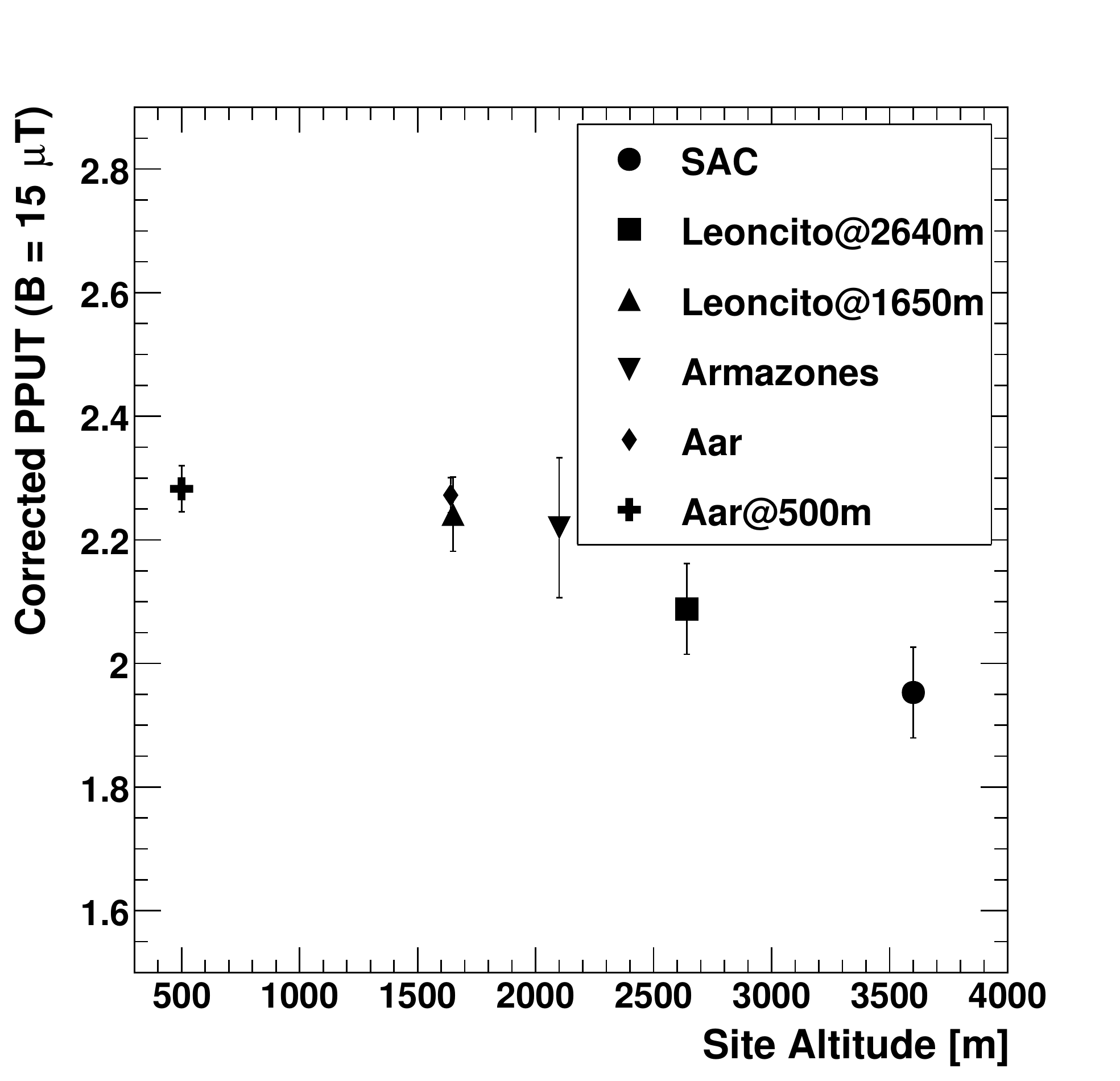}
\caption{Left: Dependence of spectrum-averaged flux sensitivity on the perpendicular component of the geomagnetic field. Right: Dependence on site altitude, after correcting for effects due to geomagnetic field. The spectrum-averaged flux sensitivity PPUT represents the geometric average of the differential sensitivity in 0.2-decade energy bands. From \cite{2017APh....93...76H}.}
\label{fig_site_MC}
\end{center}
\end{figure}

The criteria considered for the scientific site ranking were science performance, which depends on the average annual observation time (AAOT) available and on the performance of the array per unit time at a given site, as well as  costs and risks. The main factor influencing the AAOT is the fraction of cloud free night hours, but losses due to environmental conditions -- such as high winds -- are also significant for some sites. Data sources to derive AAOT included dedicated ground instrumentation -- the Atmoscopes \cite{2015JInst..10P4012F} monitoring weather, night sky brightness, sky temperature as an indicator of clouds, and providing all-sky cameras to determine cloud coverage. Since these own measurements typically covered only one year, data were combined with long-term (typically one decade) satellite data on cloud coverage and data from long-term (one decade) weather and cloud modelling, and long-term local measurements, where available. The resulting cloud-free fraction ranged between 71\% and 89\%, for the different sites. For a fixed array configuration, the performance per unit observation time depends on site altitude, geomagnetic field, characteristic aerosol optical depth and night sky background level. Simulations \cite{2017APh....93...76H} identified altitude and geomagnetic field as the primary drivers of science performance. Fig.~\ref{fig_site_MC} illustrates how performance depends on geomagnetic field and site altitude. Basically, each $\mu$T in geomagnetic field perpendicular to the shower axis costs 1\% in sensitivity, since shower particles are spread out. The dependence of spectrum-averaged sensitivity (PPUT) on site altitude is less than initially assumed, with constant sensitivity up to 2000 m and a mild degradation beyond; higher sites provide improved low-energy performance but degraded performance at the high end of CTA's energy range; the two effects tend to compensate in the spectrum-averaged sensitivity.

The estimates for the average annual observing time ranged from 1141 h to 1293 h for the Southern sites, and from 993 h to 1114 h for the Northern sites. To compare sites, a figure of merit was defined combining the average annual observing time with the performance per unit time. Resulting figures of merit ranged from 1.18 to 1.60 for the southern sites, with highest values for the sites in Namibia and Chile. For the Northern sites, figures of merit between 1.33 to 1.38 proved identical for all sites, within the errors of the estimates. The Site Selection Summary report also provided estimates for the cost of site infrastructure and the operating cost, as well as an evaluation of risks.

\begin{figure}[htbp]
\begin{center}
\includegraphics[width=6cm]{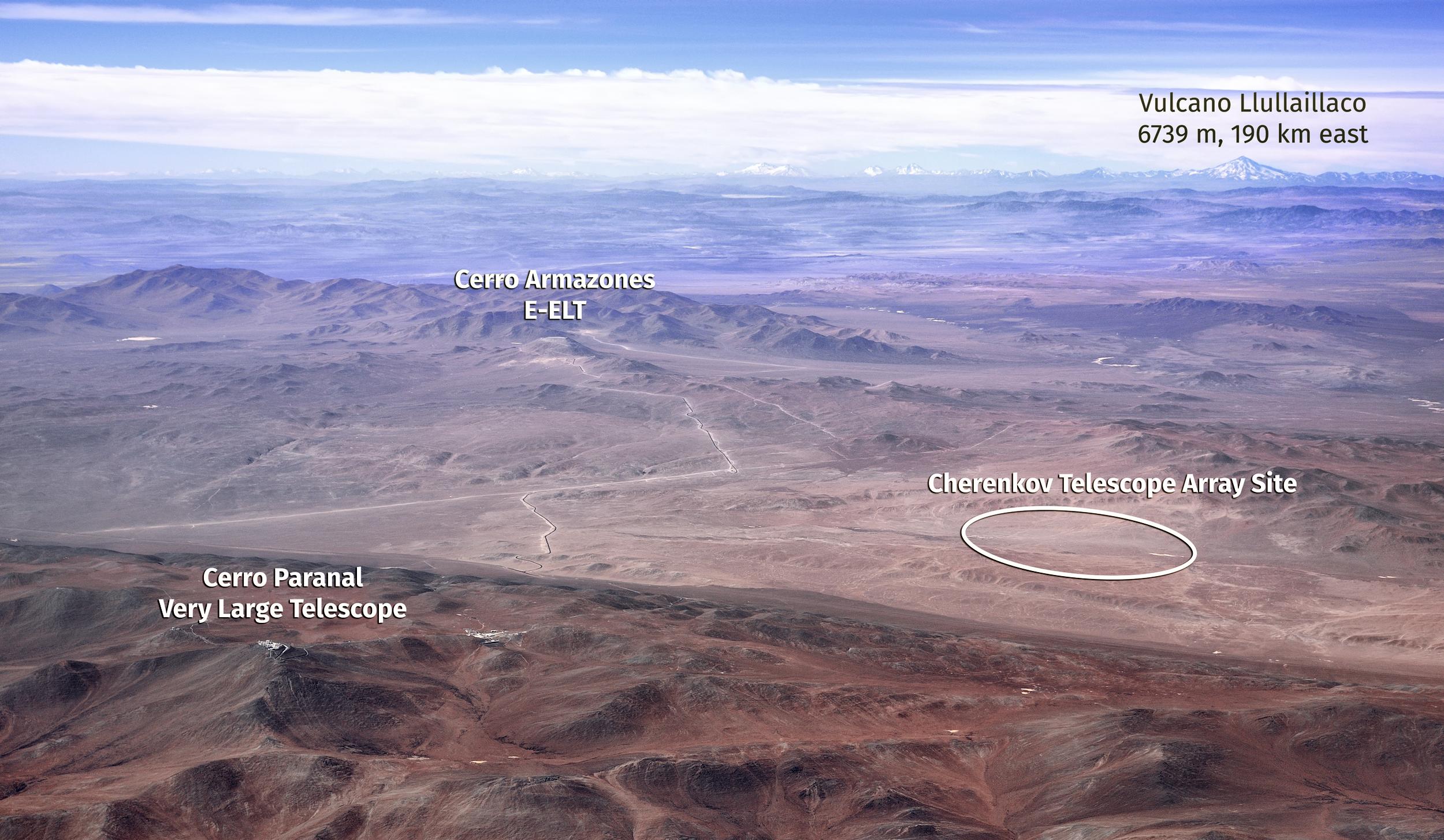}
\includegraphics[width=5.2cm]{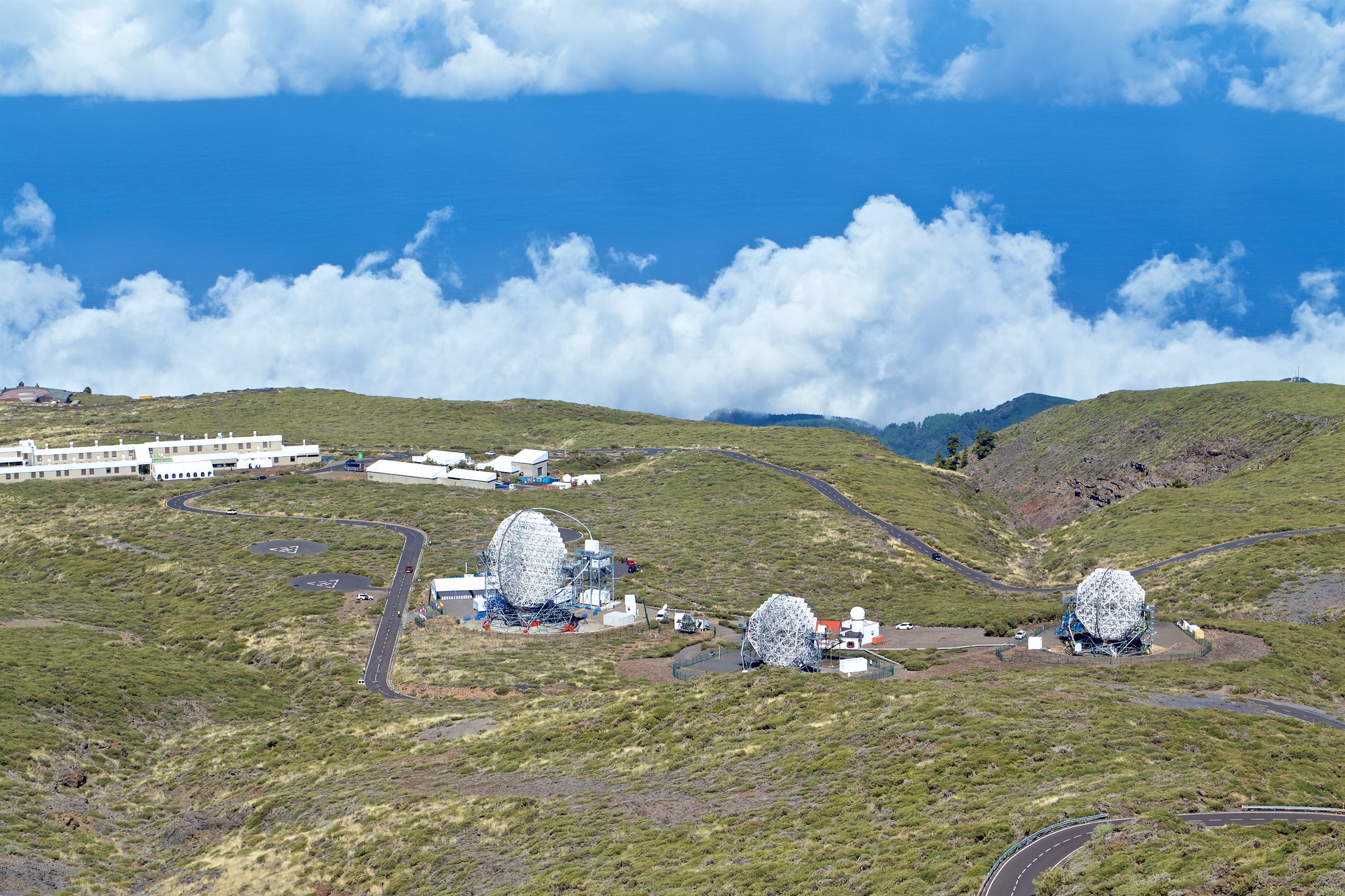}
\caption{Left: Aerial view of the southern array site, also indicating the locations of the ESO Paranal and ELT telescopes (2018). Credit: Marc-André Besel, CTAO / ESO. Right: view of the northern array site on the Roque de los Muchachos Observatory, showing the first LST telescope as well as the two MAGIC telescopes (2018). Credit: Akira Okumura.}
\label{fig_sites}
\end{center}
\end{figure}

On the basis of the recommendations by the CTA Consortium and the Site Selection Committee, and taking into account other factors including resource considerations, the CTA Resource Board decided in July 2015 to enter into detailed contract negotiations for hosting the CTAO Southern array on the European Southern Observatory (ESO) Paranal grounds in Chile and the Northern array at the Instituto de Astrofisica de Canarias (IAC), Roque de los Muchachos Observatory on La Palma, Spain. Sites in Namibia and Mexico were kept as viable alternatives. In September 2016, the CTAO Council concluded negotiations with IAC to host the Northern array on La Palma (at $28^\circ 45'$ N, $17^\circ 53'$ W, $\approx$ 2200 m asl.). The final agreement to host CTAO Southern array in Chile (at $24^\circ 41'$ S, $70^\circ 18'$ W, $\approx$ 2100 m asl. ) were signed in December 2018, after complex negotiations between CTAO, ESO and Chile. Fig.~\ref{fig_sites} shows views of the array sites.

\subsection{The alpha configuration}
\label{SubSecAlpha}

In early 2016, it became evident that the anticipated resources for the construction of the CTA would not cover the proposed 118-telescope arrays (later referred to as the `omega configuration'). The CTAO Council asked the CTA Consortium `to deliver a plan how CTA can be constructed in a staged approach, in order to start construction as soon as possible', with a cost of the first stage of about 60\% of the full cost. An extensive discussion took place if a uniform degradation across all energy ranges for both arrays is preferred for the first stage of CTA, or if one should better aim to achieve – within the financial constraints – optimum sensitivity in certain site-specific energy ranges, the two sites complementing each other. After a detailed study of configurations and science cases, a scenario was proposed where sites temporarily specialise: the northern site with 4 LSTs and 5 MSTs provides a focus on low-energy performance and hence on extragalactic science, whereas the southern site with 15 MSTs and 50 SSTs focuses on the TeV and multi-TeV energy domain and on galactic science. It was emphasised that the priority for subsequent implementation steps should be to provide LSTs for the south, and to add further MSTs in the north, creating viable coverage of the TeV energy domain.

In 2021, towards submission of the application for the CTAO ERIC, and on the basis of a revised detailed cost estimate (`CTAO Cost Book'), the starting configuration was fine-tuned, providing a matrix of in-kind contributions relating country contributions to observatory components, in particular telescopes but also calibration systems, software, or infrastructure elements. This resulted in the `alpha configuration' of CTA, with 4 LSTs and 9 MSTs in the north, and 14 MSTs and 37 SSTs in the south. This configuration already includes the desired enhancement in the number of MSTs in the north, from 5 to 9. Reflecting constraints in production capacity and funding, the northern MSTs will use NectarCAM cameras, whereas the southern MSTs will employ FlashCam cameras; the two camera types provide nearly identical performance, and have compatible interfaces. Anticipating the addition of LSTs in the south, the southern infrastructure will include excavations for LST foundations, to avoid later excavation works and related dust at the centre of the array.

\begin{figure}[htbp]
\begin{center}
\includegraphics[width=10cm]{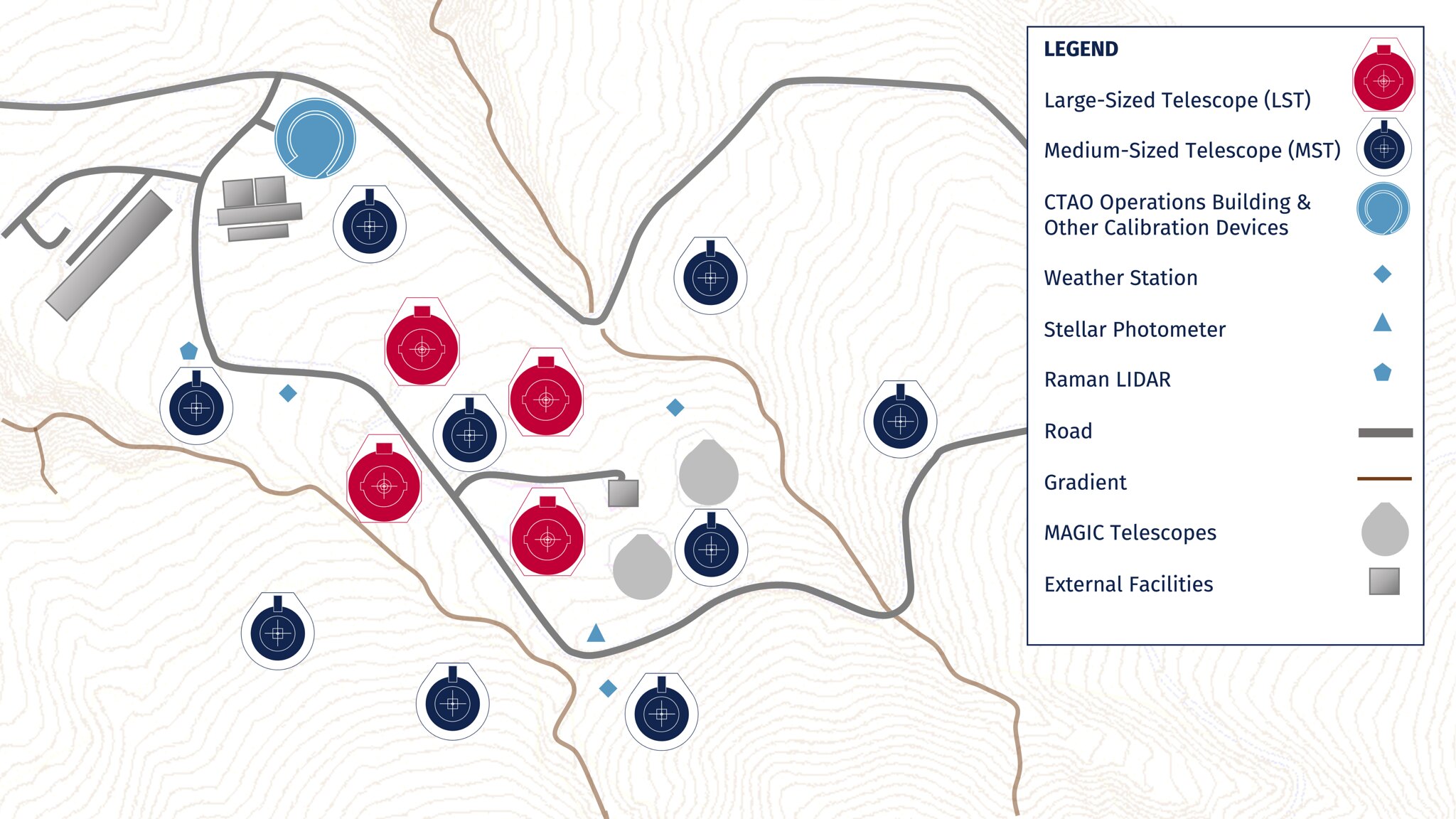}
\caption{Arrangement of the 4 LSTs and 9 MSTs of the `alpha configuration' on the Northern array site. Also indicated are the positions of monitoring instruments, and of the two MAGIC telescopes. Credit: CTAO.}
\label{fig_north_array}
\end{center}

\end{figure}

\begin{figure}[htbp]
\begin{center}
\includegraphics[width=7cm]{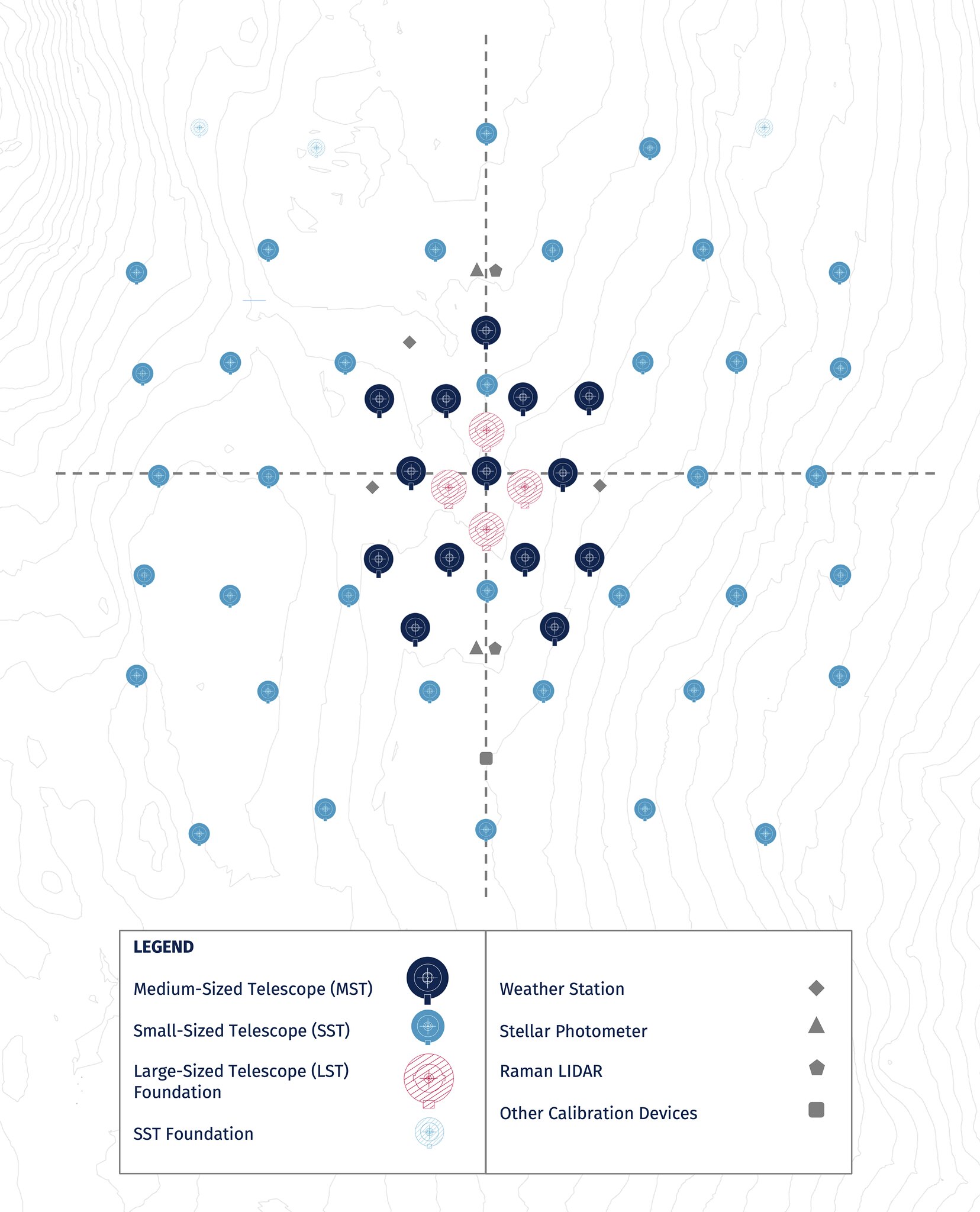}
\includegraphics[width=11cm]{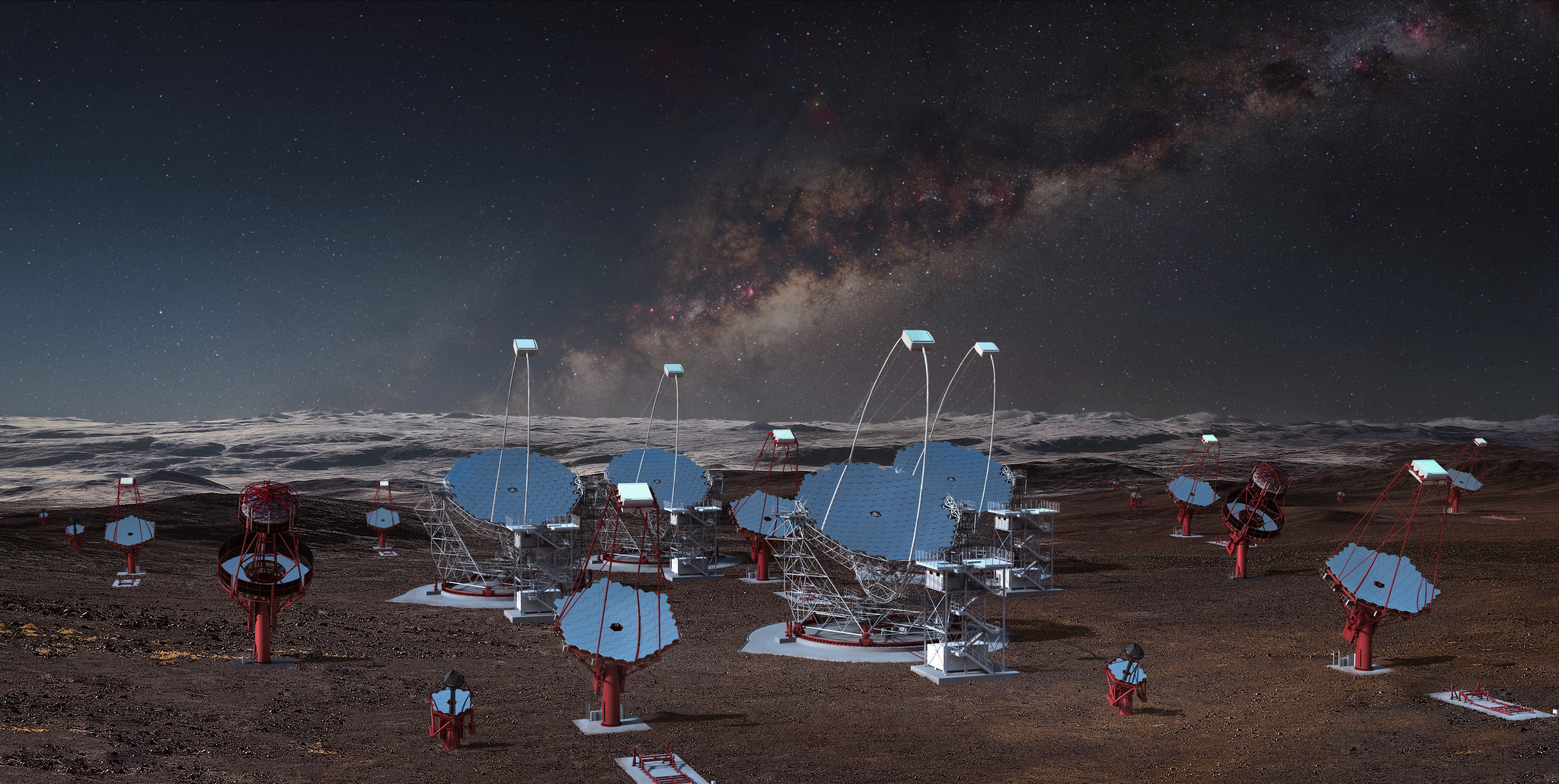}
\caption{Top: Layout of the 14 MSTs and 37 SSTs of the `alpha configuration' Southern array, and of the LST foundations.
Bottom: Artists view of the central region of the southern site, including LST and SCT telescopes of enhancement stages. Credit: CTAO.}
\label{fig_south_array}
\end{center}
\end{figure}

For the `alpha configuration', Fig.~\ref{fig_north_array} shows the arrangement of the telescopes of the Northern array, and Fig.~\ref{fig_south_array}  the preliminary layout of the Southern array. 

The performance of the `alpha configuration' arrays is discussed in detail in section~\ref{SubSecPerformance}; remarkably, the on-axis sensitivity of the arrays meets or even exceeds the sensitivity requirements, except for energies below 50 GeV in the south, reflecting the missing LSTs. This is due to more detailed simulation of telescope performance, due to improved reconstruction algorithms, but also due to a certain safety margin in the initial sensitivity requirements. Current simulations do not take into account the availability of telescopes (about 98\%), degradation of mirror reflectivity over time etc., so the long-term average performance is expected to be up to 20\% worse than predicted by the simulations.

In 2022, additional funding became available for the construction of 2-3 LSTs and 5 additional SSTs that could define the first enhancement of CTAO Southern array, restoring performance to near design specifications even for the sub-100 GeV domain.

%% file: SecObservatory.tex
Contrary to instruments such as H.E.S.S., MAGIC or VERITAS, that are operated by collaborations and that are not, or only to limited extent, open to external users, CTAO will be the first truly open VHE observatory, providing accessible data products and support services to a wide community.

This description of the CTA Observatory and the services it offers is largely based on the Scientific \& Technical Description of the observatory, which serves as auxiliary document to the CTAO ERIC statutes, and on the statutes themselves.

\subsection{Architecture and data flow}

\begin{figure}[htbp]
\begin{center}
\includegraphics[width=11cm]{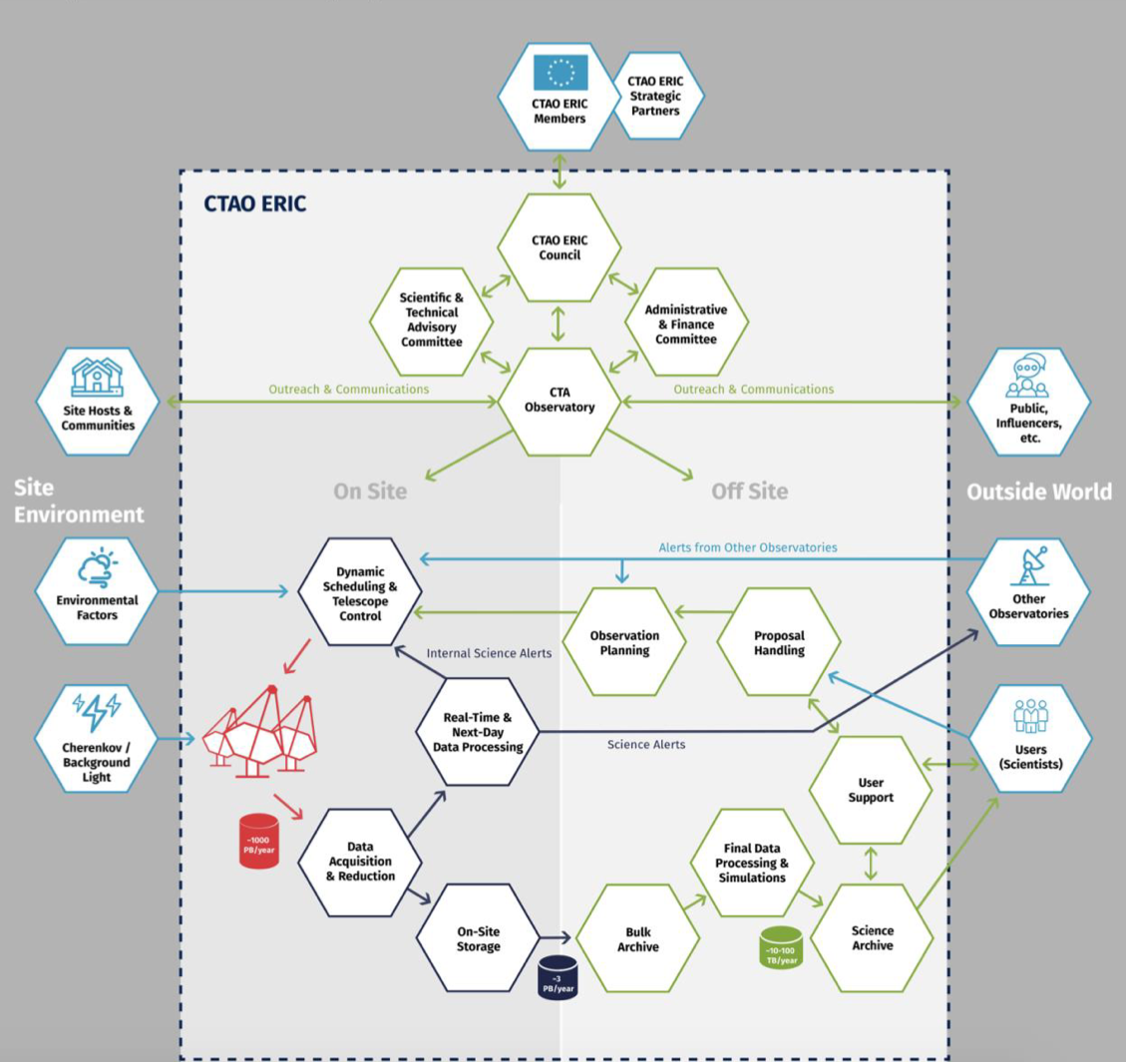}
\caption{Schematic representation of the observatory architecture, illustrating the flow and processing of data and information, the interaction with the outside world, and the observatory governance. Credit: CTAO.}
\label{fig_ObsOverview}
\end{center}
\end{figure}

Fig.~\ref{fig_ObsOverview} illustrates the observatory workflow. Observations are executed in queue mode; presence of guest observers at the array sites is not required. In proposals responding to Announcements of Opportunity (AO), observatory users specify targets as well as specific conditions e.g. regarding array configuration, atmospheric or environmental conditions, or time windows or Target of Opportunity (ToO) conditions for the execution of observations. Users can request the full telescope arrays, or subsystems such as the LST telescopes, or the SST telescopes, in case coverage of the full energy range is not required. Subsystems of few MST telescopes can e.g. be used for monitoring of variable sources, in parallel with other observations.

\begin{figure}[htbp]
\begin{center}
\includegraphics[width=10cm]{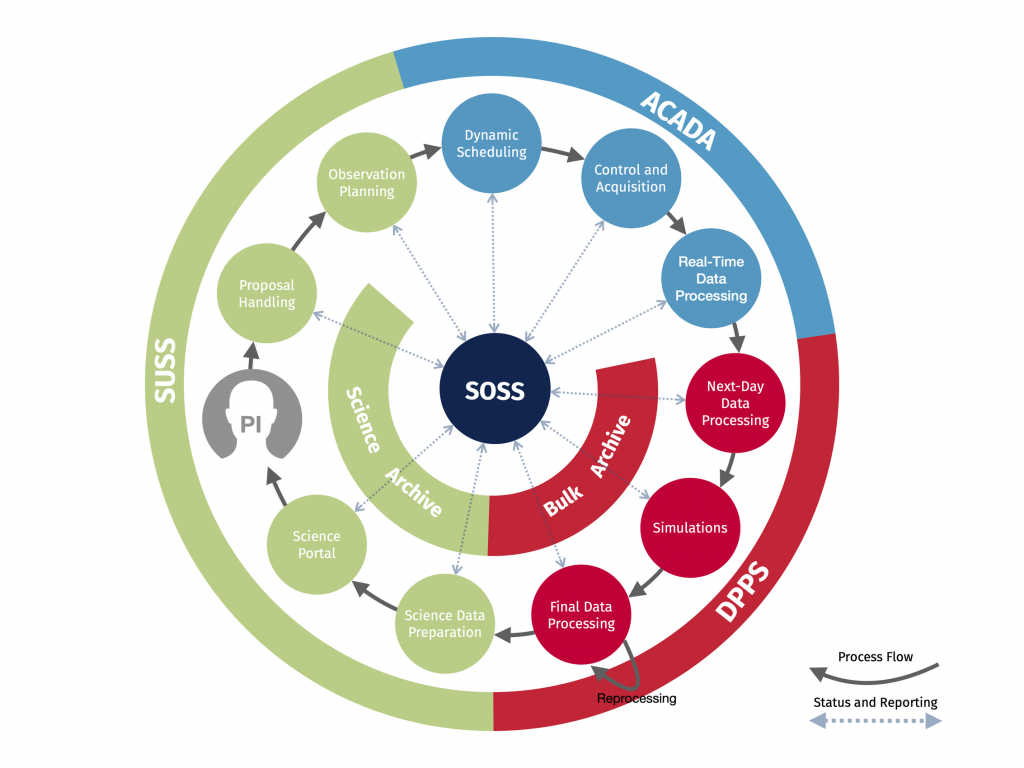}
\caption{Flow of processes in CTAO's science operations, from proposal submission to execution of observations, data processing, and data delivery to the user. Credit: CTAO.}
\label{fig_ObsOverview2}
\end{center}
\end{figure}

 Proposals are reviewed by a Time Allocation Committee (TAC), and a long-term schedule is created based on a ranked list.
 Based on the long-term schedule, observations are dynamically scheduled, accounting for the relative priority of proposals, specific observing conditions and sky conditions, and availability of the telescopes. A first on-site data processing serves for data quality control as well as monitoring of variable sources. Data are then transferred to the off-site bulk data archive, processed, and -- via the science archive -- made available to the user. In more detail, the following observatory software systems are involved:
\begin{description}
\item[Array Control and Data Acquisition System (ACADA):] the supervisory control and data acquisition system of the arrays. It will also perform the dynamic scheduling, the monitoring of the array performance, as well as the automatic generation of science alerts.
\item[Data Processing and Preservation System (DDPS):] a software system responsible for reducing the raw data (Data Level 0 and 1) and producing low-level data products appropriate for science analysis; the reconstructed air showers comprise Data Level 2. The data processing and reconstruction pipeline is Python-based \cite{CTApipe}. DPPS also includes the production of simulated data, (re)processing and long-term preservation of data products, as well as data transfer from the site to the off-site computing centres.
\item[Science \& Technical Operation Support Systems (SOSS):] a collection of software systems supporting the CTA Observatory operations, including, for instance, configuration management, issue tracking, and maintenance planning.
\item[Science User Support System (SUSS):] a software system providing for the user the point of access to the observatory for proposal submission, user support, and retrieval of high-level CTAO science data products and of the CTAO science analysis tools. SUSS provides support for proposal evaluation, for generation of the observation schedule, and for user support.
\end{description}
Users are provided fully-calibrated event-level (Data Level 3) information in FITS format, containing the reconstructed directions and energies for gamma-ray candidates, as well as the appropriate instrument response functions required to derive sky maps, gamma-ray spectra and light curves. Targets for the quality of calibration for user data products include knowledge of the gamma-ray energy scale within 10\% systematic error, and of the effective gamma-ray detection area within 5\%.
Fig. \ref{fig_ObsOverview2}  illustrates  the different systems behind the observatory’s science operations, and their interaction.

Open source tools are provided for user data analysis. Two science analysis packages were developed as prototypes of the CTAO science analysis tools: ctools \cite{ctools} and gammapy \cite{gammapy}; the latter has been chosen by the observatory. CTAO will also provide higher-level data products, such as sky maps and source catalogs (Data Levels 4 and 5) to guarantee a quick-look exploration of the data. The observatory will also release a small set of archived raw data (Data Level 0) and their subsequent data levels to allow testing of analysis algorithms. Access to larger data sets at raw/calibration data level is conditioned to the approval of a development proposal where the access must be well justified. 

An important aspect for time-domain astronomy is the ability to react to external alerts, and to issue alerts. With its dynamic scheduling of observations, CTAO will be able to react and start observations within 30 s from an incoming alert, assuming a corresponding ToO proposal. On a similar time scale, thanks to its real-time data analysis, CTAO will be able to issue alerts about gamma-ray sources; the sensitivity of the real-time analysis is, however, a factor two reduced compared to the sensitivity of the final processing happening in the off-site datacenters.

\subsection{Observatory organization and access to the observatory}

A detailed description of the organisation of, and access to CTAO is beyond the scope of this document, but the main aspects -- at the time of this writing still subject to formal approval by the European Commission -- are summarised in the following (see also Fig.~\ref{fig_ObsOverview}).

CTAO will have the legal form of a European Research Infrastructure Consortium (ERIC); Members of the ERIC are countries, usually represented by ministries or research organisations. ERIC regulations allow membership of countries that are not members of the EU; in addition, the status of a `Strategic Partner' is foreseen, with similar rights and duties as ERIC Members. ERIC statutes provide an Observer status, and allow agreements with third parties.

The CTA Observatory ERIC is governed by a Council composed of up to two delegates from each member; voting rights in Council relate to a member's contributions to CTA construction. The Council appoints the Director General who will carry out the day-to-day management of CTAO. The Director General is in particular responsible for the allocation of observing time by organising and managing a peer review process during which scientific excellence and feasibility of proposals are evaluated. Council is supported by an Administrative and Finance Committee (AFC) and a Scientific and Technical Advisory Committee (STAC). 

The CTAO ERIC will be a distributed facility with statutory seat in Bologna, a Science Data Management Center (SDMC) located in Zeuthen, Germany and the two arrays: the Northern array located at the Observatorio del Roque de Los Muchachos in La Palma, Spain, and the Southern array situated at the La Silla Paranal Observatory in Chile. The final data processing and the long-term data storage will be distributed over computing centres in four countries.

\begin{figure}[htbp]
\begin{center}
\includegraphics[width=6cm]{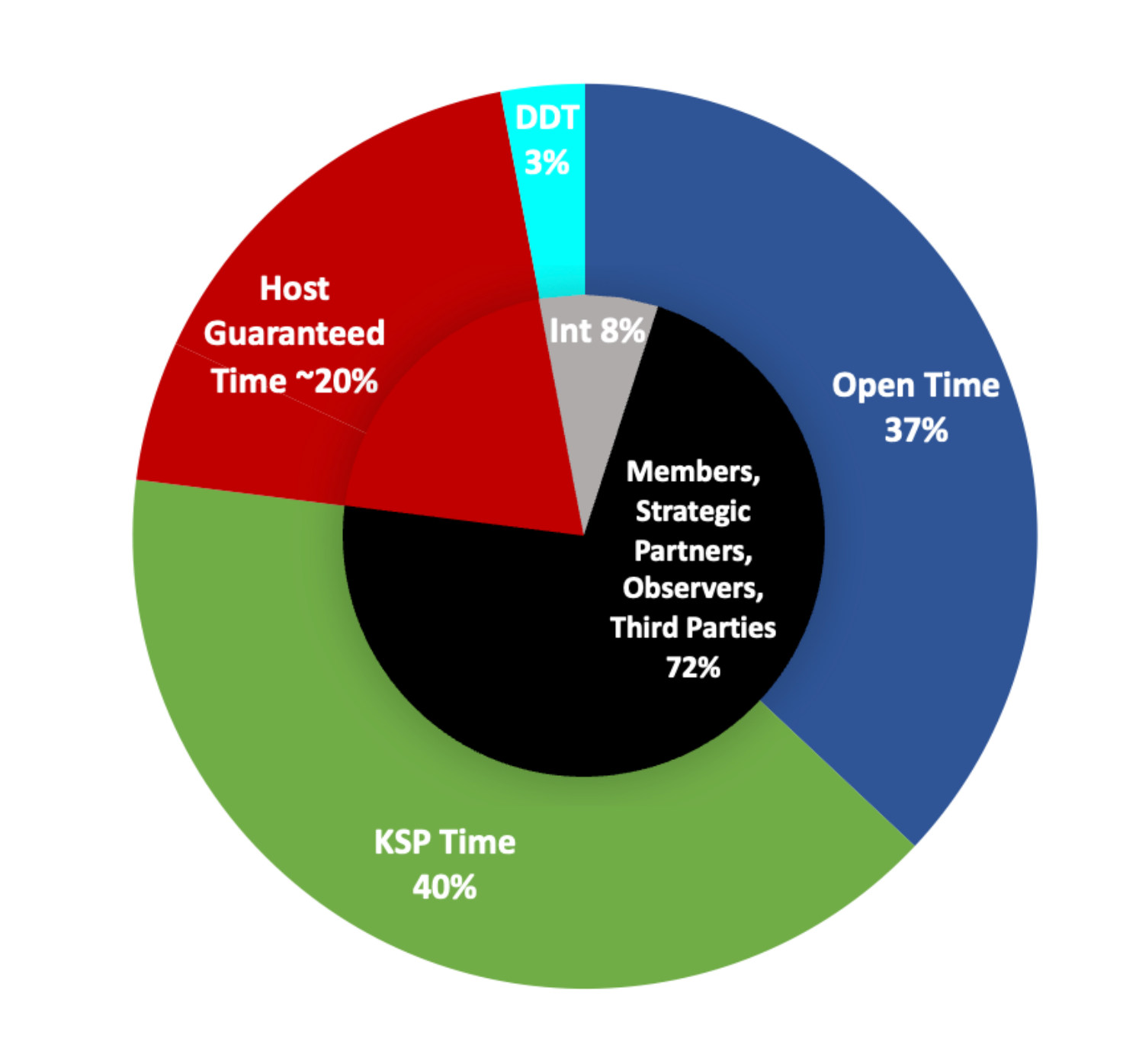}
\caption{An example of the sharing of the categories of observing time - integrated over the first 10 years of operation, including Host Guaranteed Time arising from contractual obligations of CTAO ERIC. DDT stands for Director's Discretionary Time, KSP time is used for Key Science Projects. `Int' denotes time available to international users outside the CTAO ERIC Members, Strategic Partners, Observers and Third Parties. The exact annual profile of observing time remains to be decided by the CTAO ERIC Council.}
\label{fig_obstime}
\end{center}
\end{figure}

CTAO observing time is classified into Open Time, Key Science Project (KSP) Time, observing time that arises from contractual obligations of the CTAO ERIC with the host countries / organisations -- the Host Guaranteed Time (HGT) -- and the Director's Discretionary Time (DDT) (outer ring in Fig. \ref{fig_obstime}).  KSP time consists of Guaranteed Time Observations (GTOs)
granted as reward for the in-kind and cash contributions to the CTAO construction by the CTAO ERIC Members, Strategic Partners, Observers and Third Parties. Host Guaranteed Time corresponds to 20-25\% of the total observing time of the two CTAO arrays.
The bulk of CTAO's observing time is aimed at researchers from countries/organizations that are CTAO ERIC Members, Strategic Partners, Observers and Third Parties.
Only a small fraction of the total observing time is granted to scientists independently from their country of affiliation: DDT plus a small fraction of the open time, about 5\% of the total observing time (outer ring in Fig. \ref{fig_obstime}). The definitive sharing will be decided by the ERIC Council and may be adjusted over time.

Observing time can be requested through science proposals that will be evaluated only on their scientific merit, and technical feasibility by the CTAO TAC. The planned types of science proposals are:
\begin{description}
\item[Standard Proposals:] science proposals with a specific science goals, responding to Announcement of Opportunity calls and evaluated by the CTAO TAC. They can request observations to be carried out in response to external or internal triggers, i.e. Target of Opportunity (ToO) observations. They are used to request open time, including the ICOT and most of the HGT. 
\item[Key Science Projects (KSP):] usually large and long science proposals requesting more than hundreds of hours over several observing periods that have the potential to lead to a major advance or breakthrough in the field of study for CTA and that will produce legacy data sets and data products to be publicly available through the CTAO science portal. KSP proposals can request KSP time and are reviewed by the TAC.
\item[Director General’s Discretionary Time (DDT) proposals:] scientifically outstanding proposals not subjected to the standard time allocation process. Their execution is time critical and/or offers the possibility of anticipating important discoveries which may be missed if subjected to the time delay introduced by the standard allocation process.  
\end{description}

Access to all CTAO science data products and the required supporting information will be publicly available, after a proprietary period of (typically) 12 months. CTAO data  policy will recognise the FAIR principles (findability, accessibility, interoperability, and reusability).

%% file: SubSecPerformance.tex
At the moment of this writing, the starting array configuration is not frozen. The `alpha configuration' is defined in the ERIC documentation and is funded.
Additional funding in 2022 will allow to enhance the `alpha
configuration' already in the construction phase with the addition of 2
LSTs and 5 more SSTs at the Southern array.

\begin{figure}[htbp]
\begin{center}
\includegraphics[width=8cm]{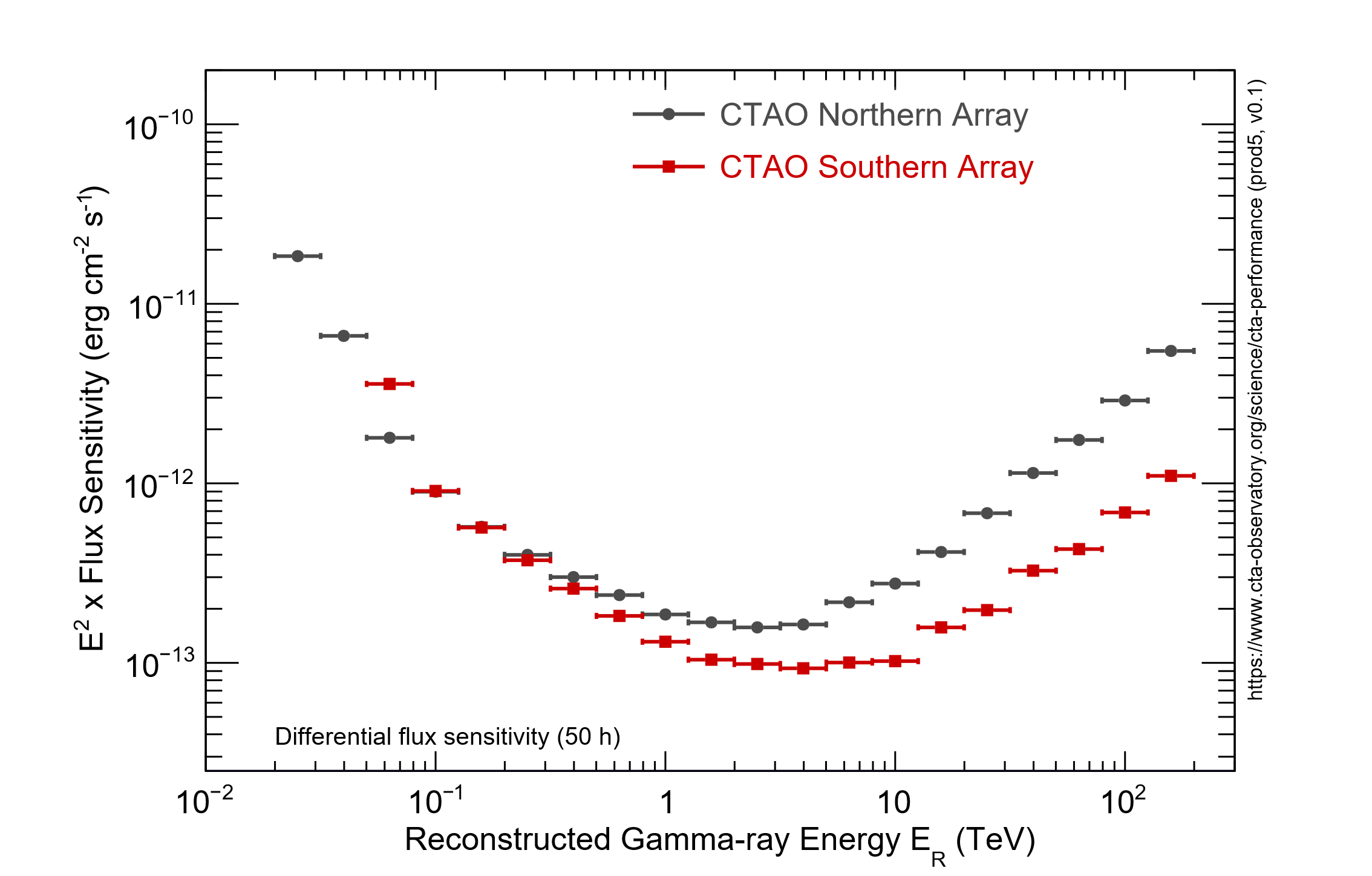}
\caption{CTA flux sensitivity for the Northern and Southern arrays, on axis, for 50 h observing time.}
\label{fig_sens}
\end{center}
\end{figure}

\begin{figure}[htbp]
\begin{center}
\includegraphics[width=8cm]{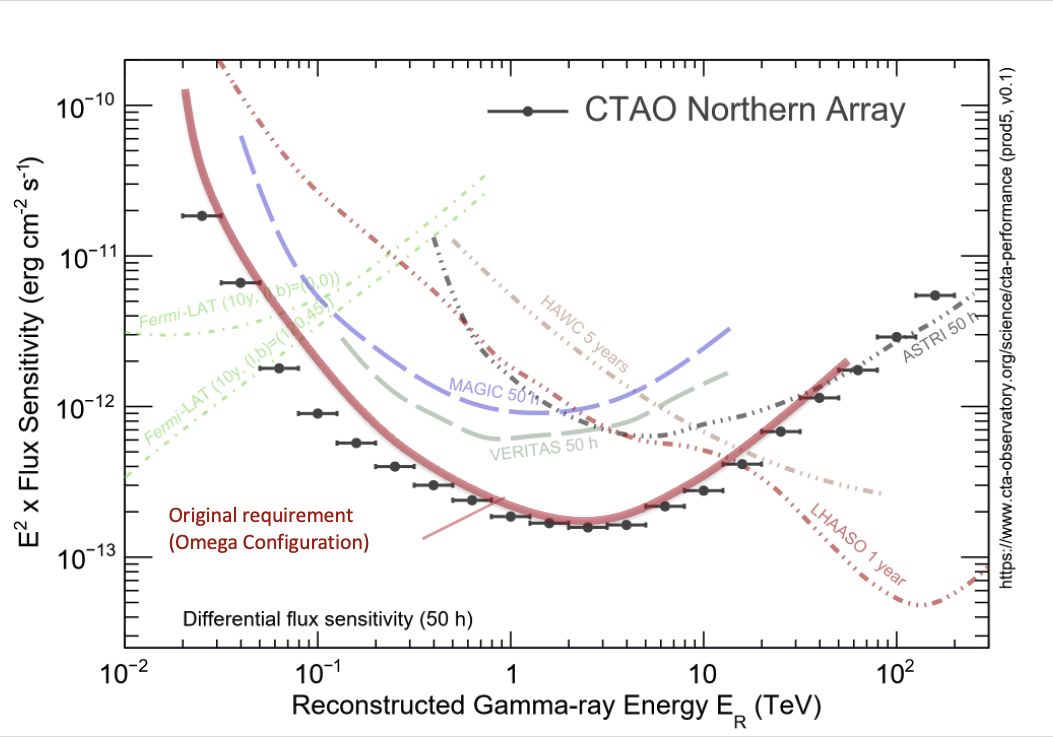}
\includegraphics[width=8cm]{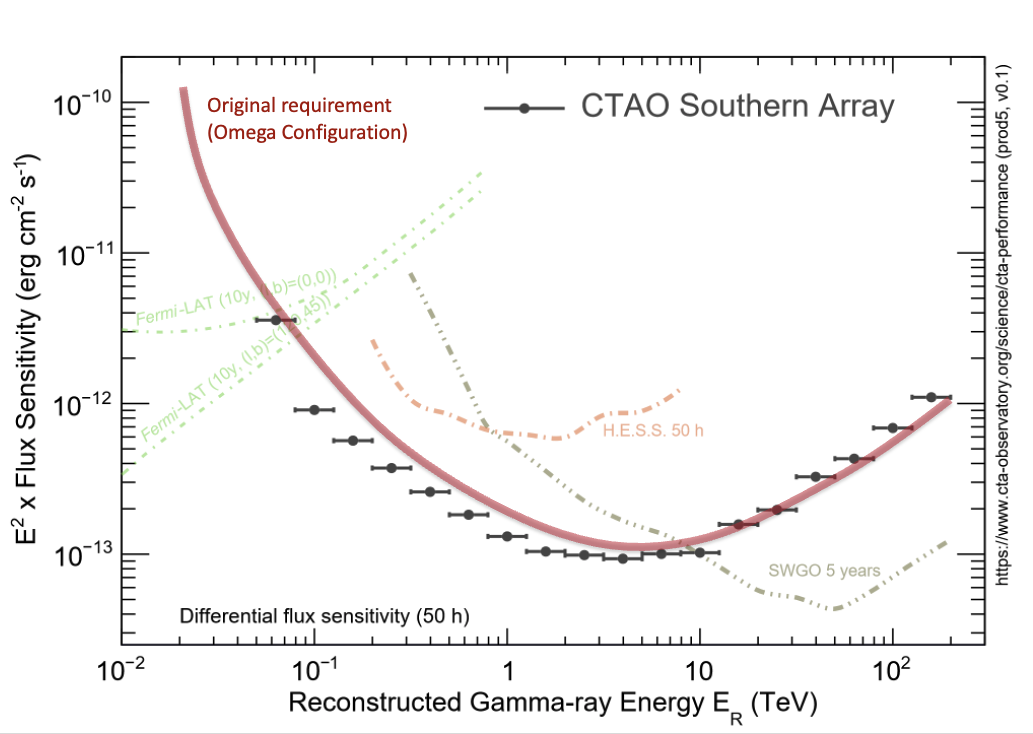}
\caption{CTA flux sensitivity for the Northern (top) and Southern (bottom) arrays, on axis, for 50 h observation time, including the sensitivity of other gamma-ray instruments. The red line indicates the sensitivity requirement.}
\label{fig_sens2}
\end{center}
\end{figure}

\begin{figure}[htbp]
\begin{center}
\includegraphics[width=7cm]{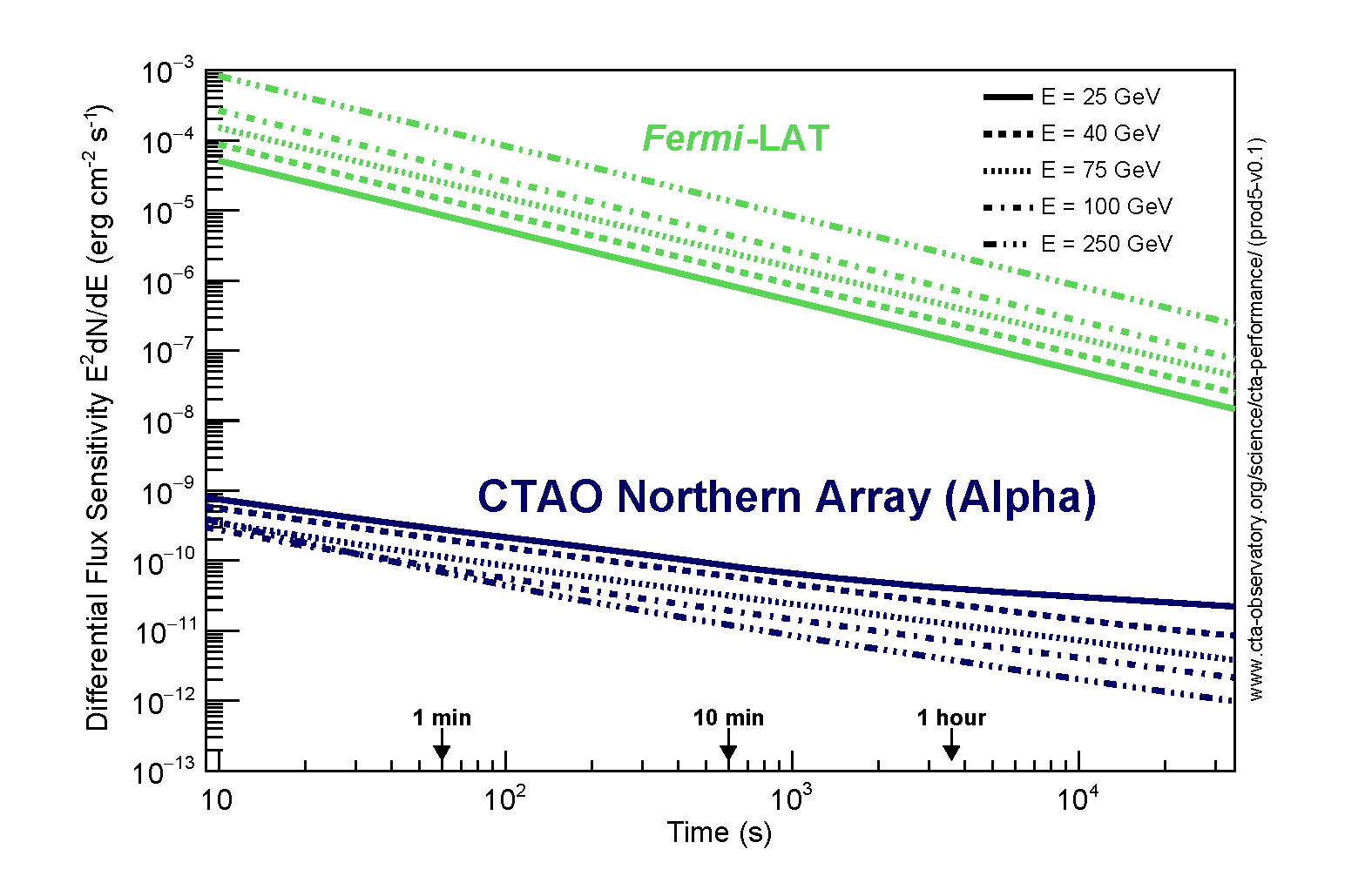}
\caption{Comparison of CTA and Fermi-LAT sensitivity at 25, 40, 75, 100 and 250 GeV, as a function of observation time.}
\label{fig_sens_time}
\end{center}
\end{figure}

\begin{figure}[htbp]
\begin{center}
\includegraphics[width=5.5cm]{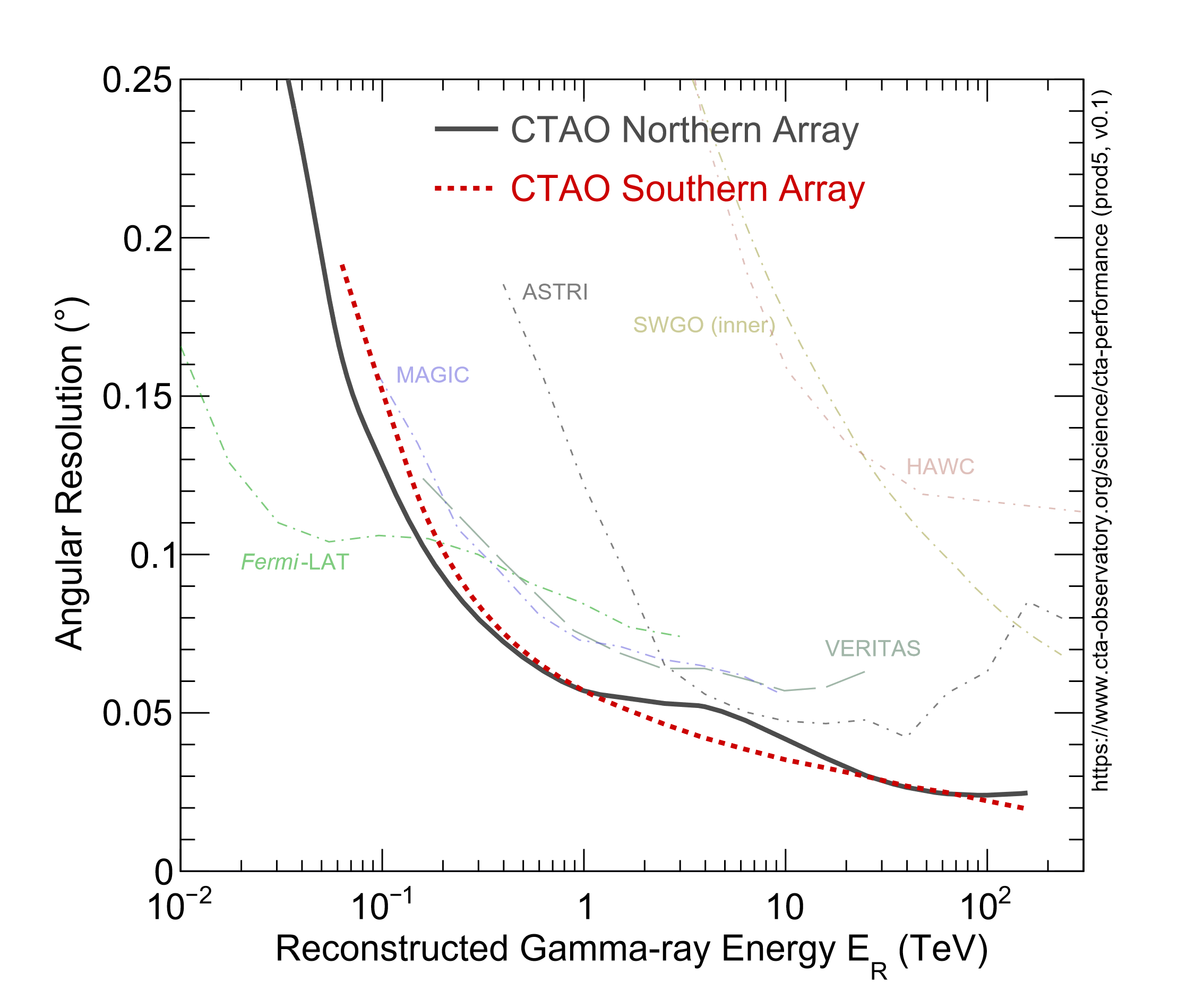}
\includegraphics[width=4.7cm]{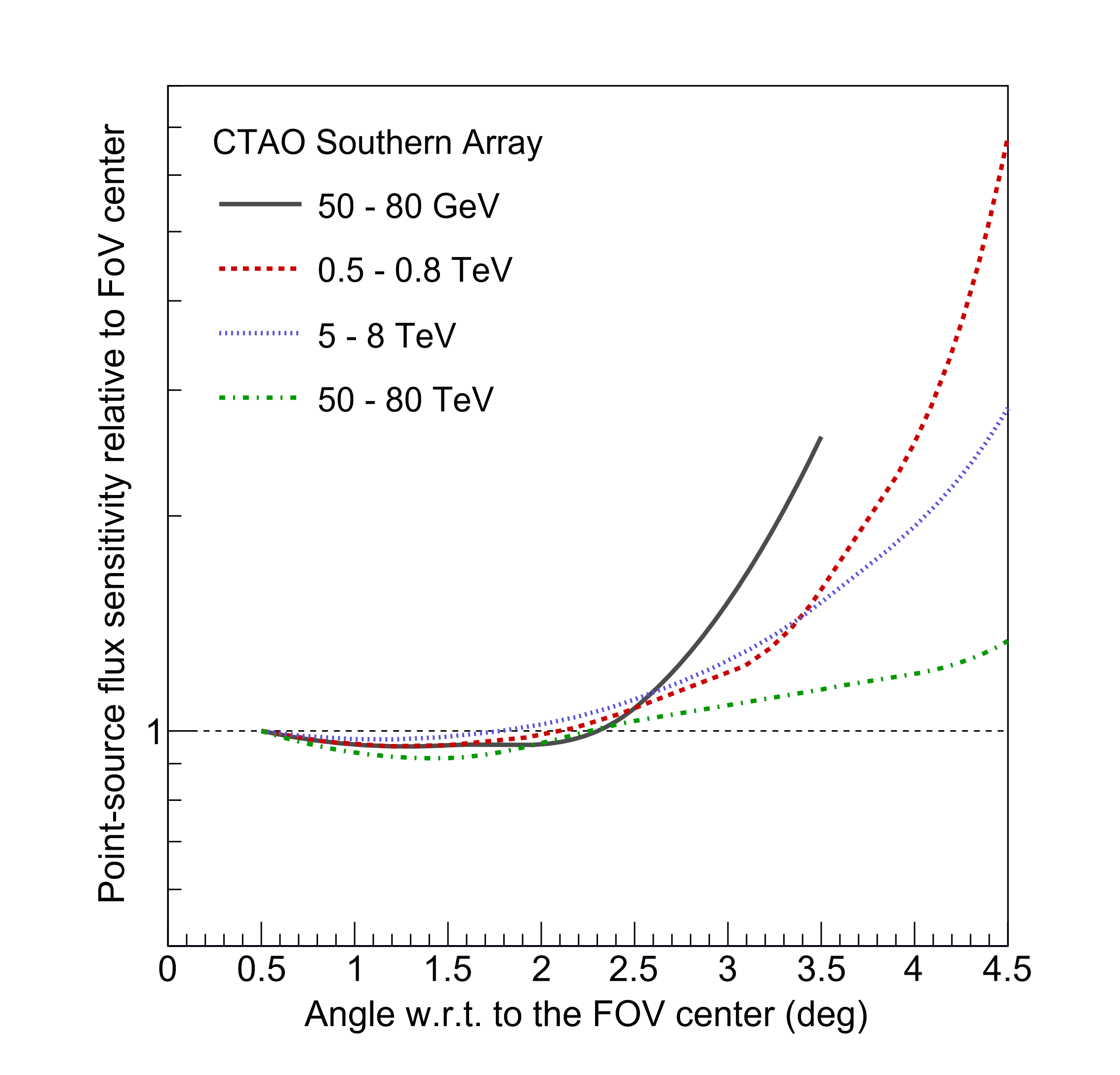}
\caption{Left: Angular resolution of the CTA Northern and Southern arrays, compared with other gamma-ray instruments. Right: sensitivity as a function of angle from the pointing direction, for the Southern array and different energies.}
\label{fig_angres}
\end{center}
\end{figure}

The instrument performance of CTA in its `alpha configuration' is summarised in Figs.~\ref{fig_sens} to ~\ref{fig_angres}.  Fig.~\ref{fig_sens} and Fig.~\ref{fig_sens2} show the differential sensitivity per 0.2 decade energy band, for the Northern and Southern arrays. The improved sensitivity of the Southern array at TeV energies reflects the larger number of MSTs, at high energies the sensitivity is dominated by the SSTs. Apart from energies below 50 GeV for the Southern array and at the highest energies where sensitivity is affected by the reduced number of telescopes, the `alpha configuration' basically meets the CTA sensitivity targets (red lines in Fig.~\ref{fig_sens2}).

Fig.~\ref{fig_sens2} compares the sensitivity of CTA with that of H.E.S.S., MAGIC and VERITAS, and of the air shower arrays HAWC and LHAASO, as well as Fermi-LAT. Compared to these  IACTs with peak sensitivities in the $10^{-12}$~erg/s cm$^2$ domain, CTA opens the  $10^{-13}$~erg/s cm$^2$ domain. At energies beyond some 10 TeV, air shower arrays and in particular LHAASO provide superior performance for steady sources (comparing 1 year of LHAASO observations with 50 h of CTA data), but CTA is complementary in that it provides significantly better angular resolution, relevant for resolving structures such as observed e.g. in the PeVatron candidate HESS J1702-420, where a compact region provides a hard spectrum, surrounded by a soft spectrum region \cite{2021arXiv210606405A}.

Below about 70~GeV, Fermi-LAT sensitivity shown in Fig.~\ref{fig_sens2} surpasses CTA sensitivity. This is, however, somewhat misleading since Fermi-LAT sensitivity refers to about a decade of LAT data whereas CTA sensitivity refers to 50 h. The sensitivity for flaring sources, with flare time scales between 10 s and $10^4$ s, is compared in Fig.~\ref{fig_sens_time}. Even at energies as low as 25 GeV, CTA improves upon Fermi-LAT by a factor of $10^4$ or more; at higher energies (250 GeV) and very short times the advantage grows to $10^6$, very relevant for time-domain gamma-ray astronomy.

Fig.~\ref{fig_angres} (left) finally illustrates the angular resolution as a function or energy, saturating at about 2 arc-min at high energy, well below the angular resolution achieved by air shower arrays.
The gamma-ray field of view of the CTAO
arrays, defined as twice the angular offset from the array 
pointing direction at which the differential point-source sensitivity
is degraded by a factor of two, is illustrated in Fig.~\ref{fig_angres} (right). The flux sensitivity is rather constant up to $2.5^\circ$ from the optical axis; at high energies, the effective field of view grows significantly.

%% file: SubSecKSP.tex
Comprehensively addressing the science themes introduced in Section~\ref{SecMotivation} requires large and coherent data sets, including sky surveys for a census of cosmic accelerators, deep observations of key objects, long-term observations of variable sources, and rapid follow-up of transient phenomena. The CTA Key Science Projects (KSPs) aim at providing such legacy data sets and data products, that will be of high value to the wider community. 
Options for CTA KSPs has been defined through a multi-year process of discussion within the CTA Consortium and in interaction with the community. These are ambitious projects with very significant scientific promise, frequently challenging in their data analysis,  and requiring considerable observation time. It is assumed that about 40\% of the total available observing time integrated over 10 yr  (see Fig.~\ref{fig_obstime}) will be granted to KSPs, i.e. the KSP Time.  KSP proposals will be peer-reviewed by the CTAO's Time Allocation Committee.

A possible set of KSPs are described in detail in \cite{CTAConsortium:2017dvg}, to which the following discussion refers. Proposed KSPs include the following surveys:
\begin{itemize}
\item{The Galactic Plane Survey} will provide a valuable legacy product to the entire astronomical community – the first sensitive VHE scan of the entire Galactic plane, with uniform coverage down to mCrab sensitivity, described in Section~\ref{SubSubSecGPS}.
\item{The Galactic Centre KSP} contains a deep exposure of the inner few degrees of our Galaxy, complemented by an extended survey to explore the regions to the edge of the bulge. In addition to the most promising Dark Matter target, the region within a few degrees of the Galactic Centre contains a wide variety of possible high-energy emitters, including the closest supermassive black hole, dense molecular clouds, strong star-forming activity, multiple supernova remnants and pulsar wind nebulae, arc-like radio structures, and the base of the Fermi bubbles.
\item{The Survey of the Large Magellanic Cloud} as a very active star-forming galaxy provides data towards population studies on supernova remnants and pulsar wind nebulae, transport of cosmic rays from their release into the interstellar medium to their escape from the system, and the search for signatures of dark matter annihilation. The LMC with its high star formation rate hosts extraordinary objects, including the starforming region 30 Doradus, the super star cluster R136, the supernova SN1987A, and the 30 Dor C superbubble.
\item{The  Extragalactic Survey} is proposed to cover 25\% of the sky. The survey is unique in that it opens a new window for the search of extragalactic sources as well as it being the first attempt for a complete log N - log S study of close-by blazars in VHE gamma rays.
\end{itemize}
KSPs could also address deep and/or long-term monitoring observations of key objects.
Topics for such KSPs are discussed in \cite{CTAConsortium:2017dvg}, and include:
\begin{itemize}
\item{The Cosmic Ray PeVatrons KSP} aims at identifying sources of PeV hadronic cosmic rays, by deep follow-up of candidate sources that do not exhibit cutoffs in their energy spectra, and by using CTA's excellent angular resolution to investigate source morphology and probe emission mechanisms.
\item{The Star Forming Systems KSP:} Cosmic rays are believed to be an important regulator of the star-formation process. The KSP aims to relate star formation and gamma ray / cosmic ray production across a wide range of scales, from Galactic star formation regions to starburst galaxies and ultra-luminous infrared galaxies.
\item{The  Transients KSP} exploits CTA's unique sensitivity for transient phenomena and short-timescale variability (see Fig.~\ref{fig_sens_time})  to follow up on gamma-ray bursts; galactic transients; X-ray, optical and radio transients detected by ``transient factory'' facilities; high-energy neutrino transients; gravitational wave transients; and serendipitous VHE transients.
\item{The Active Galactic Nuclei KSP} probes the nature of AGN jets, the mechanisms that accelerate particles, and the origin of the rapid variability of gamma-ray AGN. The AGN KSP aims to provide a comprehensive data set by long-term monitoring of the variability of selected AGN of different classes, see Section~\ref{SubSubSecAGN}.
\item{The Clusters of Galaxies KSP:} galaxy clusters represent the latest stage of structure formation, and are expected to be reservoirs of cosmic rays accelerated by structure formation processes, galaxies and active galactic nuclei. Goal of the KSP is to detect, for the first time, diffuse gamma-ray emission from a cluster of galaxies, and to study the clusters’ cosmic-ray acceleration, propagation and confinement properties.
\end{itemize}
The Dark Matter Programme, aiming to detect annihilation radiation from massive Dark Matter particles, with its characteristic spectral and spatial features, is based on data sets collected in KSPs; relevant targets are the Galactic Center, but also dwarf galaxies, the LMC, and clusters of galaxies. CTA's capabilities are illustrated below in Section~\ref{SubSubSecDM}.

\begin{figure}[htbp]
\begin{center}
\includegraphics[width=11cm]{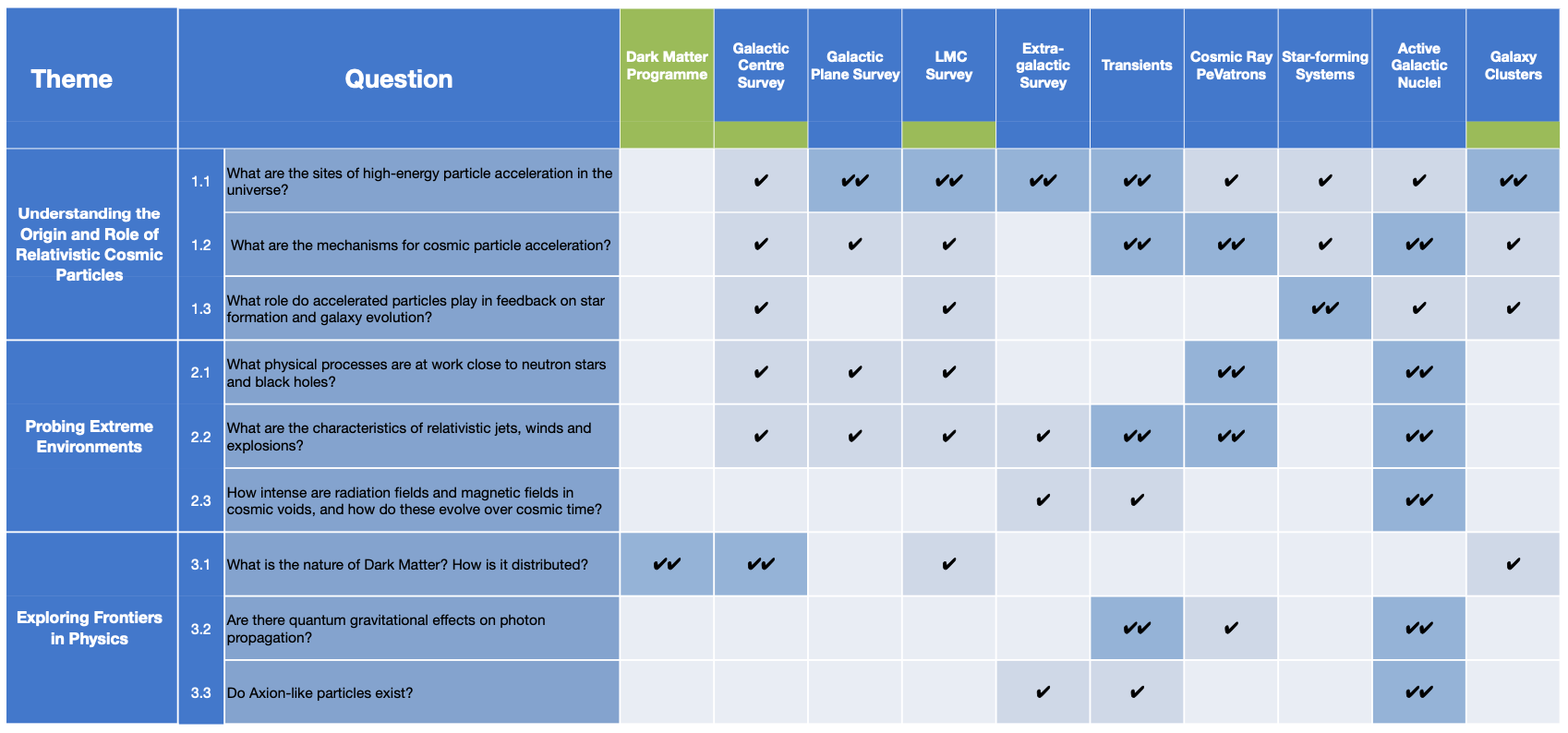}
\caption{Matrix of CTA science questions and proposed Key Science Projects (KSPs). KSPs which contribute to the programme aimed at dark matter detection are indicated in green. The check marks are intended to give a qualitative assessment of the impact of each KSP on a particular science question. From \cite{CTAConsortium:2017dvg}.}
\label{fig_KSP_matrix}
\end{center}
\end{figure}

As illustrated by the Dark Matter Programme, the discussed  KSPs consist of sets of observations addressing multiple science questions within the CTA themes, as shown in the matrix of CTA science questions and Key Science Projects described in \cite{CTAConsortium:2017dvg} (Fig.~\ref{fig_KSP_matrix}). For example, data from the AGN KSP also enable  measurements of the Extragalactic Background Light through the detection of spectral features of astrophysical sources placed at cosmological distance (see Section~\ref{SubSubSecEBL}), as well as the estimation of the intergalactic magnetic field. 

Of course, KSPs are not static; they will be updated, both when formally proposed to CTAO and likely also during their multi-year execution, reflecting (i) the evolution of the field in general, (ii) the improved understanding of CTA’s capabilities, (iii) the capabilities and results from other multi-wavelength and multi-messenger facilities. Most KSPs rely on extensive data from such other facilities (see e.g. \cite{BarresdeAlmeida:2019pui}), for target selection and Target of Opportunity (ToO) alerts, as well as for characterisation of the objects and determination of their wide-band spectral energy distributions.

%% file: SubSubSecGPS.tex
\subsubsection{Surveying the Galactic Plane}
\label{SubSubSecGPS}

\begin{figure}[htbp]
\begin{center}
\includegraphics[width=11cm]{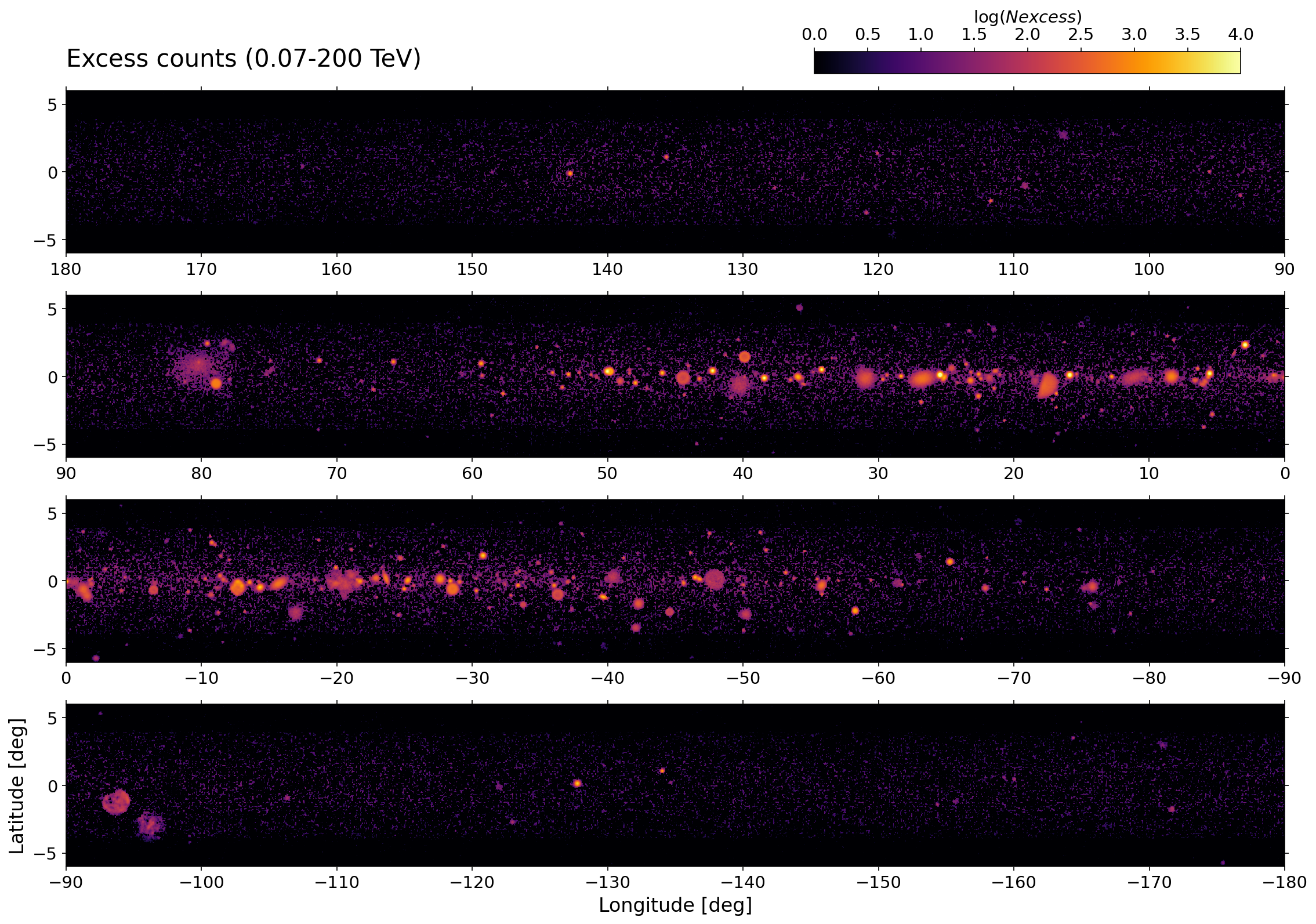}
\caption{Simulated excess counts above background for the CTA Galactic Plane Survey, in the energy range 0.07 - 200 TeV. From \cite{CTA_GPS}}
\label{fig_GPS1}
\end{center}
\end{figure}

\begin{figure}[htbp]
\begin{center}
\includegraphics[width=6.0cm]{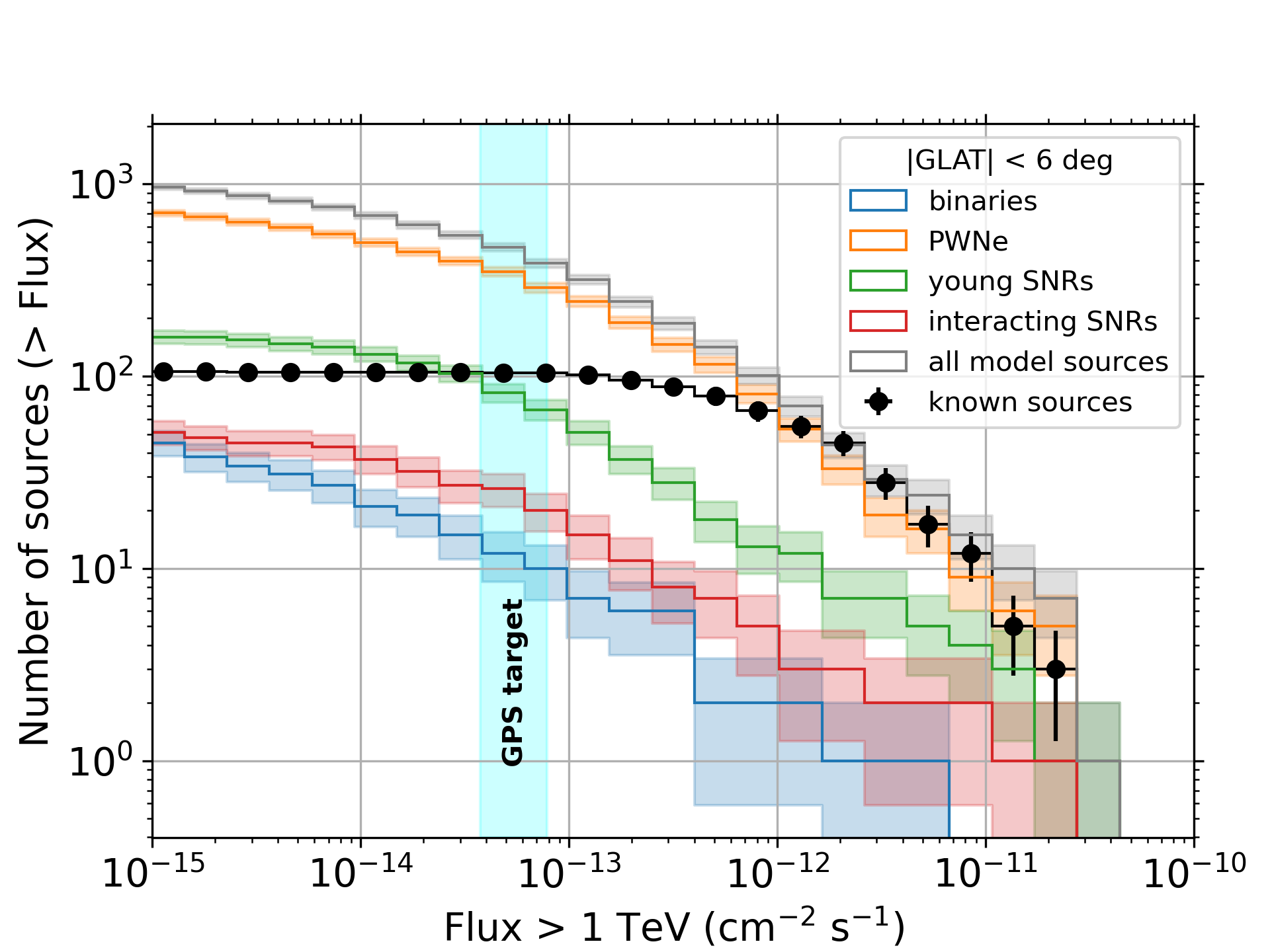}
\includegraphics[width=5.5cm]{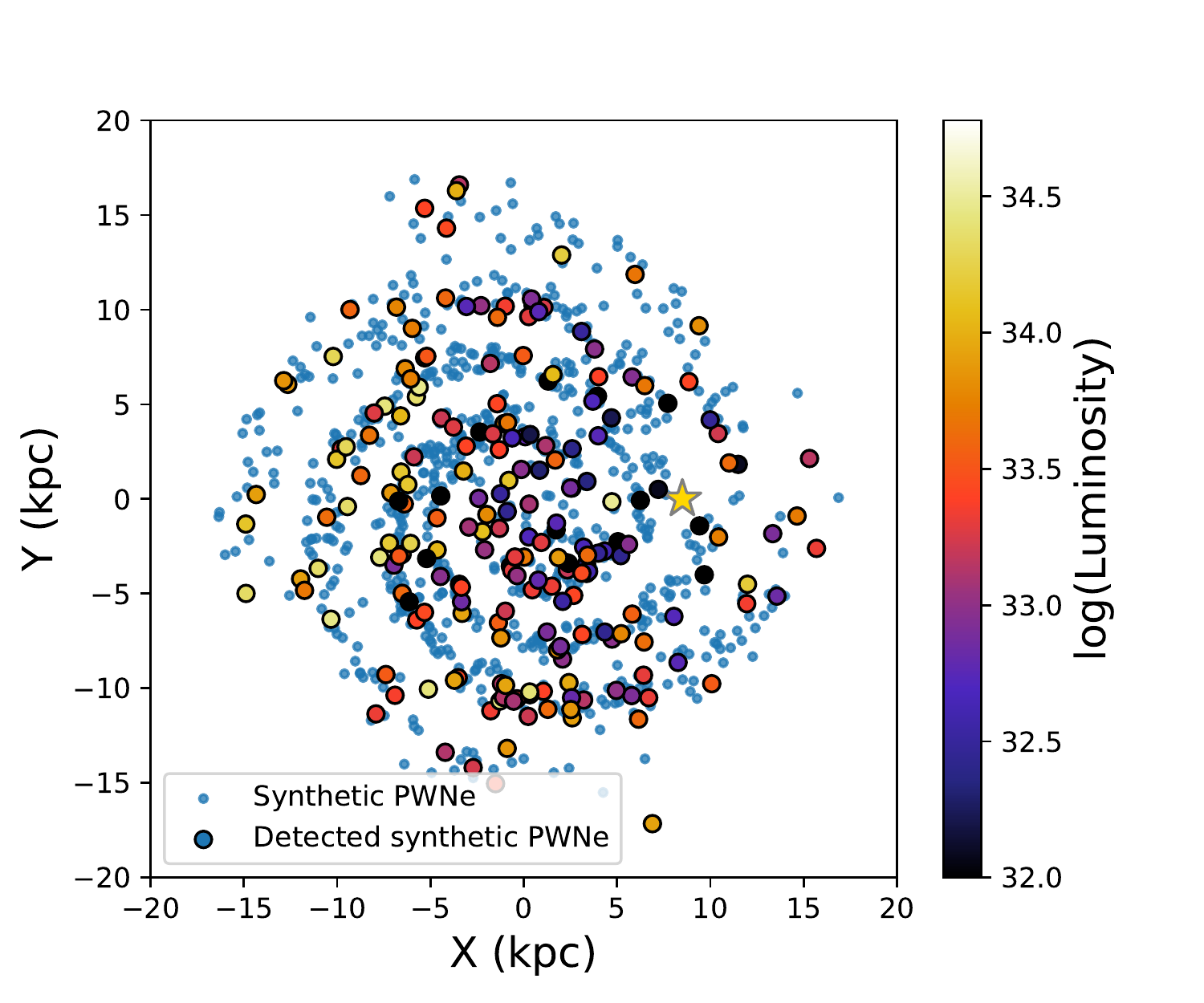}
\caption{Left: Cumulative number of sources as a function of source flux, showing known emitters and the synthetic source populations from the sky model for different source classes. The vertical band shows the target sensitivity for the CTA survey. Right: The synthetic PWN population and corresponding detections. From \cite{CTA_GPS}.}
\label{fig_GPS2}
\end{center}
\end{figure}

Surveys provide the basis both for studies of ensembles of sources, and for identification of interesting targets for deeper follow-up observations. Here, the Galactic plane survey is presented in more detail. Surveys in the TeV domain by Cherenkov telescopes such as H.E.S.S. or ground-based arrays such as HAWC reach point-source sensitivities approaching 1\% of the Crab flux, and have revealed about 100 sources in the Galactic plane. None of the individual surveys covers the whole Galactic plane in a uniform fashion. The goal of CTA is to survey the whole Galactic plane at a sensitivity approaching the milli-Crab level. The proposed survey would require a total of 1620 hours split between the northern and southern arrays, to be carried out over several observing periods. A thorough study showed that a non-equilateral double row pattern with steps in longitude of 2.25$^{\circ}$, spaced in latitude by 1.95$^{\circ}$, provides the pointing grid with the best sensitivity in the Galactic plane, up to $4^\circ$ in latitude. 
Detailed simulations of the survey were carried out \cite{CTA_GPS}, using a very elaborate source model matched to existing data. Fig. \ref{fig_GPS1} illustrates a simulated survey, showing the gamma-ray excess counts above background.   About 500 sources are detected, as also shown in Fig. \ref{fig_GPS2} (left), where the cumulative number of simulated sources is shown as a function of flux, levelling off at about 100 sources for the $\approx 10^{-12}$ cm$^{-2}$s$^{-1}$ sensitivity of current surveys, and reaching about 5 times that number for the CTAO target sensitivity. In particular, CTA is able to detect objects like Pulsar Wind Nebulae (PWN) and Supernova Remnants (SNRs) across the full Galaxy, important for proper population synthesis (Fig. \ref{fig_GPS2} (right)). The Galactic plane survey and the resulting source catalogs will be among the key CTAO data products.

%% file: SubSubSecAGN.tex
\subsubsection{Understanding Active Galactic Nuclei}
\label{SubSubSecAGN}

\begin{figure}[htbp]
\begin{center}
\includegraphics[width=11cm]{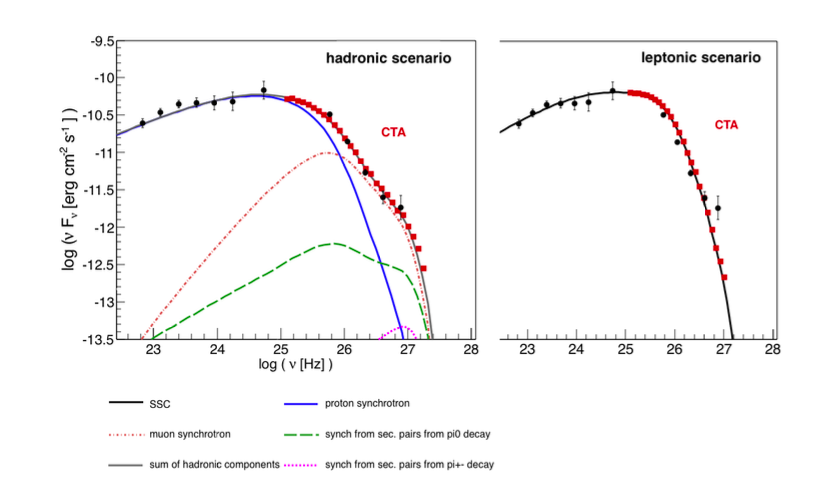}
\caption{A comparison of the expected CTA spectra for two specific (simple) emission models for the blazar PKS2155-304. A hadronic scenario, where high-energy emission is caused by proton- and muon-synchrotron photons and secondary emission from proton-photon interactions, is shown on the left, and a standard leptonic synchrotron self-Compton (SSC) model on the right. The exposure time assumed for the simulations (33h) is the same as the live time for the H.E.S.S. observations (black data points above $3 \cdot 10^{25}$ Hz). The statistical uncertainties in the CTA data points are smaller than the red squares. From \cite{CTAConsortium:2017dvg}.}
\label{fig_AGN1}
\end{center}
\end{figure}

\begin{figure}[htbp]
\begin{center}
\includegraphics[width=11cm]{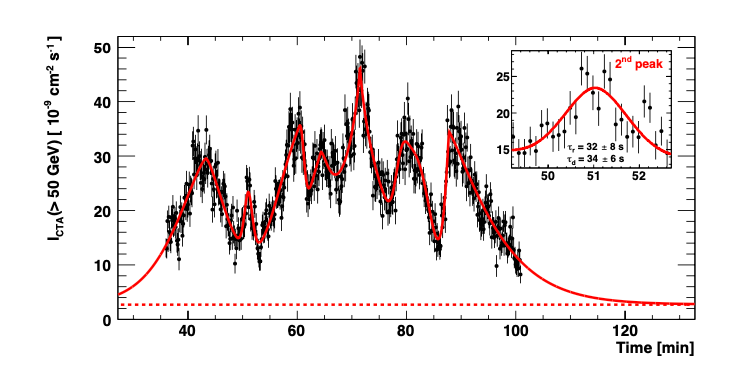}
\caption{Simulated light curve based on an extrapolation of the power spectrum of the 2006 flare from PKS 2155-304. The doubling rise time and decay time are indicated for the second peak in insert. From \cite{CTAConsortium:2017dvg}.}
\label{fig_AGN2}
\end{center}
\end{figure}

About half of the known VHE gamma-ray sources are Active Galactic Nuclei, where a pair of counter-propagating relativistic jets is launched in the vicinity of a supermassive black hole (SMBH). Some of these sources show extreme variability, on time scales much shorter than the light-crossing time of the SMBH. The exact mechanism of jet launching, the connection between jets and accretion onto the black hole, the composition of the jets, the mechanisms of particle acceleration in the jets, and the origins of the variability of gamma-ray emission are not fully understood. Phenomenological models postulate particle acceleration within a uniform region in the jet (a ``blob'' of certain size), where gamma rays are produced in interactions of accelerated particles either with self-generated (synchrotron) photon fields, or external fields. High-quality gamma-ray spectra obtained for different AGN classes (UHBL, HBL, IBL, LBL, FSRQs, and radio galaxies) will help to pin down the nature of the emitting particles, and their spectra (see Fig. \ref{fig_AGN1} for an example). About 500 h of observations are proposed to provide legacy data sets. AGN variability is probed by a combination of unbiased long-term monitoring of key objects representative of AGN classes, requiring close to 180 h per year, and the follow-up of flaring AGN, see Fig. \ref{fig_AGN2} for an examples of CTA's capability of resolving AGN flares. CTA data taking should be supported by observations at other wavelengths, to obtain high-quality wide-band simultaneous SEDs of AGNs. The AGN data will also be used to probe photon propagation, with possible effects due to quantum gravity or coupling to axions, and will allow probing the density of extragalactic background light (EBL, see Section \ref{SubSubSecEBL}) and the level of intergalactic magnetic fields (see  \cite{CTAConsortium:2017dvg,2021JCAP...02..048A} for details).

%% file: SubSubSecEBL.tex
\subsubsection{Measurement of the EBL intensity}
\label{SubSubSecEBL}

\begin{figure}[htbp]
\begin{center}
\includegraphics[width=8cm]{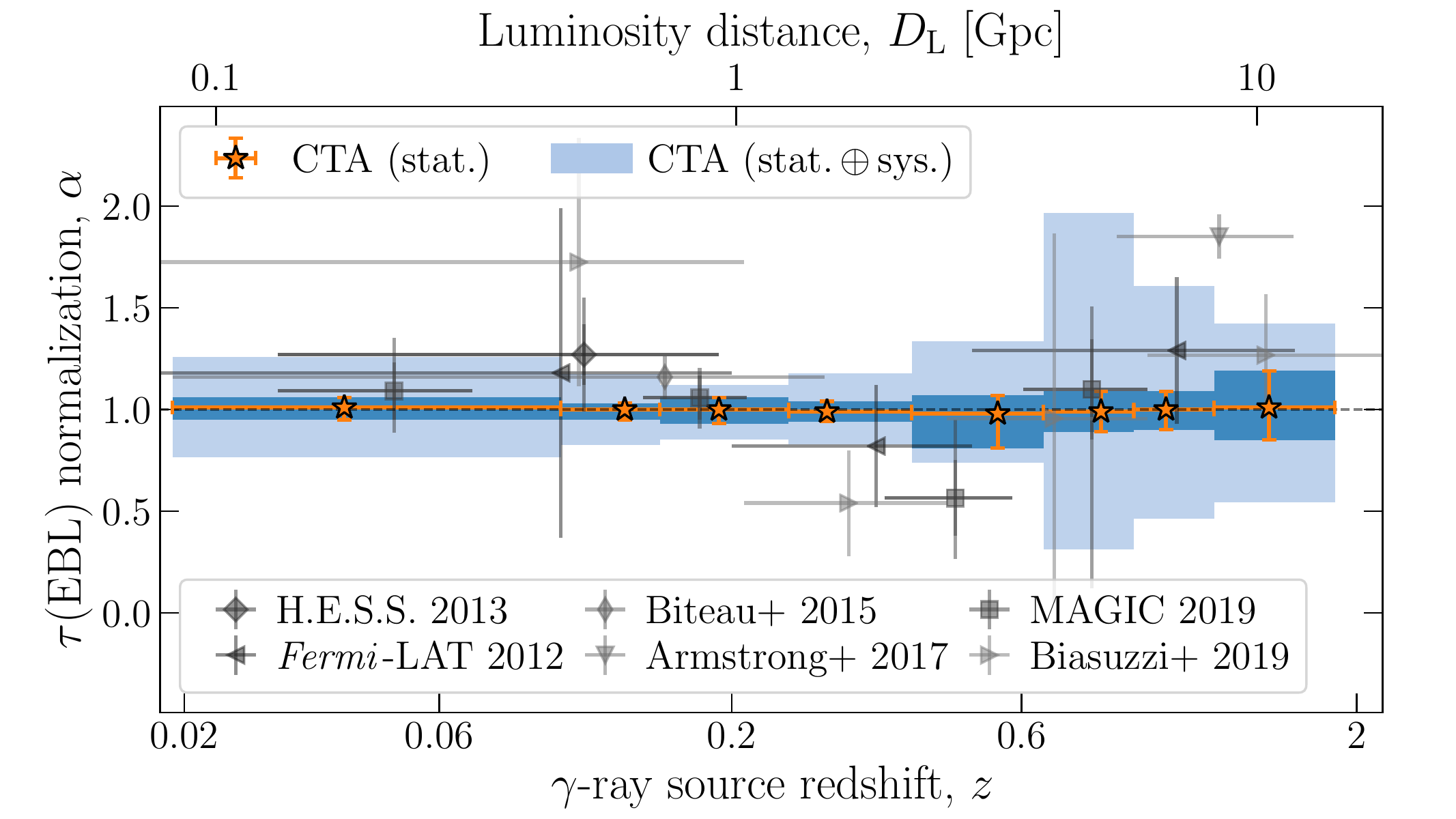}
\caption{Projected CTA constraints on the EBL scale factor (relative to a nominal EBL) as a function of redshift of the gamma-ray sources. The median reconstructed scale factor and the 16–84\% quantiles of the distribution in each redshift bin are shown as orange stars and error bars (also dark blue shaded regions), respectively. The accumulated effect of statistical uncertainties and systematic uncertainties resulting from changes in the energy scale and effective area are illustrated with the light blue shaded regions. From \cite{2021JCAP...02..048A}.}
\label{fig_EBL}
\end{center}
\end{figure}

The universe is not fully transparent for gamma rays; pair production on the CMB and on infrared and visible photon fields limits the range of gamma rays. The threshold for pair production is approximately $E_\gamma \mbox{[TeV]} \approx  1/E_T \mbox{[eV]} $ where $E_T$ is the characteristic energy of the target photons. While severely limiting gamma-ray escape from radiation-intense environments, the attenuation of gamma rays travelling through intergalactic space allows probing the level of extragalactic background light (EBL), which is difficult from the Earth due to solar-system foregrounds. The level of EBL is a very interesting cosmological quantity, since it reflects the (red-shifted) light from all stars since the birth of the Universe. AGNs at cosmological distances are used as reference sources. To identify the threshold and level of attenuation, a model for the intrinsic spectra of sources is required, usually based on the extrapolation from lower (GeV) energies where attenuation is negligible, and on characteristics from nearby sources. Fig. \ref{fig_EBL} illustrates the precision that can be reached with CTA in determining the redshift evolution of the EBL up to redshifts of 2 \cite{2021JCAP...02..048A}. Vice versa, for a known EBL level, the data can be used to measure source distance, and in particular to constrain the Hubble constant.

%% file: SubSubSecDM.tex
\subsubsection{Search for Dark Matter annihilation}
\label{SubSubSecDM}

\begin{figure}[htbp]
\begin{center}
\includegraphics[width=6cm]{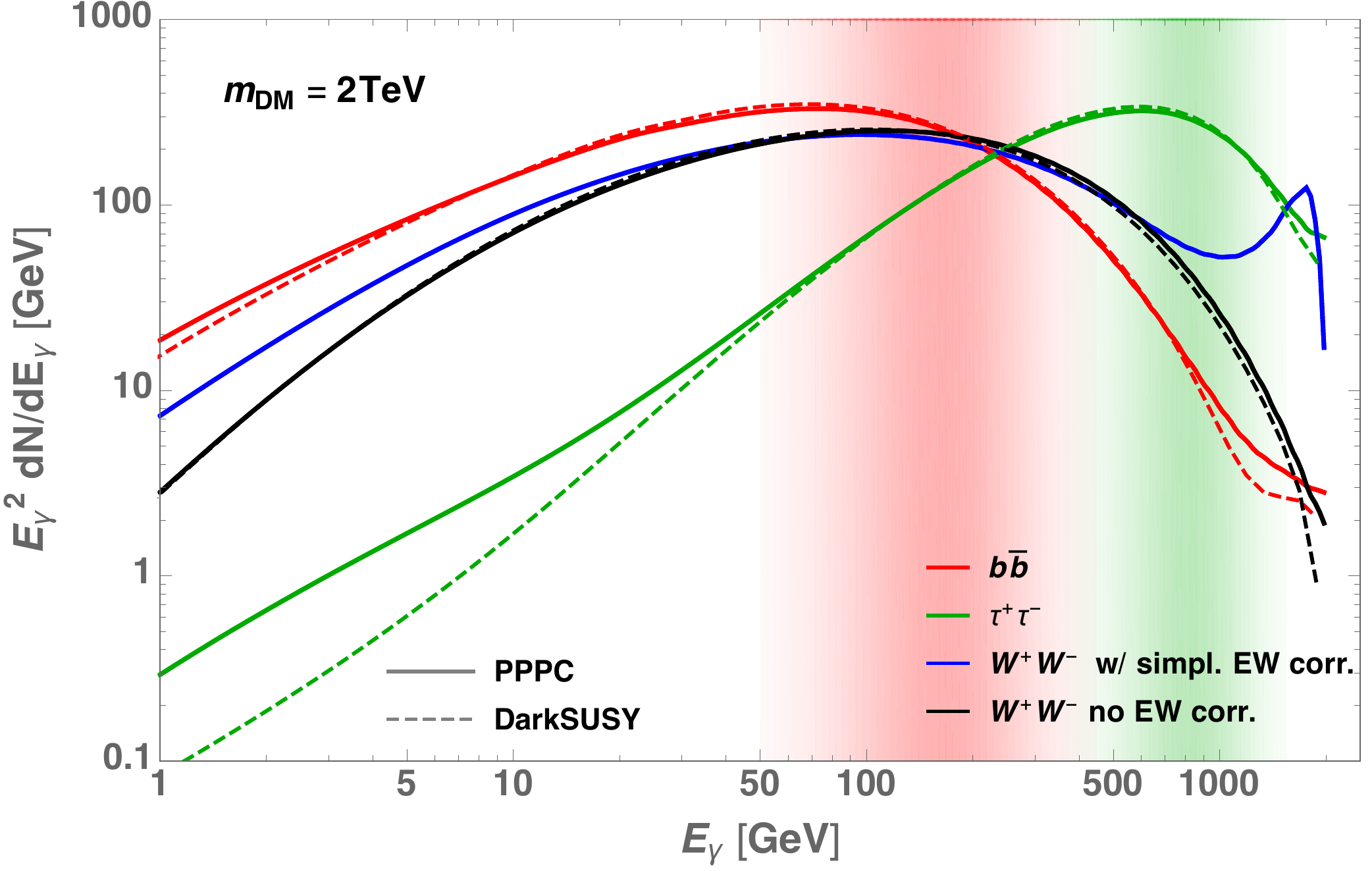}
\caption{Summary of benchmark DM spectra for various final states, indicated by solid lines with different colours. Dashed lines show the corresponding spectra obtained with an alternative event generator. From \cite{2021JCAP...01..057A}.}
\label{fig_DM_spectra}
\end{center}
\end{figure}

\begin{figure}[htbp]
\begin{center}
\includegraphics[width=5.5cm]{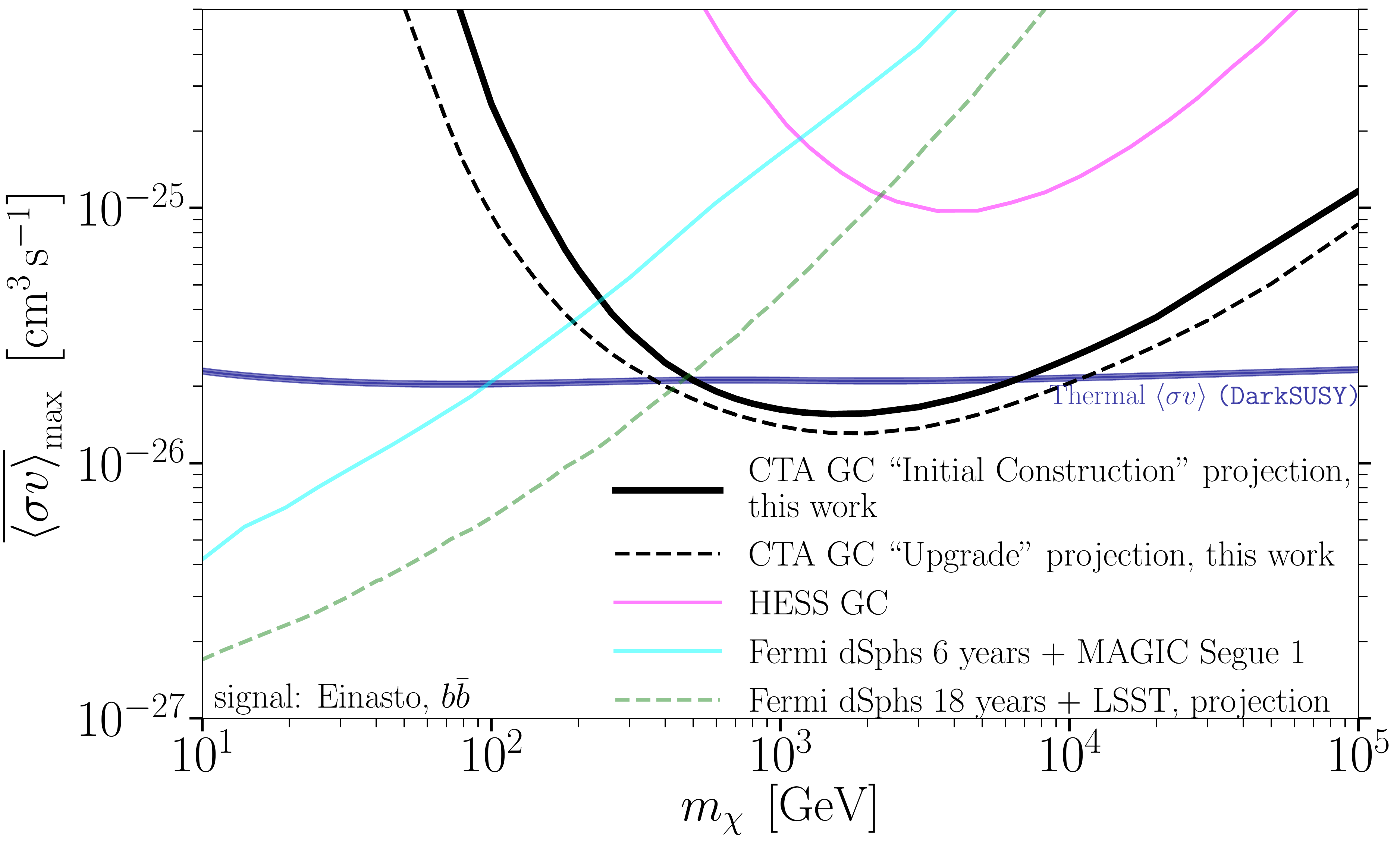}
\includegraphics[width=5.5cm]{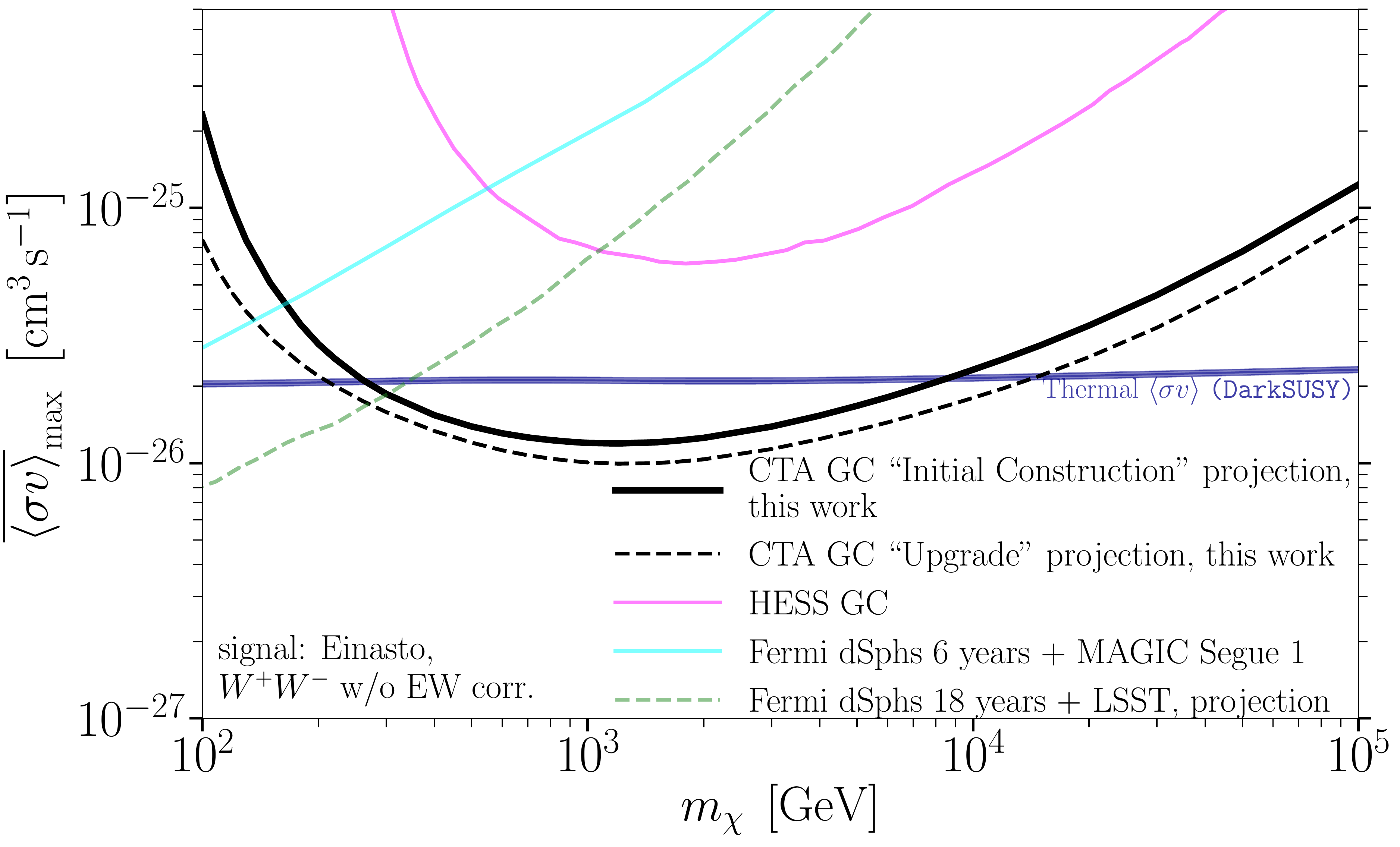}
\caption{CTA sensitivity for dark matter annihilation at the Galactic centre, for the $b \bar{b}$ and $W^+W^-$ annihilation modes, assuming an Einasto dark matter profile. Also shown are limits from current instruments. The full lines labeled `CTA Initial Construction' refer to a telescope configuration that is similar to the `alpha configuration', the black dashed lines refer to the `omega configuration'. From \cite{2021JCAP...01..057A}.}
\label{fig_DM_limits}
\end{center}
\end{figure}

An important aspect of the CTA science programme concerns fundamental physics, addressing questions like (a) what is the nature of dark matter, (b) are there quantum gravitational effects on photon propagation, or (c) do axion-like particles exist? Here, the focus is on one of these topics, the search for WIMP-like dark matter annihilating into standard model particles. Dark matter annihilation rates are proportional to the square of the dark matter density, hence objects with peaks in dark matter density are suitable to search for annihilation signatures. Among galaxy clusters, dwarf or satellite galaxies and the Galactic centre, the latter is by far the most promising source, due to its combination of dark matter enhancement and proximity.  Predictions for the annihilation rate evidently scale with the annihilation cross section, but are significantly dependent on the poorly known dark matter profiles at the centre of our Galaxy. The gamma-ray spectra from annihilations depend on the annihilation channel (see Fig.~\ref{fig_DM_spectra}), but for a given channel are reasonably well predicted. An expected (velocity-weighted) annihilation cross section of $< \sigma v> \approx 2 \cdot 10^{-26}$ cm$^3$/s can be derived from the condition that the right amount of dark matter survives in the early Universe. To derive the sensitivity of CTA to dark matter annihilation  \cite{2021JCAP...01..057A}, a template likelihood analysis was used to distinguish the predicted dark matter annihilation gamma-ray flux against the background, in bins of energy and distance from the Galactic centre. The analysis also considered systematic uncertainties, e.g. due to uncertainties in instrument response or in the determination of the background. Fig. \ref{fig_DM_limits} shows the resulting upper limits on the annihilation cross section for two different annihilation modes, $b \bar{b}$ and $W^+W^-$, together with other limits (as of 2021), based on about 800 h of observations of the Galactic centre region. CTA will, for the first time, probe the predicted annihilation cross section for TeV WIMP mass scales.
 
%% file: SecConclusions.tex
\footnote{Adapted from the Executive Summary of `Science with CTA' \cite{CTAConsortium:2017dvg}}  The Cherenkov Telescope Array, CTA, will be the major global observatory for very high energy gamma- ray astronomy starting from the late 2020ies, for a decade and beyond. The scientific potential of CTA is extremely broad: from understanding the role of relativistic cosmic particles to the search for dark matter. CTA is an explorer of the extreme universe, probing environments from the immediate neighbourhood of black holes to cosmic voids on the largest scales. Covering a wide range in photon energy from 20 GeV to 300 TeV, CTA will improve on all aspects of performance with respect to current instruments. Wider field of view and improved sensitivity will enable CTA to survey the gamma-ray sky hundreds of times faster than previous TeV telescopes. The angular resolution of CTA will approach 1 arc-minute at high energies — the best resolution of any instrument operating above the X-ray band — allowing detailed imaging of a large number of gamma- ray sources. A one to two order-of-magnitude collection area improvement makes CTA a powerful instrument for time-domain astrophysics, three orders of magnitude more sensitive on hour timescales than Fermi-LAT at 30 GeV. The observatory will operate arrays on sites in both hemispheres to provide full sky coverage and will hence maximise the potential detection of rare phenomena such as very nearby supernovae, gamma-ray bursts or gravitational wave transients. With at least 51 telescopes on the southern site and 13 telescopes on the northern site, flexible operation will be possible, with sub-arrays available for specific tasks.
CTA will have important synergies with many of the new generation of major astronomical and astroparticle observatories. Multi-wavelength and multi-messenger approaches combining CTAO data with those from other instruments will lead to a deeper understanding of the broad-band non-thermal properties of target sources, elucidating the nature, environment, and distance of gamma-ray emitters. 

The CTA Observatory will be operated as an open, proposal-driven observatory, with all data available on a public archive after a pre-defined proprietary period (of typically one year). Scientists from institutions worldwide have combined together to the CTA Consortium that has prepared a proposal for a core programme of KSPs \cite{CTAConsortium:2017dvg}. The science cases have been prepared over several years by the CTA Consortium, with community input gathered via a series of workshops connecting CTA to neighbouring communities. Well before the observatory formally starts operations, this scientific collaboration will be re-organised, with governance and membership reflecting the countries' contributions to the CTAO construction, and will be preparing, proposing and analysing the KSPs. 
A major element of the science programme is the search for dark matter via the annihilation signature of weakly interacting massive particles (WIMPs). The strategy for dark matter detection places the expected cross-section for a thermal relic within reach of CTA for a wide range of WIMP masses from $\approx$200 GeV to 20 TeV. This makes CTA extremely complementary to other approaches, such as high-energy particle collider and direct-detection experiments. CTA will conduct a census of particle acceleration over a wide range of astrophysical objects, with quarter-sky extragalactic, full-plane Galactic and Large Magellanic Cloud surveys planned. Additional KSPs can be focused on transients, acceleration up to PeV energies in our own Galaxy, active galactic nuclei, star-forming systems on a wide range of scales, and the Perseus cluster of galaxies. All provide high-level data products which will benefit a wide community, and together they will provide a long-lasting legacy for CTAO.
Finally, while designed for the detection of gamma rays, CTAO has considerable potential for a range of astrophysics and astroparticle physics based on charged cosmic-ray observations and the use of its Cherenkov telescopes for optical measurements.

%% file: Acknowledgements.tex
This overview of CTA reflects and summarises the work of well over a decade of the CTA Consortium and -- more recently -- of the CTA Observatory. Material presented is taken from numerous papers and documents authored by the CTA Consortium as a whole, by individuals within the Consortium, or by CTAO staff. In particular, part of the material presented is based on the `Science with CTA' book \cite{CTAConsortium:2017dvg}. Predictions of CTA performance and science capabilities presented in this summary have made use of the CTA instrument response functions provided by the CTA Consortium and CTA Observatory.

The authors acknowledge the contributions of Rene Ong during the conceptional phase of the paper.  The authors thank both CTA's Speakers and Publication Office (SAPO) -- in particular Oleh Petruk --  and Gerd P\"uhlhofer for the careful review of the manuscript.

The authors gratefully acknowledge financial support of CTA from the agencies and organizations listed here: https://www.cta-observatory.org/consortium\_acknowledgments/